\documentclass[12pt,preprint,flushrt]{aastex}

\newcommand{\beq}{\begin{equation}}
\newcommand{\eeq}{\end{equation}}
\newcommand{\e}{$^{-1}$}
\newcommand{\calm}{{\cal M}}
\newcommand{\ee}{$^{-2}$}
\newcommand{\eee}{$^{-3}$}
\newcommand{\ellp}{\ell_{P}}
\newcommand{\ellpc}{\ell_{\rm pc}}

\newcommand{\pmax}{P_{\rm max}}
\newcommand{\pmin}{P_{\rm min}}

\newcommand{\pth}{P_{\rm th}}
\newcommand{\pthave}{P_{\rm th,\,ave}}
\newcommand{\sth}{\sigma_{\rm th}}
\newcommand{\ag}{\mbox{ \raisebox{-.4ex}{$\stackrel{\textstyle >}{\sim}$} }}
\newcommand{\al}{\mbox{ \raisebox{-.4ex}{$\stackrel{\textstyle <}{\sim}$} }}
\newcommand{\tmin}{T_{\rm (min)}}
\newcommand{\nmin}{n_{\rm min}}
\newcommand{\gs}{G_{0}^\prime}
\newcommand{\zd}{Z_{\rm d}^\prime}
\newcommand{\zg}{Z_{\rm g}^\prime}
\newcommand{\zio}{\zeta_{t}^\prime}
\newcommand{\phied}{\phi_{{\dot\epsilon}}}
\newcommand{\gturb}{\Gamma_{\rm turb}}
\newcommand{\muh}{\mu_{\rm H}}

\slugcomment{Accepted for Publication in the Astrophysical Journal}

\shortauthors{Wolfire et al.}
\shorttitle{Neutral Atomic Phases of the ISM}
\received{2002 July 2}
\begin{document}

\title{Neutral Atomic Phases of the ISM in the Galaxy}

\author{
Mark\ G. Wolfire}
\affil{Department of Astronomy, University of Maryland,
College Park, MD 20742-2421}
\email{mwolfire@astro.umd.edu}

\author{
Christopher F.\ McKee}
\affil{Physics Department and Astronomy Department,
University of California at Berkeley, Berkeley, CA 94720}
\email{cmckee@astron.berkeley.edu}

\author{
David Hollenbach}
\affil{NASA Ames Research Center, MS 245-3, Moffett Field,
 CA 94035}
\email{hollenbach@ism.arc.nasa.gov}
\and

\author{
A. G. G. M. Tielens}
\affil{Kapteyn Astronomical Institute, PO Box 800, 9700 AV Groningen,
The Netherlands}
\email{tielens@astro.rug.nl}

\begin{abstract}

        Much of the interstellar medium in disk galaxies is in the
form of neutral atomic hydrogen, H~I.  This gas can be in thermal
equilibrium at relatively low temperatures, $T\al 300$ K
(the cold neutral medium, or CNM) or at temperatures somewhat
less than $10^4$ K (the warm neutral medium, or WNM).
These two phases can coexist over a narrow range of pressures,
$\pmin\leq P\leq\pmax$.
We determine $\pmin$ and $\pmax$
in the plane of the Galaxy as
a function of Galactocentric radius $R$ using recent determinations
of the gas heating rate and the gas phase abundances of
interstellar gas.
We provide an analytic
approximation for $P_{\rm min}$ as a function of metallicity,
far-ultraviolet radiation field, and the ionization rate of
atomic hydrogen. Our analytic results show that the existence
of $\pmin$, or the possibility of a two-phase equilibrium, generally
requires that ${\rm H^+}$ exceed ${\rm C^+}$ in abundance
at $\pmin$. The abundance of ${\rm H^+}$ is set by EUV/soft X-ray
photoionization and by recombination with negatively charged PAHs.
In order to assess whether thermal or pressure equilibrium is
a realistic assumption,
we define a parameter $\Upsilon\equiv t_{\rm cool}/t_{\rm shk}$
where $t_{\rm cool}$ is the gas cooling time and $t_{\rm shk}$ is the
characteristic shock time or ``time between shocks in a turbulent
medium''. For $\Upsilon < 1$ gas has time to reach
thermal balance between supernovae induced shocks. We find that
this condition is satisfied
in the Galactic disk, and thus
the two-phase description of the interstellar H~I is approximately
valid even in the presence of interstellar turbulence.
Observationally, the mean density $\langle n_{\rm H\, I} \rangle$
is often better determined
than the local density, and we cast our results in terms of
$\langle n_{\rm H\, I} \rangle$ as well.
Over most of the disk of the Galaxy, the
H~I must be in two phases: the weight of the H~I in the
gravitational potential of the Galaxy is large enough to
generate thermal pressures exceeding $\pmin$, so that turbulent
pressure fluctuations can produce cold gas that is
thermally stable; and the mean density of the H~I is too
low for the gas to be {\em all} CNM. Our models predict
the presence of CNM gas to $R\simeq 16-18$ kpc, somewhat
farther than previous estimates.  An estimate for the
typical thermal pressure in the Galactic plane for
3~kpc$\al R\al 18$~kpc
is $\pth/k\simeq 1.4\times 10^4\exp(-R/5.5$~kpc) K cm\eee.
At the solar circle, this gives $\pth/k \simeq 3000$ K cm\eee.
We show that this pressure is consistent with the
${\rm C~I^*/C~I_{tot}}$ ratio observed by \cite{jen01}
and the CNM temperature found by \cite{hei02}.

We also examine the potential impact of turbulent heating on our results
and provide parameterized expressions for the heating rate as a
function of Galactic radius. Although the uncertainties are large,
our models predict that including turbulent heating does not
significantly change
our results and that thermal pressures remain above
$P_{\rm min}$ to $R\simeq 18$ kpc. 

\end{abstract}


\section{INTRODUCTION}

        The interstellar medium (ISM) has a complex
thermal and ionization structure.
Much of the neutral atomic gas is observed to
be either warm neutral medium (WNM)
with $T\sim 10^4$ K or cold neutral medium (CNM)
with $T\sim 100$ K \citep{kul87,dic90}.
Some of the warm gas is partially ionized,
the warm ionized medium (WIM), which
also has $T\sim 10^4$ K \citep{mck77,rey83,haf99}.
A small mass fraction of the gas is in the form of
hot ionized medium (HIM) with $T\sim 10^6$ K \citep{cox74,mck77}.
Inside the solar circle, about half the interstellar gas is
molecular \citep{sco87,bro88,bron00}.

        A significant simplification occurs if one focuses
on the neutral atomic gas,  the CNM and WNM.
Some decades ago, \cite{fie69} demonstrated that
the CNM and WNM could coexist in pressure equilibrium,
so that the neutral atomic gas could be considered to be
a two-phase medium.  They assumed that cosmic rays dominate
the heating, but it was subsequently realized that
UV starlight dominates the heating due to photoelectric
emission from the dust grains in the gas \citep{wat72}.
Using the photoelectric heating rates calculated by
\cite{bak94}, Wolfire et al.\ (1995; hereafter WHMTB)
investigated the thermal balance of the WNM and CNM phases
in the local ISM and showed
that the two-phase model is in good agreement
with a wide variety of data on the ISM in the solar vicinity.

What is the evidence for a two-phase medium
elsewhere in the Galaxy?
In the inner Galaxy, \cite{gar89} found that there is H~I emission
(which can originate from both CNM and WNM) at all velocities
allowed by Galactic rotation.  On the other hand, they found
that absorption (which originates only from CNM at their
sensitivity) is somewhat less pervasive, particularly within
2 kpc of the Galactic Center. \cite{liszt93} suggested that the 
H~I absorption in the inner Galaxy at $R >  2$ kpc is much higher
than that reported by \cite{gar89}. \cite{kol01}
recently repeated  the earlier H~I absorption  study and confirmed 
the presence of cold gas in the inner Galaxy with an absorption 
coefficient at $R=5$ kpc approximately 5 times higher than reported by
\cite{gar89}. 
In the outer Galaxy, the
presence of WNM is reasonably well
established \citep{kul87}, whereas that of a widely distributed
CNM is less so. Carilli, Dwarakanath, \& Goss (1998) have measured the
temperature of the WNM in absorption
features seen towards Cygnus A at distances of 9 kpc and 12 kpc
(and $z$ height of $\sim 1$ kpc) using the Westerbork radio telescope.
They find gas temperatures of $\sim 6000$ K and $\sim 4800$ K respectively,
which are consistent with the low pressure
and low UV field models of WHMTB for atomic gas above
the plane. Several high-velocity absorption components have
been observed in H~I \citep{col88} and Na~I \citep{semb94} that
arise from CNM clouds at Galactic radii $R\al 14$ kpc.
\cite{kol01} show H~I absorption to $R\al 17$ kpc.
Molecular clouds,  which presumably
form from the CNM phase, are traced to at least $R\sim 20$ kpc
\citep{wou89,hey01} with an extremely distant H~II region
and molecular cloud complex at $R=28$ kpc \citep{dig94}.
\cite{wou90} and \cite{bron00} show that the molecular surface density
can be fit by a radial exponential in the outer Galaxy to  $\sim 18$ kpc
\cite[see also][]{wil97}.   In their study of the Perseus arm,
\cite{hey98}, however, find that the molecular gas disk
is effectively truncated at $R\sim 13.5$ kpc.  These two results
could be consistent if molecular gas extends to greater
radii in directions other than those studied by \cite{hey98} or if 
isolated molecular clouds extend to distances much greater than 
the molecular surface density can be reliably measured from CO surveys.

Thus direct observations of cold H~I or molecular gas extend to
at least $R\sim 18$ kpc. Star formation provides an indirect test 
for the presence of CNM,
since the gas that forms stars presumably goes through the stage of being
cold H~I.  In the Galaxy, near-infrared sources, IRAS sources, 
and H~II regions
\citep{wou90,rud96, kor00,snell02} are seen out to $R\sim 17-20$ kpc, 
suggesting that CNM extends out to at least that distance. 

What can be learned from observations
of H~I in other galaxies?  Two-phase atomic gas has been observed
using 21 cm absorption techniques in several extragalactic systems
including M31 \citep{dic93, braun92}, M33 \citep{dic93}, the
LMC \citep{meb97}, and the SMC \citep{dic00}.  \cite{braun97}, using
the VLA,  examined the neutral hydrogen emission in 11 nearby spirals.
By associating high brightness, narrow emission components with cold gas,
he finds that the fraction of cold gas remains relatively constant
until the B band surface brightness falls to $\mu_B\sim 25$ mag
arcsec$^{-2}$, i.e., the $R_{25}$ radius.  At larger radii, the
fraction drops,
although in some systems more than
10\% of the H~I is in the form of cold gas out to
$(1.5-2) \times R_{25}$.
Since the  extinction-corrected radius of the Galaxy is
$R_{25} = 12.25$ kpc \citep{devau83},
cold gas in the Galaxy could extend out to
$1.5-2\times R_{25}$, or $R\sim 18.4-24.5$ kpc.
\cite{sel99} interpret the line width of H~I in the outer parts of
galaxies as being due to CNM that is being stirred by the
magnetorotational instability.
Evidence for recent star formation in
the outer disk of M31 is presented by \cite{cuil01},
who find a population of B stars at $1.7 \times R_{25}$, pointing
towards the presence of cold gas in the outer parts of galaxies.

The fact that the neutral atomic gas in the Galactic
ISM is in  two phases is a powerful
result, since two phases can coexist only
over a narrow range of pressure,
$\pmax > P >\pmin$ with $\pmax\al 3\pmin$
\citep[see \S~\ref{sec:Results}]{fie69}.
It is thus possible to estimate the thermal pressure
of the H~I with reasonable accuracy---when
it is in two phases---from knowing the
gas phase abundances, the dust properties, and the intensity of the
radiation field.
We used this property to study gas in the Galactic halo
and constrain the properties of the High Velocity Clouds
in \cite{wol95b}.  \cite{cor88} have argued that achieving
the condition for a two-phase equilibrium  is a necessary step in
initiating star formation in young galaxies,
while \citet{par88,par89} has suggested that
two-phase equilibria play a key role in regulating
the rate of star formation in disk galaxies.

    The primary goal of this paper is to predict the average thermal
pressure of the ISM as a function of position in the Galaxy using the
two-phase criteria. To do this,  we shall extend the models of WHMTB
to the inner and outer Galaxy.  In light of the observational 
evidence that cold gas exists in the
outer Galaxy, we shall carry out our model calculations for Galactic radii
between 3 kpc and 18 kpc. 
Knowing the thermal pressure allows
one to predict the intensities of the dominant cooling lines
of the gas, such as that of C~II 158 \micron, and examine
the heating and cooling routes which determine the energy budget.
Locally, the thermal pressure in the ISM
is measured through ultraviolet absorption line
studies \citep{jen83,jen01}.   In the near future, telescopes
such as ALMA, SOFIA, and Herschel will provide additional measurements
of the thermal pressure and dominant cooling lines
throughout the Galaxy and in other galaxies. These will test
our model for the gas thermal balance and check the importance
of thermal instability.

We also can calculate whether the ISM could exist as
pure WNM at various positions in the Galaxy
by comparing the weight of the H~I layer with $\pmax$.
The problem of determining the phase structure of the
H~I in the outer Galaxy has been considered previously by
\cite{elm94}, who find a transition to mainly WNM at $R\ag 12-14$
kpc. Our results are compared with theirs in \S~\ref{sub:prev}.

Although the focus of this paper is on the determination
of the thermal pressure in Galactic H~I, it is
well known that the thermal pressure is only a small
part of the total pressure in the gas; in particular,
the turbulent pressure is considerably greater than
the thermal pressure \citep{bou90}.  In \S~\ref{sec:tur}, we discuss
the relation between the turbulent pressure and the
thermal pressure and determine the conditions under
which it makes sense to consider multi-phase equilibria
in a turbulent medium.  We also discuss in Appendix~\ref{appen:turbheating}
the dissipation of turbulent energy in the ISM and its potential 
effects on  our results.
In \S~\ref{sec:gasdust}
we discuss the distribution of gas and dust
in the Galaxy between these radii,
together with the abundances we have adopted.  The heating
and ionization in the gas are governed by energetic
photons and particles, which are discussed in 
\S~\ref{EnergeticPhotons}. The thermal and
chemical processes in our model are slightly modified from those
discussed by WHMTB; the differences are briefly described
in \S~\ref{sec:thermal}.
The results of our calculations are presented in \S~\ref{sec:Results}.
We then construct a simple analytic
model of a two-phase equilibrium that shows how the
properties of the equilibrium scale with the input parameters
(\S~\ref{sec:toy}).
We compare our model with local and extragalactic observations in
\S~\ref{sec:comp} and
discuss our results in \S~\ref{sec:disc}.

\section{Turbulent Pressure in the Multiphase Interstellar Medium}
\label{sec:tur}

        The ISM is observed to be highly turbulent \citep{lar79}.
When averaged over the vertical structure of the galactic disk,
the turbulent velocity dispersion exceeds the thermal
velocity dispersion, and correspondingly the
turbulent pressure exceeds the thermal pressure \citep{bou90}.
Theoretical arguments 
(Spitzer 1968, 1978; McKee \& Ostriker 1977) 
and numerical simulations
\citep{bre99,kor99} show that these turbulent motions
can be accounted for by the injection of energy by supernovae.

        Insofar as the disk of the Galaxy is in approximate
hydrostatic equilibrium, the total pressure in
the midplane of the disk must balance the weight of the
material above it \citep{par69,bou90}, and the turbulent
motions are an important contributor to the total pressure.
In the solar neighborhood, the total pressure at the midplane is about
$P/k\simeq 2.8\times 10^4$ K cm\eee\ \citep{bou90},
about 10 times greater than the median thermal pressure
of $\sim 3000$ K cm\eee\
\citep{jen01}.\footnote{The mean pressure quoted
by \cite{jen01} is  $P/k = 2240$ K cm$^{-3}$ based on
data from the Space Telescope Imaging Spectrograph. In
\S~\ref{sec:comp} 
we find that corrections for gas temperature
and atomic constants raise the pressure derived from the observations
to $\sim 3000$ K cm$^{-3}$.}
A similar situation
occurs in molecular clouds, where the total pressure
can also be about an order of magnitude greater than the
thermal pressure \cite[depending on scale---see][]{lar81}.
If the thermal pressure is such a small fraction of
the total, why are arguments based on thermal pressure
equilibrium relevant in the interstellar medium?

        The answer is that {\it the turbulent motions
determine the temporal and spatial structure of the
thermal pressure.}  A key feature of turbulence is that
the motions are spatially correlated, so that
the rms velocity increases with scale---i.e., there
is a relation between the line width and the size of
a region,
\beq
\sigma=\sigma(1)\ellpc^q,
\eeq
where $\sigma(1)$ is the 1D turbulent velocity on
the scale of 1 pc and $\ellpc$ is the scale in units of parsecs.
Larson's (1979) data on H~I clouds give
$\sigma(1)=0.64$~km~s\e\ and $q=0.37$
over the range $1\al\ellpc\al 1000$.
(Larson 1981 subsequently found a similar
line width--size relation for molecular clouds.)
We can estimate $\sigma(1)$ from a more homogeneous data
set by using the recent H~I study by \cite{hei02}.
They find that the mass-weighted velocity dispersion is
7.1 km s\e\ for CNM clouds and 11.4 km s\e\ for WNM clouds.
The typical line of sight in this survey is at a Galactic
latitude slightly greater than 30\arcdeg, so the typical
path length is about twice the half-height of the disk---i.e.,
the full disk thickness.  From Dickey \& Lockman's (1990)
model for the vertical distribution of the H~I,
we infer that the FWHM of the CNM is 212 pc,
whereas that for the WNM is 530 pc.
We assume that the outer scale of the turbulence is greater
than the disk thickness, so that the power-law behavior
extends over this length scale; this assumption is consistent with
the results of \cite{lar79}, who did not find a break
in the power-law behavior out to 1 kpc in the Galaxy,
and with the results of \cite{laz00},
who did not find a break between 40 pc and 4 kpc in the data
for the Small Magellanic Cloud.
We shall set
$q=1/3$, as expected for subsonic
turbulence. On
scales large enough for the turbulence to be supersonic,
$q$ is expected 
to be somewhat larger: \cite{lar79} suggested that
$q$ would approach $\frac 12$, whereas \cite{bold02}
finds $q\simeq 0.37-0.38$ (where $\sigma$ is interpreted
as an rms velocity).
For the CNM, we then find $\sigma(1)=1.2$ km~s\e,
whereas for the WNM we find $\sigma(1)=1.4$ km~s\e.
The value of $q$ is likely to be closer to the assumed
1/3 on the smaller length scales associated with the CNM,
so we shall adopt $\sigma(1)=1.2$ km s\e\ for the H~I
in the solar neighborhood.  Although this value was
determined from the velocity dispersion of different H~I clouds, 
we shall assume that it applies within individual
H~I clouds as well.
We note that this 
assumption is consistent with
the results of \cite{lin96}, who found that the velocity
dispersion within the H~I toward 
the nearby star $\alpha$ Cen is about 1.2 km s\e,
corresponding to $\sigma(1)=1.1$ km s\e.
The use of a single turbulent velocity law in both
the CNM and the WNM is 
undoubtedly an oversimplification; in particular,
there is a range of scales over which the turbulence
is supersonic in the CNM and subsonic in the WNM, and
$q$ might be expected to have different values within
individual clouds of CNM and WNM over this range.

    Let $\ellp$ be the scale on which the turbulent
pressure begins to dominate the thermal pressure;
on scales less than $\ellp$ the gas typically will be
in approximate thermal pressure equilibrium.
To estimate $\ellp$,
we equate the thermal velocity dispersion, $\sth=0.80 T_2^{1/2}$ km s\e,
with the turbulent velocity dispersion, so that
\beq
\ellp=\left[\frac{\sth}{\sigma(1)}\right]^{1/q}~{\rm pc}\;
        \rightarrow\; 0.3T_2^{3/2}~{\rm pc},
\eeq
where $T_2\equiv T/(100$~K) and
where the numerical evaluation is for our fiducial case.
Because $\ellp$ depends on the cube of the uncertain
quantity $\sigma(1)$, the numerical value of $\ellp$ is
quite uncertain.  Bearing this in mind, we find that
the CNM should be in approximate thermal pressure equilibrium on
scales $\ell<\ellp$(CNM)$\sim$ 0.3 pc.
For the WNM, with a typical temperature of
about 8000 K, we find $\ellp(\rm WNM)\simeq 215$ pc.
This is somewhat larger than the size of the turbulent cells in the warm
gas, 60 pc, found in the 3D
numerical simulations of the ISM by \cite{kor99}.

In a multiphase medium, the CNM is embedded
in warm gas, either predominantly neutral (WNM) or ionized (WIM),
which in turn may be embedded in hot gas,
the HIM.  Since the sound speed in the WIM exceeds that
in the WNM, $\ellp$(WIM) is somewhat larger than
$\ellp$(WNM); both are much larger than
the typical size of a CNM cloud, which is
$\sim 1-2$ pc
\citep{mck77}.  As a result, the CNM clouds should typically
be embedded in a medium that is spatially isobaric.
The surface layers of the cloud [i.e., those
layers within a distance $\ellp$(CNM) of
the surface] should typically have the same thermal pressure
as the ambient warm medium, whereas the inner regions
of the cloud should have the time-averaged value
of the ambient thermal pressure. \cite{elm97}
shows that in a 1D simulation of interacting magnetized clouds the
gas maintains approximate phase equilibrium with the intercloud
medium.

This description of the relative roles of thermal and turbulent pressures
in the ISM is consistent with the three-phase model of the ISM by
\cite{mck77}. In their model, the supernova-generated HIM is pervasive
and sets the thermal pressure in the embedded clouds of CNM and WNM.
The ISM is viewed as the superposition of many supernova remnants; as
a result there are substantial fluctuations in the thermal pressure, in
qualitative agreement with observations \citep{jen83,jen01}. The turbulent
motions provided by supernovae are consistent with observations of
H~I velocities in the Galactic plane. Turbulent pressure is the single
largest contributor to the support of the ISM in the gravitational field
of the disk \citep{bou90}, and \cite{mck90} has argued that the turbulent
motions that produce this pressure are produced by supernovae. \cite{slav93}
and \cite{cox95} have argued that magnetic fields in the ISM limit the
size of supernova remnants so that they occupy only a small fraction of
the volume. Such a model cannot account for the observed level of
turbulence, however.

Under what conditions is the turbulence in the ISM weak enough
that a two-phase description of the H~I is valid?  We first consider
the cooling time $t_{\rm cool}$ for
an element of gas to return to thermal equilibrium after
a significant perturbation. From our numerical results
presented in \S~\ref{sec:Results}, we find the cooling rate
at a constant thermal pressure of $P_{\rm th}/k = 3000$ K ${\rm cm^{-3}}$
is given by
\beq
   \Lambda = 5.7\times 10^{-26} T_4^{0.8}\; {\rm erg~cm^{3}~s^{-1}},
\label{eq:Lambda}
\eeq
where $T_4\equiv T/(10^4\,{\rm K})$ and where the fit
is accurate to within a factor $1.35$ for temperatures between  
$T = 55~{\rm K}$ and $T = 8500~{\rm K}$. 
With this cooling rate, $t_{\rm cool}$ is given by
\beq
t_{\rm cool}=\left(\frac{\frac 52 1.1nkT}{n^2\Lambda}\right)
       \simeq 7.7\times 10^6 T_4^{1.2}
       \left(\frac{\pth/k}{3000\;{\rm K~cm^{-3}}}\right)^{-0.8}
       ~~~~{\rm yr},
\label{eq:tcool}
\eeq
where $n$ is the hydrogen nucleus density 
($n = n_{\rm H\, I} + n_{\rm H^+} + 2n_{\rm H_2}$). The fit to the 
pressure term results in an overall accuracy 
within a factor $1.5$ for temperatures between 
$T = 55$ K and $T = 8000$ K and thermal
pressures between $P_{\rm th}/k = 1000$ K ${\rm cm^{-3}}$ 
and $P_{\rm th}/k = 6000$ K ${\rm cm^{-3}}$. 
In order for a two-phase description to be valid, this
time must be less than the characteristic time $t_{\rm shk}$
for a shock to induce motions of order $\sth$ or, 
equivalently (for an isothermal shock), to double the 
pressure.\footnote{In fiiting equations (\ref{eq:Lambda}) and (\ref{eq:tcool})
  we have assumed ionization equilibrium holds although the recombination
  time in WNM gas is found to be comparable to, or up to 2 times
  greater than the cooling time. Since the cooling time in the WNM
  is proportional to  $1/n_e$ for Ly$\alpha$ cooling, and proportional
  to $1/n_e^{0.6}$ for electrons recombining on to positively
  charged grains (see \S~\ref{sec:thermal}), any lingering
   ionization after the passage of a shock will tend to decrease the
   cooling  time. Thus, our fits provide a maximum cooling time in WNM
   gas.}
One method of estimating $t_{\rm shk}$ is based 
on observations of interstellar turbulence.
In order to induce motions of $\sth$, the velocity 
of an isothermal shock must
be $\surd 2\sth$.
The time interval between shocks in a turbulent medium is then
$t_{\rm shk}\sim\ellp/\surd 2 \sth$, where,
as discussed above, the value of $\ellp$ is quite uncertain.
Defining
a dimensionless parameter $\Upsilon$ that measures the strength of
the turbulence, we obtain
\beq
\Upsilon\equiv\frac{t_{\rm cool}}{t_{\rm shk}}
        \sim 0.3 T_4^{0.2}
       \left(\frac{\pth/k}{3000\;{\rm K~cm^{-3}}}\right)^{-0.8}
        \left[\frac{\sigma(1)}{1.2\;{\rm km~s^{-1}}}\right]^3.
\label{eq:upsilon}
\eeq
So long as $\Upsilon\al 1$, a two-phase description of the H~I
is approximately valid, since the gas has time to reach
thermal balance between shocks.  This condition is
well satisfied for the CNM ($\Upsilon \simeq 0.1$), but only marginally
satisfied for the WNM ($\Upsilon \simeq 0.3$).

     We can also estimate $t_{\rm shk}$ and $\Upsilon$ analytically
under the assumption that the shocks are produced by supernova
remnants, although this estimate is necessarily 
uncertain.  Let $M_{\rm shk}(>v_s)$ be the mass of
interstellar gas per supernova that suffers
a shock with a shock velocity greater than $v_s$.  
Since the shock must have a velocity $\surd 2\sth$ in order
to induce motions of $\sth$, we have
\beq
t_{\rm shk}\simeq\frac{\Sigma_{\rm WNM}}{\dot\varsigma_{\rm SN,\, eff}
       M_{\rm shk}(>\surd 2 \sth)},
\label{eq:tshk}
\eeq
where $\Sigma_{\rm WNM}$ is the surface density of WNM gas,
and $\dot\varsigma_{\rm SN,\, eff}$ is the effective supernova
rate per unit area.  Allowing for the clustering of supernovae
in associations, \cite{mck89} estimated that the effective
supernova rate is only about 36\% of the actual rate.  For
a two-phase ISM, he also estimated $M_{\rm shk}=2460 n_{\rm WNM}^{-0.1}
(v_s/100$~km~s\e$)^{-9/7}\;M_\odot$,
where $n_{\rm WNM}$ is the H nucleus density in the WNM and
where we have assumed that a typical
supernova remnant has an energy of $10^{51}$ erg.  
(The results for a three-phase ISM 
with a substantial WNM filling factor should not differ qualitatively
from the two-phase results.)
At the solar circle,
the supernova rate per unit area is about $3.8\times 10^{-11}$ SN
pc\ee\ yr\e\ \citep{mck97}, so
this gives $t_{\rm shk}\simeq 5.3\times 10^6$ yr for a density
$n_{\rm WNM}=0.3$ cm\eee.  The corresponding value of
the turbulence parameter for the WNM is 
$\Upsilon\simeq 0.9$, somewhat larger than the value estimated above
(and indicative of the uncertainty in this parameter).
Both estimates of $\Upsilon$ suggest that
the WNM will often exhibit non-equilibrium temperatures.
\cite{hei02} find that about half the WNM is at temperatures
below the minimum equilibrium value, which is qualitatively consistent
with our estimate of $\Upsilon$.

     Recent papers by  \cite{mac01}, \cite{vaz00}, and \cite{vaz02}
argue that thermal instability is of less importance in 
determining the distribution of phases than is turbulent dynamics followed 
by cooling towards thermal equilibrium. 
To some extent, the difference between these conclusions and
those in the present paper are a matter of perspective:
these authors focus on the time-dependent aspects
of a turbulent medium, whereas we emphasize the
utility of the equilibrium aspects. (We also note that 
\cite{mac01} did not include heating sources in their calculations.)
In terms of the turbulence parameter $\Upsilon$
introduced above, we choose to approximate
the case $\Upsilon\sim 1$ for the WNM with the
two-phase $\Upsilon\ll 1$ results, whereas they
prefer to emphasize the case $\Upsilon\gg 1$. Also,
in contrast to Mac Low et al., who concentrate on
evaluating the thermal pressure distribution in the solar
neighborhood, the focus of our work is on determining
the mean (volume-averaged) thermal pressure in the ISM throughout
the Galaxy, $\pthave$.
Unless $\pthave$ is in, or close to, the range $\pmin\al\pthave\al
\pmax$, the gas will be almost all CNM or all WNM,
and the complex interplay between cold and warm gas
seen by these authors will not occur.  By determining
$\pmin$ and $\pmax$, we can determine the range of
pressures in which cold and warm gas can coexist;
furthermore, we shall argue that most of the Galactic
disk {\it must} have a thermal pressure such that this
is the case.

\section{Gas and Dust in the Milky Way}
\label{sec:gasdust}

\subsection{Distribution of H~I}\label{HIDistribution}
\label{sub:HI}

We require the azimuthally averaged
H~I surface density distribution, $\Sigma_{\rm H\,I}(R)$,
the half width to half maximum of the H~I emission above the plane,
$H_z^{\rm H\,I}(R)$, and the mean H~I density in the Galactic plane,
$\langle n_{\rm H\,I}(R)\rangle \propto
\Sigma_{\rm H\, I}(R)/H_z^{\rm H\, I}(R)$.
These will be used to trace the opacity in order to
determine the distribution of energetic
photons and particles in the Galaxy as discussed in
\S~\ref{EnergeticPhotons}.

Several surveys have been conducted of the Galactic
H~I distribution \cite[e.g.,][]{ww73,bur85,kerr86,sta92,hart97}.
\cite{loc02} discusses several
limitations in interpreting the observations including:
velocity crowding which renders $\Sigma_{\rm H\, I}(R)$ highly  sensitive
to the adopted rotation curve, optical depth effects which introduce
uncertainty in the volume and surface densities, and a low
dynamic range in emission, which makes small variations in the brightness
temperature difficult to measure. Thus, the distribution in
$\Sigma_{\rm H\, I}(R)$ is not well determined. In addition, the derived
scale height $H_z^{\rm H\, I}$ may depend on the method of analysis
which could be preferentially sensitive to the CNM or WNM  component.
With these caveats in mind, we use the published H~I data to derive
our distributions.

The distribution of H~I in the outer Galaxy was presented by
\cite{wou90} who
combined the northern hemisphere \citep{kerr86,bur85} and
southern hemisphere \citep{kerr86} data.
They provide plots of
the H~I surface  density, $\Sigma_{\rm H\,I}(R)$, from the 2nd and 3rd
Galactic quadrants (northern and southern data) as well as the
average surface density and find that the
average radial distribution at $R > 13$ kpc is well
fit by an exponential,
$\Sigma_{\rm H\,I}(R)\propto \exp(-R/H_R^{\rm H\,I})$, with
$H_R^{\rm H\,I}=4$ kpc.
A relatively flat rotation curve was used,
similar to that derived by  \cite{brand93}.

\cite{dam93} and \cite{loc88} showed
that minor ($\sim 2$\%) differences in the
rotation curve can greatly affect the surface density
derived from the data with differences in the surface density amounting
to $\sim 50$\%.  Based on the northern hemisphere
data, \cite{dam93} found a peak in surface density near 12 kpc
for a flat rotation curve, but a nearly constant surface density out
to 17 kpc when using the slightly  rising rotation curve of \cite{kul82}.
\cite{wou90} attributed the peak in the H~I surface density to gas
associated with the
Perseus arm and show that the average of the northern and southern data
partially smoothes the 12 kpc peak.

In addition to the surface density enhancement at 12 kpc,
the \cite{wou90} plots of the outer Galaxy show a dramatic rise
as $R$ decreases towards the solar circle,
reaching $\Sigma_{\rm H\, I} = 8.6$ $M_\odot$ pc$^{-2}$
at $R=9$ kpc. (Note that in our notation $\Sigma_{\rm H\,I}$ does not
include the mass associated with helium or the metals.)
This surface density is inconsistent with the value
of $\Sigma_{\rm H\,I}(R_0)=5$
$M_\odot$ pc$^{-2}$ at the solar circle determined by
\cite{dic90}. Furthermore, the average  surface density in the
outer Galaxy does not seem to join smoothly onto the inner Galaxy
where $\Sigma_{\rm H\,I}$ is also  5 $M_\odot$ pc$^{-2}$.
The inferred pile up of H~I towards the edge of the solar circle may be an
artifact of a flat rotation curve with strictly circular velocities,
and non-circular motions may alleviate this problem \citep{blitz91}.
An example of the H~I surface density retaining the peak at $R_0$ is
shown in \cite{oll98}, who also argue for  $R_0 \sim 7.1$ kpc.
In our post-Copernican world, we believe that a narrow density
enhancement centered at $R = 8.5$ kpc is unrealistic, and have smoothly
joined the \cite{wou90} surface densities at $R > 10$ kpc onto
the distribution for $R < 8.5$ kpc.
We are mainly concerned with
the disk properties in the inner and outer Galaxy, and thus errors less
than a factor 2 in a thin region between $8.5$ and $10$ kpc are not critical.

The observational evidence suggests that the H~I surface density
is constant in the inner Galaxy between 8.5 kpc and $\sim 4$ kpc
and then drops by a factor $\sim 3$ by 1.5 kpc
\citep{dic90,lis92,dam93}.
Our piecewise analytic fit to the H~I surface density data
is shown in Figure~\ref{HIsurf} and is given by
\begin{eqnarray}
\Sigma_{\rm H\,I}(R) & = & 1.4 R_k - 0.6 \,\,\,\,\,
           M_\odot \, {\rm pc}^{-2} \,\,\,\,\,
                   \,\,\,\,\,~~~~~~~~~~~~~~~~~~  (3\le  R_k < 4),
\label{eq:h1surf1} \\
   & = & 5 \,\,\,\,\,
           M_\odot \, {\rm pc}^{-2} \,\,\,\,\,
           \,\,\,\,\, ~~~~~~~~~~~~~~~~~~~~~~~~~~~~~~~~ (4\le  R_k < 8.5),\\
   & = & -1.12 + [6.12\times (R_k/8.5)] \,\,\,\,\,
             M_\odot \, {\rm pc}^{-2}\,\,\,\,\,
         \,\,\,\,\,\,  (8.5 \le R_k < 13),\\
    & = & 8.24 \,{e}^{-(R_k-13)/4}\,\,\,\,\,
              M_\odot \, {\rm pc}^{-2}  \,\,\,\,\,
         \,\,\,\,\  ~~~~~~~~~~~~~(13 \le R_k < 24),
\label{eq:h1surf4}
\end{eqnarray}
where $R_k\equiv R/(1$~kpc) and where
we use the conversion 1 $M_\odot$ pc$^{-2} = 1.25\times 10^{20}$
H~I cm$^{-2}$. \cite{dam93} found that using the H~I surface density given
in \cite{lis92}, the H~I mass at $R < R_0$ (excluding the Galactic
center) is $M_{\rm H\, I}\sim 1.7\times 10^{9}$ $M_\odot$ while
our distribution gives a mass $M_{\rm H\, I}\sim 1.0\times 10^9$ $M_\odot$
in this same range. A simple estimate of the maximum H~I mass
is $M_{\rm H\, I} < \pi R^2 \Sigma_{\rm H\, I}(R_0) \al 1.2 \times 10^9$
$M_\odot$ and our lower estimate seems reasonable. We find an H~I mass
of $M_{\rm H\, I}\sim 4.3\times 10^9$ $M_\odot$ in the range $8.5 < R < 18$ kpc
consistent with \cite{wou90} who report
$M_{\rm H\, I} = 5.3\times 10^{9}$ $M_\odot$  between $R=8.5$ kpc and 24 kpc.

The H~I half height, $H_z^{\rm H\, I}(R)$, at $R > R_0$
is taken from the data of \cite{wou90}.
As for
the surface density distribution, we assume that the data
at 10 kpc should be smoothly joined to that at 8.5 kpc. With this assumption
we find that the variation in the height over the entire
range from 8.5 kpc to 18 kpc can be reasonably well fit by an
exponential of scale length $\sim 6.7$ kpc.
From \cite{dic90} we assume the half height at
$R < R_0$ is approximately constant
and given by  $H_z^{\rm H\, I}(R < R_0 ) = 115$ pc.
We note that \cite{mal95} finds a constant half height of
$\sim 118$ pc  in the inner ($R < 5.1$ kpc) Galaxy which rises
to $\sim 260$ pc at the solar circle; a height which
is nearly identical to that of the WNM ($\sim 265$ pc)
found by \cite{dic90}.  Note that \cite{mal95} fit both the
midplane height and the height of the H~I above the midplane and thus
possible affects due to disk corrugation are removed from her results.
The dramatic rise in height at the solar circle might
be partly  attributed to the correction for saturated
H~I emission by \cite{dic90} which was not
accounted for by Malhotra. Including this correction
increases the intensity of the CNM component and
tends to weight the height more towards the CNM height
than the WNM height.

Figure~\ref{HIzfig} shows the H~I height distribution
with an analytic fit given by
\begin{eqnarray}
H_z^{\rm H\, I}(R) & = & 115 \,\,\,\,\, {\rm pc} \,\,\,\,\,
   \,\,\,\,\,~~~~~~~~~~~~~~~~  (3\le R_k < 8.5),\\
 & = & 115\,e^{(R_k-8.5)/6.7} \,\,\,\,\, {\rm pc}\,\,\,\,\,
    \,\,\,\,\,\, (8.5\le R_k \le 18).
\label{eq:hiz}
\end{eqnarray}

The H~I surface density and height are used to scale the
mean midplane density $\langle n_{\rm H\,I}(R)\rangle
\propto \Sigma_{\rm H\,I} (R)/H_z^{\rm H\,I}(R)$. \cite{lis92} notes
that with a density of 
$\langle n_{\rm H\,I} \rangle\sim 0.4$ cm$^{-3}$ and a
single temperature
$T_{\rm spin} = 135$ K, both the emission and absorption characteristics
of the H~I in the plane can be simultaneously modeled.
On the other hand, he points out that a higher density is
required to match the observed H~I surface density of \cite{dic90}
who find  $\langle n_{\rm H\,I} \rangle\simeq 0.57$
cm$^{-3}$. With the mean density of hydrogen nuclei
as derived from extinction studies
$n_{\rm H\,I} + 2n_{{\rm H_2}}= 1.15$ H-nuclei
cm$^{-3}$ \citep{boh78}, the \cite{dic90} mean H~I density
requires a molecular density of 2$n_{{\rm H_2}}\sim 0.58$ cm$^{-3}$,
roughly consistent with the observations of \cite{bro88} who find
2$n_{{\rm H_2}}\sim 0.5$ cm$^{-3}$.  We adopt a local mean 
H~I density of  $\langle n_{\rm H\,I} \rangle =0.57$ cm$^{-3}$ 
and show the run of mean
midplane H~I density with $R$ in Figure \ref{fig:HImean}.

Our H~I distribution differs from that adopted by \cite{fer98} for her
global models of the Galactic ISM. Ferri\`ere  adopted a constant surface
density $\Sigma_{\rm H\,I}(R)  = 5$ $M_\odot \, {\rm pc}^{-2}$
to $R=20$ kpc with a scale height that increases linearly
with R at $R > 8.5$ kpc and a mean midplane density that decreases
as $1/R$. Our surface density is thus equal to, or higher than \cite{fer98}
out to a radius of $R=15$ kpc and then drops to lower values. Our
midplane density is slightly higher until $R=14$ kpc and then
drops to a factor of $\sim 4$ lower at $R=18$ kpc.

\subsection{Distribution of ${\rm H_2}$}
\label{sub:H2}

The distribution of the ${\rm H_2}$ surface density
in the Galaxy is required primarily as a cosmic ray opacity source.
We shall also use the distribution of ${\rm H_2}$ height to
estimate the height of OB stars in the Galactic plane.
We use a surface density distribution from \cite{bron00}
that is a Gaussian in
Galactic radius in the inner Galaxy and a radial exponential in the
outer Galaxy. The distribution has a peak value of 4.5 $M_\odot$ pc$^{-2}$
centered on $R=4.85$ kpc, a full width at half maximum extent
equal to 4.42 kpc, and a radial exponential of scale length
$H_R^{\rm H_2}=2.89$ kpc beyond
$R \ge 6.97$ kpc. The ${\rm H_2}$ surface density at the solar circle is
$\Sigma_{\rm H_2}(R_0 ) = 1.4$  $M_\odot$ pc$^{-2}$. These values
are derived from an average of the data from both
northern and southern Galactic quadrants. \cite{bron00} find
an H$_2$ mass of $M({\rm H_2})=6.1\times 10^8$ $M_\odot$ for
$1.7 < R < 8.5$ kpc,
and $M({\rm H_2})=2\times 10^8$ $M_\odot$ for $8.5 < R < 14.5$ kpc.
By comparison,
\cite{wil97} adopted a simpler, purely exponential distribution for the
${\rm H}_2$ and found $M({\rm H_2}) = 7.1\times 10^8$ $M_\odot$ for
$R$ between 1.7 kpc and 8.5 kpc (their quoted value of
$1.0\times 10^9$ $M_\odot$
included the He mass).
Figure~\ref{HIsurf} shows the ${\rm H_2}$ surface density
distribution.

In the inner Galaxy, \cite{bron00} find the ${\rm H_2}$ height to
half maximum density is roughly constant
at $H^{\rm H_2}_z(R \le R_0) \simeq 59$ pc. In the outer Galaxy we
simply scale the ${\rm H_2}$ height by the ${\rm H~I}$ height of \cite{wou90}
(eq.\ [\ref{eq:hiz}]).

\subsection{Distribution of Dust and Metals}
\label{sub:Metallicity}

The photoelectric heating rate depends on the abundance of the smallest
dust particles, which may primarily consist of polycyclic aromatic
hydrocarbons or ``PAHs'',
while the radiative cooling rate depends mainly on
the gas phase abundances of carbon and oxygen.
The photoelectric heating rate has been studied in detail by \cite{bak94}
 using a realistic model for the PAH and grain size
distribution.  For the grains (i.e., particles with $a>$ 15 \AA), they
adopted the MRN grain size distribution for spherical grains (i.e.,
$n[a]da\propto a^{-3.5} da$; Mathis et al.\  1977).  The PAH
molecules ($a<$ 15 \AA) are assumed to be small disks.  Their
distribution is given by
$n(N_{\rm C})dN_{\rm C} = 1.15\times 10^{-5} N_{\rm C}^{-2.25} dN_{\rm C}$ 
(with
$N_{\rm C}$ the number of C-atoms in the molecule), which ensures 
that the total
volume in disks between 12 and 275 C-atoms is equal to that of the MRN
spherical grain size distribution between 3 and 15 \AA.  WHMTB have shown
that this heating rate can adequately heat the local WNM and CNM phases.
The characteristics of the PAH distribution
adopted by WHMTB correspond to
a total fraction of C in the form of PAHs of $14\times 10^{-6}$ relative
to H.  Recent analysis of ISO observations of the PAH emission in the
galaxy conclude that this fraction is actually $22\times 10^{-6}$
relative to H \citep{tie99}.
We note that this corresponds to a total PAH abundance,
$n_{\rm PAH}/n$,
of $6\times 10^{-7}$ as compared to $4\times 10^{-7}$ in
WHMTB.  We have adopted the former, higher value, which enters in the
photoelectric heating rate and the chemical
network.  As will be discussed in \S~\ref{sec:comp}, these new values still
reproduce the observed [C~II] 158 $\mu$m cooling rates and the thermal
pressures in the local ISM.

We use a local gas phase abundance of metals consistent with the
{\em Hubble Space Telescope} observations of \cite{sof97}, \cite{car96},
and \cite{mey98}
with  $n_{\rm C}/n=1.4 \times 10^{-4}$ for carbon and
$n_{\rm O}/n=3.2\times 10^{-4}$ for oxygen. The gas phase abundances
of these elements are seen to be independent of the physical conditions
in the diffuse medium, showing relatively constant values for fractional
${\rm H_2}$ abundances between
$\log n_{{\rm H_2}}/[n_{{\rm H_2}}+n_{\rm H\,I}]=-5.0$ and $-0.2$.
Although of relatively minor importance in these calculations, we
also include Si, S, Mg, and Fe and, with the exception of S,
increase their depletions with gas density as in \cite{jen87} and
\cite{van88} (see also WHMTB).

The optical and infrared line diagnostics seem to be converging on
a value for the oxygen abundance gradient in the Galaxy that is roughly
consistent with the early findings of \cite{sh83}, which were based on
radio observations of the temperatures of H~II regions
\cite[e.g.,][]{torr77,si95,af96,af97,gumm98,roll00}.  We will use oxygen
as our basis for the metallicity gradient $Z(R)$ and assume that both the
elemental abundances and the gas phase abundances of all elements
scale similarly with $Z(R)$. Since both carbon and oxygen are primary
elements,  the elemental carbon abundance gradient is seen to
closely follow the oxygen gradient \citep{roll00}. We further assume
that the grain size distribution does not
vary significantly with Galactic radius, but that the total dust abundance
scales with $Z(R)$.
\cite{af97} note that they do not observe a jump in the nitrogen to
oxygen abundance ratio
at $R \le 6$ kpc as suggested by \cite{si95}.
In addition, \cite{roll00} find that their data can be fit with
a single slope of
$d[{\rm O/H}]/dR=-0.067\pm 0.008$ dex kpc$^{-1}$
without discontinuities \citep{twa97} or a flattening
of the gradient in the outer Galaxy as suggested by \cite{fi91}.
We take a constant gradient of $d[{\rm O/H}]/dR=-0.07$ dex kpc$^{-1}$
in the range 3 kpc $\le R \le 18$ kpc. This slope corresponds to a
radial exponential scale length of $H_R^Z = 6.2$ kpc.

\subsection{Distribution of Ionized Gas}
\label{sub:wim}

        Although the ionized gas in the Galaxy does not enter
into our analysis directly, we include a brief discussion of
it for completeness.
Most of the ionized gas in the Galaxy is
produced by photoionization.  \cite{tay93} identified three components of
ionized gas:
a diffuse component that extends out to $R\ga 20$ kpc,
an annular component in the inner Galaxy centered at 3.7 kpc,
and a component
associated with spiral arms. \cite{hei96} suggested that
there are actually only two separate
components, since the annular component is most likely
due to spiral arms in the inner Galaxy.

    \cite{tay93} infer that the diffuse component has a mean
electron density in the Galactic plane at the solar circle
of 0.019 cm\eee, a vertical
scale height of 0.88 kpc, and a radial distribution proportional
to ${\rm sech}^2(R_k/20)$.  Assuming that the ionization of He is similar
to that of H \citep{slav00}, this corresponds to a surface
density of diffuse ionized gas of $\Sigma_{\rm H\,II,\,diff}
=0.89\,{\rm sech}^2(R_k/20)\; M_\odot$~pc\ee\ (not including He); at the 
solar
circle, this is $0.75\;M_\odot$~pc\ee.  In the \cite{wol95b} model 
of the Galactic halo,
the surface density of H in the collisionally ionized, hot interstellar
medium (HIM) at the solar circle is $0.26\; M_\odot$~pc\ee, which accounts
for about 1/3 of the total diffuse ionized gas. We note that the HIM,
however, has a much larger vertical scale height.

The photoionized gas in spiral arms
is associated with the H~II regions produced by
OB associations.  These H~II regions typically have
dense cores, which appear as radio H~II regions,
and much lower density envelopes, which absorb
a significant fraction of the ionizing photons
\citep{ana85,mck97}.  Most of the mass of the photoionized
gas is in the envelopes, which have a surface density
$\Sigma_{\rm H\,II,\, env}=3.5\exp(-R_k/3.5)\, M_\odot$~pc\ee\
\citep{mck97}, or 0.31 $M_\odot$~pc\ee\ at the solar circle.

\section{Distribution of Energetic
             Photons and Particles}
\label{EnergeticPhotons}

The distributions of far-ultraviolet (FUV, 6 eV $< h\nu < 13.6$ eV)
radiation, extreme ultraviolet (EUV, 13.5 eV $< h\nu \al 100$ eV)
radiation, soft X-ray (100 eV $ \al h\nu \al 1$ keV) radiation,
and cosmic rays are required to calculate the ionization fraction in the
gas, the charge on grains, and the grain photoelectric
heating rate. In the next three subsections we describe our adopted
distributions.

\subsection{Far-Ultraviolet Radiation}\label{sec:FUVradfield}

The FUV radiation field strength enters the photoelectric
heating rate in two ways. First, it provides the total photon energy
available for gas heating, and second, it governs the grain charge and
thus the efficiency at which photon energy is converted into gas heating
\citep{wat72}.
We calculate the far-ultraviolet radiation field by carrying out a simple
radiative transfer calculation in an inhomogeneous medium.
Note that we shall be scaling the numerical
results to the measured value in the solar neighborhood and thus
we need calculate only the variation from the local value. In addition,
we are concerned with the {\it radial} variation of the mean intensity
in the Galactic midplane and not the variation with height above the plane.
And finally, we note that the vertical scale height of the H~I
and diffuse dust distribution, $H_z^{\rm H\, I}$,
is always much greater than that of the OB
stars that contribute to the FUV field. With these considerations,
we assume that the FUV emissivity and opacity
have a constant value between the midplane and the scale height of
OB stars above the plane $H_z^{\rm OB}(R)$, and that the emissivity is zero
at heights greater than $H_z^{\rm OB}(R)$. The mean intensity in the Galactic
midplane at radius R is given by
\begin{equation}\label{Jequation}
 4\pi J^{\rm FUV}(R) =  \int_0^{2\pi}\, dl \int_0^{\pi/2}2 I^{\rm FUV}(R,l,b)
 \cos b\, db
\end{equation}
where $b$ and $l$ are Galactic latitude and longitude and $I^{\rm FUV}(R,l,b)$
is the far-ultraviolet intensity in direction $(l,b)$
at Galactocentric radius $R$. The expression
for the intensity in direction $(l,b)$ is given by
\begin{equation}\label{Iequation}
    I^{\rm FUV}(R,l,b) = \int_0^{\tau_{\rm max}}\frac{j(R')}{\kappa(R')}
      e^{-\tau '}\, d\tau '
\end{equation}
where $j$ and $\kappa$ are the FUV emissivity and opacity respectively and
$\tau_{\rm max}$ is the maximum optical depth along
the line of sight ($l,b$) (see the discussion below eq.~[\ref{jeq}]).
The FUV emissivity is determined by the distribution of OB stars in
the Galactic plane while the opacity is provided by dust, mainly in the
diffuse atomic phases. Normalized to the
local value in the solar neighborhood, the FUV emissivity scales as
\begin{equation}\label{eq:emissiv}
     \frac{j(R)}{j(R_0)} =
\frac{\Sigma_{\rm OB}(R)}{\Sigma_{\rm OB}(R_0)}
\frac{H_z^{\rm OB}(R_0)}{H_z^{\rm OB}(R)}
\end{equation}
where $\Sigma_{\rm OB}(R)$ is the surface density of OB stars.
We assume the vertical height ${H_z^{\rm OB}(R)}$ is given by the
${\rm H_2}$ height in the inner Galaxy i.e.,
${H_z^{\rm OB}(R \le R_0)} = 59$ pc, and that it scales with the ${\rm H_2}$
height at $R$ greater than $R_0$.

To proceed further we need to find how $\Sigma_{\rm OB}(R)$ scales with
$R$, or equivalently, we require the radial exponential scale length,
$H_R^{\rm OB}$, of the OB star surface density. \cite{mck97} examined the
\cite{smith78} catalog of giant radio H~II regions and found the surface
density of the exciting OB  associations
between 3 kpc $ < R < 11$ kpc
can be fit with a scale length of roughly 3.5 kpc. They found no evidence
for giant radio H~II regions beyond 11 kpc and within 3 kpc
(other than at the Galactic Center).
\cite{bron00} recently obtained
a scale length of only $\sim 1.8$ kpc for the surface density of
embedded OB stars in the outer Galaxy ($8.5 \al R  \al 17$ kpc). We note
that the \cite{bron00} data can be fitted well
by an $H_R^{\rm OB} = 3.5 $ kpc
scale length between 8.5 kpc and $\sim 13$ kpc, with a steeper slope
at greater radii.
The Bronfman et al. results are based on observations of far-infrared
radiation from deeply embedded OB stars, but the embedded stage
may last a shorter time in the outer Galaxy because molecular
clouds are smaller there: \cite{sol87} found no
GMCs at $R>10$ kpc (after correcting their distances to
a Galactocentric distance of 8.5 kpc); \cite{hey01} found no
clouds more massive than $10^5 M_\odot$ at $R>11.6$ kpc, whereas
half the molecular gas inside the solar circle is in clouds
with $M> 10^6 M_\odot$ according to \cite{wil97}.
Thus, it is conceivable that the total OB star distribution extends beyond
13 kpc without a break in slope.

   What do observations of other galaxies tell us about the radial
distribution of OB stars?
In M31, \cite{cuil01} have found B stars out to $\,\sim\! 33$ kpc, or
$1.7\times R_{25}$,
a distance corresponding to $\,\sim\! 21$ kpc in the Galaxy.
Further evidence for star formation at large galactocentric distances
is provided by
\cite{wan97}, who found that the radial distribution of SNe in
disk galaxies is exponential (with an average radial scale length
of 3.5 kpc) with no evidence for an outer cutoff in the distribution.
\cite{ferg98} found that star formation in several disk
galaxies extended out to at least $2R_{25}$ (corresponding to
about 24.5 kpc in the Galaxy).
They did find a break in the star formation rate, but the
scale-height in the outer parts of the disk ($R_{25}<R<2 R_{25}$)
averages about $0.3 R_{25}$, similar to the value found
by \cite{mck97} for the Galaxy between 3 and 11 kpc.
The H~II regions in the outer parts of these galaxies were
substantially smaller than in the inner regions, consistent
with the lack of giant H~II regions in the Galaxy beyond 11 kpc.
On the other hand, \cite{mar01} analyzed a larger sample
of galaxies and found that most galaxies exhibit
a strong cutoff in their star formation at a radius determined
by the Toomre criterion; this cutoff radius is generally of
order $R_{25}$.
In view of the uncertainties in the distribution of star
formation at large distances, we shall extend our analysis
only out to 18 kpc, corresponding to about
$1.5 R_{25}$.  Between 4 and 18 kpc, we shall adopt a scale
length for the OB star surface density of $H_R^{\rm OB} = 3.5$ kpc,
noting that the actual distribution beyond 13 kpc
is uncertain and, if anything, is less than our adopted distribution.
Between 3 kpc and 4 kpc,  we shall adopt a constant OB-star surface
density; since the
H~I and ${\rm H_2}$ surface densities appear to drop towards
the Galactic center in this range, it
is unlikely that the OB star surface density
would continue to rise at radii less than 4 kpc.

Using $H_R^{\rm OB} = 3.5$ kpc (at $R\ge 4$ kpc) and
substituting $H_z^{\rm H_2}$ for
the OB star vertical height, equation~(\ref{eq:emissiv}) can be simplified to
\begin{equation}\label{jeq}
      \frac{j(R)}{j(R_0)} = {e}^{-(R-R_0)/H^{\rm OB}_R}
\frac{H_z^{\rm H_2}(R_0)}{H_z^{\rm H_2}(R)}\, ,
\end{equation}
With the adopted cutoff in the OB star
distribution, the maximum path of integration $\tau_{\rm max}$ in equation
(\ref{Iequation}) extends to
$R = 18$ kpc in the plane or until the line of sight reaches the
perpendicular height of
$z=H_z^{\rm OB}(R') = H_z^{\rm OB}(R_0)[H_z^{\rm H_2}(R')/H_z^{\rm H_2}(R_0)]$.

The opacity to the FUV radiation $\kappa_{\rm FUV}(R)$ depends on the
dust abundance,
which in turn is proportional to the mean gas density and
metallicity.  The run of mean density $\langle n(R)\rangle$
and metallicity $Z(R)$ are discussed in \S\S\
\ref{HIDistribution} and \ref{sub:Metallicity} respectively.
Dust in the molecular phases is to a large extent
shielded from the FUV radiation. Recent Orpheus I
observations \citep{dix98} confirm
the earlier Copernicus result that the H to H$_2$ transition in diffuse
gas occurs at $E(B-V) \approx 0.1$. For $R_V = 3.1$ this corresponds to
$A_V\approx 0.3$ or $A_{FUV}\approx 0.6$. Thus, the FUV field
responsible for gas heating does not penetrate deeply into the molecular
layer. In addition, the small volume filling factor of molecular gas
($f\sim 0.1$\%) means that the molecular component of the FUV opacity
can be safely neglected.

The opacity at visual wavelengths is found from the
local observed extinction
[$A_V = N/(2\times 10^{21}$ cm$^{-2}$)] and mean
H~I density ($\langle n_{\rm H\,I}[R_0] \rangle=0.57$ cm$^{-3}$) from
\cite{dic90},  yielding
$\kappa_V(R_0 ) = 0.88$ kpc$^{-1}$ in the midplane. The FUV
opacity for photoelectric
heating is taken from the photodissociation region models
of \cite{tie85} and is based on the  radiation
transfer results of \cite{flann80}. We use
$\kappa_{FUV}(R_0) = 1.8 \kappa_V(R_0) = 1.6$ kpc$^{-1}$.
The FUV opacity is then given by
\begin{equation}
  \kappa_{FUV}(R) = 1.6 \frac{\langle n_{\rm H\,I}(R) \rangle}
          {\langle n_{\rm H\,I}(R_0) \rangle}
          {e}^{-(R-R_0)/H_R^Z} \, \, \, {\rm kpc^{-1}}
\end{equation}
with the radial metallicity scale length $H_R^Z=6.2$ kpc.
Since the scale height of
diffuse gas is much greater than that of OB stars, we can consider the
gas density in our integrations to be independent of height and a
function of only the radial distance $R$.
Figure \ref{fig:Opac} shows the FUV opacity as a function of $R$.

Locally, the FUV (6 eV $< h\nu < 13.6$ eV) intensity
has a measured strength of approximately
$4\pi J(R_0) = 2.7 \times 10^{-3}$ erg cm$^{-2}$ s$^{-1}$
\citep{dra78}. This is a factor of 1.7 higher than the
integrated field of \cite{hab68},
often used as a unit ($G_0=1$) of flux in models
of  photodissociation regions.  Since the {\em local} FUV flux is observed,
we find the distribution of flux in the Galaxy  by
scaling
$j(R_0)$ in
equation (\ref{Iequation}) so that the value of the
integral at  $R=R_0$ is given by the Draine field. Results
are shown in  Figure \ref{fig:FUVf}.
The drop off in intensity near 18 kpc is due to the abrupt cut off in the
OB star population that we have imposed in the outer Galaxy. The true
intensity
distribution should fall off more gradually.

We find that the calculated mean intensity
at $R\ge 4$ kpc can be fitted by  an exponential
with a radial scale length of $H_R^J = 4.1$ kpc,
\begin{equation}
    4\pi J^{\rm FUV}(R) = 4\pi J^{\rm FUV}(R_0){ e}^{-(R-R_0)/H_R^J}  
         \, \, \,\,\,\,R\ge 4\,{\rm kpc},
\end{equation}
whereas for $R$ between 3 kpc and 4 kpc it is constant,
\begin{equation}
  4\pi J^{\rm FUV}(R) = 4\pi J^{\rm FUV}(R_0){e}^{-(4-R_0)/H_R^J}  
         \, \, \,\,\,\,3 \ge R < 4\,{\rm kpc}.
\end{equation}
These expression are good  to within 10\% for $R$ between 3 kpc and 17 kpc.

\subsection{Cosmic Rays}

The low-energy ($E\al 100$ MeV)
cosmic rays that contribute to ionizing the
CNM and WNM do not travel far from their point of origin \citep{kul71,spi75}.
We obtain the cosmic ray ionization
rate as a function of position in the Galaxy by scaling the local
primary rate
(taken to be $\zeta_{\rm CR} = 1.8\times 10^{-17}$ s$^{-1}$)
by the production rate of cosmic rays per unit area (sources) divided by
the mass per unit area (sinks). For the distribution of sources
we use the surface density of OB stars (\S~\ref{sec:FUVradfield}),
and for the distribution of sinks we use the total surface density of molecular
and neutral atomic (CNM+WNM) gas (\S\S~\ref{HIDistribution} and
\ref{sub:H2}). This scaling differs from that
adopted by \cite{hun97} and \cite{bert93},
who made the assumption that the cosmic ray intensity is
proportional to the  surface density of gas alone;
however, these authors
were studying Galactic gamma-ray emission, and
were therefore interested in cosmic rays with higher energies
than those that dominate the ionization. Such high energy cosmic
rays can travel more freely in the Galaxy and therefore acquire
a fairly homogeneous distribution.

The surface
density of WIM can potentially affect our cosmic-ray ionization rate
by providing an additional sink for cosmic rays
(the energy loss
rate for cosmic rays in ionized gas is several times that
in neutral gas---Ginzburg \& Syrovatskii 1964).
In order to
simplify our model, we have chosen to neglect the effect of the
WIM surface density in calculating the distribution of sinks of cosmic
rays;
note that this effect enters only insofar as the distribution of
the WIM differs from that of the rest of the gas.
Our neglect of the effect of the WIM
will not strongly influence our results since EUV/X-ray  radiation 
generally dominates the ionization and FUV radiation dominates the
heating.

The resulting primary cosmic ray ionization rate is given by:
\begin{equation}
    \zeta_{\rm CR}=1.8\times 10^{-17}
   {e}^{-(R-R_0)/H_{R}^{\rm OB}} \left [ \frac{\Sigma_{\rm neutral}(R_0)}
   {\Sigma_{\rm neutral}(R)} \right ]\,\,\, {\rm s^{-1}}
\end{equation}
where $H_{R}^{\rm OB}=3.5$ kpc,
$\Sigma_{\rm neutral}(R) =\Sigma_{\rm H_2}(R) +\Sigma_{\rm H\,I}(R)$ and
$\Sigma_{\rm neutral}(R_0) = 6.4$ $M_\odot$ pc$^{-2}$. The variation
in the primary cosmic ray ionization rate with Galactocentric radius $R$ is
shown in Figure \ref{CosR}.

\subsection{Soft X-ray and EUV Radiation}

As demonstrated in WHMTB, the ionization of H and He by soft X-rays
and EUV photons (13.6~eV$<h\nu\al 10^3$~eV) provides
a source of electrons in the local WNM amounting to an electron
fraction of $n_e/n\sim 2$\% (at $n = 0.3$ cm$^{-3}$), a fraction 
that depends on the column $N_w$
of warm absorbing gas and dust traversed by the X-rays.
These electrons neutralize the positive charging of grains caused by the
FUV photoelectric effect, and thereby help maintain a high
photoelectric heating efficiency in the WNM phase. The local X-ray field
arises from the Local Bubble and the Galactic halo plus an
extragalactic background, with the low energy (50~eV$\al E\al 100$ eV)
emission from the Local Bubble dominating the ionization at typical
columns $N_w\approx 10^{19}$ cm$^{-2}$. WHMTB used a fit to the
observed X-ray
intensity from \cite{gar92} to generate the X-ray spectrum
incident on a WNM ``cloud''. The fit consisted of the temperature and emission
measure in the Local Bubble and halo components, an absorbing layer through
which the halo is seen, and an extragalactic component and absorbing layer.
The ionization rate at a column $N_w$ into the WNM cloud was then calculated
by computing the attenuation of the incident X-ray flux by the column $N_w$.

The \cite{gar92} fit was based on pre-ROSAT data.
\cite{sno98} and \cite{kuntz00} evaluated the ROSAT data to analyze
the origin and distribution of the diffuse X-ray background.
They find somewhat lower temperatures than \cite{gar92} for the
Local Bubble and argue that the 1/4 keV and 3/4 keV
band observations can be fit only if the halo is emitting at two or more
temperatures. Using the observed ratios in the ROSAT R1, R2, R3, and
R4 bands,  \cite{kuntz00} find
log $T_{b} = 6.11$ for the bubble and log $T_{h1} = 6.06$, log $T_{h2}=6.46$
for the two halo temperatures as compared
with log $T_b = 6.16$ and log $T_h = 6.33$ from \cite{gar92}.
(Note that there is an error in Table 1 of WHMTB in that the emission
measures for the bubble and halo components are reversed).

The fit to the ROSAT data provides the {\em local} soft X-ray flux.
However, we know that the Sun is located in a bubble of hot gas that
dominates the low energy radiation and hence the ionization, and the
local emission may differ from the average emission at the solar circle.
We derive the average X-ray emission in the disk from the calculations of
\cite{slav00}. They define the emissivity per unit area
$\langle \epsilon_{\nu A}\rangle$ as
\begin{equation}
    \langle \epsilon_{\nu A}\rangle = \int_{-\infty}^{\infty} dz\,
    \langle \epsilon_\nu \rangle
\end{equation}
where $\langle \epsilon_\nu \rangle$ is the volume emissivity and the
integration is carried out perpendicular to  the disk.
The average value of the mean intensity is found from
$\langle \epsilon_{\nu A}\rangle$ as
\begin{equation}\label{jslavin}
\langle J_{\nu} \rangle = \frac{(1-\eta_\nu)}{\tau_{\nu ,{\rm d}}}\left(
             \frac{\langle \epsilon_{\nu A} \rangle}{4\pi }\right)
\end{equation}
where $\tau_{\nu ,{\rm d}}$ is the opacity through the disk, and $\eta_\nu$
is the mean escape probability of soft X-rays out the Galaxy.
The opacity $\tau_{\nu ,{\rm d}}$ is taken to be that of a cloudy
medium of column density through the disk $N_{\rm WNM,\, d}$
with clouds of typical column density $N_{\rm cl}({\rm H~I})$.
(These parameters appear as $\tau_{0\nu}$, $N_{\rm H^0\perp}$,
and $N_{\rm H^0c}$ respectively in the Slavin et al.\  notation).
We consider WNM clouds only; the small filling factor of the CNM
means that it does not contribute much to the opacity of very soft
X-rays.
The opacity is then given by
$\tau_{\nu ,{\rm d}} =
[N_{\rm WNM,\, d}/N_{\rm cl}({\rm H~I})][1-\exp(-\tau_{\nu, \rm cl})]$
where $\tau_{\nu ,{\rm cl}}$ is the optical depth through a single WNM 
cloud, and $N_{\rm WNM,\, d}$ is the WNM column through the disk.

Slavin et al.\ determine the average emissivity produced
by supernova remnants
$\langle \epsilon_{\nu}^{\rm SNR} \rangle$, which accounts for
the time and space averaged emissivity produced as supernova
remnants evolve into
the interstellar medium of ambient density $n_a$, although
Slavin et al.\  show
that the average emissivity is very insensitive to $n_a$.
Slavin et al.\  applied
their model to  a specific line of sight towards
a high latitude cloud and calculated the ionizing flux as a function
of height in the disk and found that they could successfully
match the observed fractional ionization.

In addition to the soft X-rays emitted by the supernova remnants,
stellar EUV radiation also contributes to the ionization rate in the
WNM. We use the stellar EUV spectrum shown by \cite{slav98}, which is
derived from {\em Extreme Ultraviolet Explorer} observations \citep{vall96}
and corrected for extinction by the
local interstellar cloud (taken to be
$N({\rm H\,I}) = 9\times 10^{17}$ cm$^{-2}$.)
While keeping the stellar EUV spectral shaped fixed, we adjust the
level of the stellar EUV emissivity $\langle \epsilon_{\nu}^{ *} \rangle$
so that the total emissivity per unit area
$\langle \epsilon_{A}\rangle = \int d\nu \,dz\,
[\langle \epsilon_{\nu}^{\rm SNR}\rangle + \langle \epsilon_{\nu}^* \rangle ]$
provides the observed ionizing photon flux of the Galactic disk
outside H~II regions, as
deduced from H$\alpha$ observations
\cite[$\langle \epsilon_A \rangle
\approx 4\times 10^{6}$ ${\rm ph\, cm^{-2}\, s^{-1}}$;][]{rey84,rey95}.
We find approximately 43\% of the EUV photon flux comes from
stars;
this is consistent with the results of Slavin et al.\ (2000), who
suggest that hot gas in SNRs produces about half the total number
of ionizing photons in the diffuse ISM.

We adopt parameters for the total column density of WNM
through the disk
appropriate for the Galactic average at the
solar circle.
\cite{dic90} fit the vertical distribution of H~I in the solar
neighborhood with three distinct components.
Following \cite{kul87}, we identify the two components with
the largest scale heights with the WNM.
The total WNM column density through
the disk is then
$N_{\rm WNM,\, d}(R_0) = 3.45\times 10^{20}$ cm$^{-2}$, or
$\sim 56$\% of the total H~I column at $R_0$. We also set the
mean escape probability equal to zero ($\eta_{\nu} = 0$). For our
standard model we set
$N_{\rm cl}({\rm H~I}) = 1\times 10^{19}$ cm$^{-2}$,
comparable to the column densities of the WNM clouds observed
along the line of sight toward the halo star HD 93521 \citep{spi93}
and disk star $\gamma^2$ Vel (Fitzpatrick \& Spitzer 1994; this line
of sight contains 4 WNM clouds ranging in H~I column density from 
$2.7\times 10^{18}$ cm$^{-2}$ to $4.5\times 10^{19}$ cm$^{-2}$ with
an average value of $1.5\times 10^{19}$ cm$^{-2}$). 
Note that in our formalism, $N_{\rm cl}$ is the WNM column density
in a typical cloud that provides the opacity
for the EUV and soft X-ray radiation. Note also that the phase diagrams
and thermal processes presented in \S~\ref{sec:Results} apply to the WNM/CNM
boundary within a cloud and thus radiation incident upon the cloud
must pass through an additional column
$N_{\rm cl}$ (or $N_w$ in WHMTB notation).  

Using the flux at $z=0$, we find that
at a cloud column of $N_{\rm cl}({\rm H~I})=10^{19}$ cm$^{-2}$,
the primary EUV plus X-ray ionization rate of hydrogen is
$\zeta_{\rm XR} = 1.6\times
10^{-17}$ s$^{-1}$, a factor $\sim 1.6$ lower than that used by WHMTB.
(This rate is approximately equal to the primary cosmic ray
ionization rate, $1.8\times 10^{-17}$ s$^{-1}$. The total
ionization rate from either cosmic rays or EUV/soft X-rays is larger
than the primary rate due to the effects of secondary ionizations.
The secondary rate increases with the energy of the primary ejected 
electron and with decreasing ionization fraction. 
In the WNM at a cloud column of
$N_{\rm cl} = 10^{19}$ cm$^{-2}$ and density $n\sim 0.3$ cm$^{-3}$
the total EUV/X-ray rate is $\sim 5.3\times 10^{-17}$ s$^{-1}$,
about 1.5 times higher than that from 
cosmic rays, while in the CNM at a density of $n\sim 33$ cm$^{-3}$ 
the total EUV/X-ray rate is $\sim 7.5\times 10^{-17}$ s$^{-1}$,
or $\sim 2.7$ times higher than that from cosmic rays.)
The ionizing photon intensity incident on clouds is $\sim 9.4 \times 10^3$
cm$^{-2}$ s$^{-1}$ sr$^{-1}$.
This is a factor 2.2 larger than that obtained
by Slavin et al.\ (2000) because (1) we are modeling
a typical region of the ISM, which has a significant flux
of stellar ionizing photons and (2) we have adopted
a somewhat smaller cloud column density.

Having determined the value for the soft X-ray ionization rate at $R=R_0$,
we must now scale the results for other Galactic radii. The ultimate
energy source for the
hot gas that produces the X-ray flux is supernova explosions, while the
opacity arises from the surface density of (WNM) H~I gas.
Thus, we assume that the X-ray ionization rate per hydrogen in the
Galactic midplane scales as the OB star
distribution divided by the H~I surface density,
$\zeta_{\rm XR}(R)\propto \Sigma_{\rm OB}(R)/\Sigma_{\rm H\,I}(R)$,
as shown in Figure
\ref{CosR}. We assume the stellar EUV photoionization rate scales
the same as the soft X-ray ionization rate. Because of the increased
gas opacity to the EUV/X-ray radiation compared to FUV radiation,
the mean free path for the EUV/X-ray photons are much shorter
than for FUV. Thus, the dust in CNM clouds dominates the FUV opacity,
but the gas in WNM clouds dominate the EUV/X-ray opacity.
Since the fraction of WNM column to the total is unknown outside
the solar neighborhood we have assumed, for the purposes of estimating
the X-ray distribution, that the fraction is always given by the value in
the solar neighborhood. If the WNM fraction were to increase in the
outer Galaxy, for example, the X-ray ionization rate would be lower than
our adopted rates. We also include ionization by the extragalactic X-ray and
EUV radiation field of \cite{ster02} which is based on the work
of \cite{haa96} and \cite{chen97}. The extragalactic
field passes through an absorbing column given by one half of the total
WNM column density. At the solar circle the extragalactic field provides
$\sim 1$\% of the total ionization rate (for $N_{\rm cl} = 10^{19}$
cm$^{-2}$), while at $R=18$ kpc the extragalactic rate rises to
$\sim 12$\% of the total rate.

\section{Chemical  and Thermal Processes in the Neutral Phases}
\label{sec:thermal}

The chemical and thermal processes included in this work are slightly
modified from those discussed in WHMTB. The main changes involve the
PAH reaction network and PAH reaction rates. These rates are important
because reactions with PAHs affect the PAH charge state and
electron abundance which in turn affects the photoelectric heating
rate. We have simplified the PAH
chemical reaction network compared to that
used in WHMTB. Here we use rates appropriate for a single PAH size containing
$N_{\rm C} = 35$ carbon atoms, the mean size in the distribution between
3 \AA\ and 15 \AA. We have also dropped the adsorption reactions used
in WHMTB since these were found to be not important in determining
the PAH charge.

For the PAH reaction rates we use the
photoionization/photodetachment rates of \cite{bak94} and a modified form
of the \cite{draine87} formalism for the interaction between ions and
electrons with neutral and charged PAHs in which
we multiply all of the collisional rates by a factor $\phi_{\rm PAH} = 0.5$.
There is considerable uncertainty in applying a classical treatment
of the interaction between atoms and grains to the molecular regime.
In particular, collision rates (using $\phi_{\rm PAH} = 1$) consistently
overestimate the laboratory measured rates for electron
attachment to neutral PAHs and for electron recombination with
${\rm PAH^+}$ (Allamandola, Tielens, \& Barker 1989; Salama et al.\ 1996;
Weingartner \&
Draine 2001a and references therein). We will consider $\phi_{\rm PAH}$
to be a parameter and rely on observation to guide
us in its appropriate value.
As noted by \cite{lepp88} and recently by
\cite{bak98}, \cite{wel01}, and \cite{wei01b},
ion recombination on small grains and PAHs can be important in determining
the neutral fraction of metals.  (In Appendix~\ref{appen:toymodel} 
we also elucidate
the conditions under which reactions of ${\rm C^+}$ and ${\rm H^+}$
with PAHs can affect the
electron abundance.) Specifically, we find that the abundance of
neutral carbon is sensitive to the rate of ${\rm C^+}$ recombination
with ${\rm PAH^-}$. We find that the observed C~I/C~II column density
ratio in diffuse clouds ($\al 3\times 10^{-3}$; Welty \& Hobbs 2001; 
Jenkins \& Tripp 2001) imply $\phi_{\rm PAH} = 0.5$ and we adopt this
value for our standard model.
In \S~\ref{sub:PnR} we discuss the effect of higher and lower values of
$\phi_{\rm PAH}$ on the phase equilibrium of the ISM. 

We have also updated the gas phase
chemical reaction rates according to the list of \cite{mil97} and modified
the H$_2$ formation and dissociation rate according to the discussion in
\cite{kauf99}. These additional chemical changes mainly affect the molecular
pathways included in our network and are of minor consequence for the
atomic phases discussed in this paper. Note, however, that the adopted
gas phase carbon and oxygen
abundances have been modified from their WHMTB values
(\S~\ref{sub:Metallicity}). The results
of these changes are discussed in \S\S~\ref{sub:PnR} and \ref{sub:thermal}.

The dominant heating process at $R=R_0$ in both the
CNM and WNM phases is grain photoelectric heating. We use the
heating rate determined by
\cite{bak94}, modified by the higher PAH abundances as explained in
\S~\ref{sub:Metallicity}, and modified by the parameter
$\phi_{\rm PAH}$ which scales the  electron-PAH collision rates.
For the same density, electron
abundance, temperature,
and incident FUV field, the higher abundances  result in an increase in the
heating rate by a factor 1.3. The cooling rate due to
electron recombination with PAHs increases similarly by
a factor 1.3. In \cite{bak94} the heating/cooling rates were calculated
self consistently with the PAH ionization state as a function
of the photo and  collision rates. To maintain this self consistency,
the heating/cooling rates (and the fit to the rates) must be modified
for the parameter $\phi_{\rm PAH}$. Including the correction for
higher PAH abundances, the  heating rate per unit volume is given by
\beq
n\Gamma_{\rm pe} = 1.3 \times 10^{-24} n \epsilon G_0
\,\,\,\,\,\,\,\,\, {\rm erg\,\,cm^{-3}\,\, s^{-1}}\, ,
\label{eq:gphe1}
\eeq
where $n$ is the hydrogen nucleus density, and 
the heating efficiency $\epsilon$ is given by
\beq
\epsilon = \frac{4.9\times 10^{-2}}
{1 + 4.0\times 10^{-3}\left(\frac{G_0 T^{1/2}}{n_{\rm e} \phi_{\rm PAH}}
  \right)^{0.73}} + \frac{3.7\times 10^{-2} (T/10^4)^{0.7}}
{1 + 2.0\times 10^{-4}\left(\frac{G_0 T^{1/2}}{n_{\rm e} \phi_{\rm PAH}}
  \right)}\, .
\label{eq:eps}
\eeq
The cooling rate per unit volume is given by
\beq
n^2\Lambda = 4.65\times 10^{-30} T^{0.94} \left(
     \frac{G_0 T^{1/2}}{n_{\rm e} \phi_{\rm PAH}}\right)^{\beta}
      n_{\rm e}\phi_{\rm PAH} n
\,\,\,\,\,\,\,\,\, {\rm erg\,\,cm^{-3}\,\, s^{-1}}\, ,
\label{eq:gco1}
\eeq
with $\beta = 0.74/T^{0.068}$.

The dominant cooling process in the CNM is radiative line cooling
in the [C~II] 158 $\mu$m fine-structure transition. Cooling in the WNM is
provided by several processes: radiative line cooling by [C~II] 158 $\mu$m,
[O~I] 63 $\mu$m, and Ly$\alpha$, and by electrons recombining onto
grains (refer to details in WHMTB.)
In this paper we adjusted the collision rate for the excitation
of C$^+$ by impacts with  e$^-$ to that of \cite{blu92}.

\section{Results}
\label{sec:Results}

\subsection{Phase Diagrams} \label{sub:PnR}

Utilizing the results from the previous sections on the FUV
intensity and ionization rates in the Galaxy as functions of $R$,
we calculate phase diagrams -- gas thermal
pressure
$P$ versus
hydrogen nucleus density $n$ -- for several Galactocentric distances $R$. The
curves are generated by calculating, at constant $n$, the chemical
equilibrium abundances and the thermal equilibrium temperature. We then step
through $n$
and plot the calculated thermal pressure, $P=\Sigma n_i kT$,
where $i$ ranges over all chemical species. The calculations are carried
out for various WNM cloud columns $N_{\rm cl}$ and apply to the WNM/CNM
boundary.
An appropriate range is $3\times 10^{18}$
cm$^{-2}$ $\le N_{\rm cl} \le 1\times 10^{20}$ cm$^{-2}$. The upper
limit is set by the size scale at which turbulent pressure
begins to dominate. As
discussed in \S~\ref{sec:tur}, we obtain $\ellp$(WNM)$\sim 215$ pc,
or $N_{\rm cl}\sim 2\times 10^{20}$ cm$^{-2}$ for a typical
density $n\sim 0.3$ cm$^{-3}$. The lower limit is set by the
requirement that there be a substantial neutral fraction in
the cloud.  We note that the smallest neutral column density
in the warm clouds along the line of sight to the disk star $\gamma^2$ Vel
is $2.7\times 10^{18}$ cm$^{-2}$ \citep{fitz94}.  The column
density of the ionized clouds in the MO model of the ISM
is $2.2\times 10^{18}$ cm$^{-2}$; since this model agrees with
observations of H$\alpha$ and pulsar dispersion measures in the disk,
it is difficult to have predominantly neutral
clouds that are smaller than this.
We shall use $N_{\rm cl}=10^{19}$ cm$^{-2}$
as a standard and demonstrate the effects of both higher
($1\times 10^{20}$ cm$^{-2}$)
and lower ($3\times 10^{18}$ cm$^{-2}$) columns.

We present in Figure~\ref{whmPnRfig} phase diagrams for Galactic radii
$R=3$, 5, 8.5,
11, 15, and 18 kpc. The model parameters are given in 
Tables~\ref{tbl:modelparam1} and \ref{tbl:modelparam2}.
As demonstrated by \cite{field65},
the region of thermal stability in $P$ versus
$n$ phase diagrams lies in the range where $d P/d n >0$.
Where $d P/d n < 0$ the gas is thermally unstable to
isobaric perturbations.
If the pressure curve has a characteristic shape
shown in Figure~\ref{whmPnRfig}, two thermally stable phases may
coexist in pressure equilibrium within a range of gas pressures,
$P_{\rm min}$ to $P_{\rm max}$. At thermal pressures greater
than $P_{\rm max}$
only the cold phase (CNM) may be present, while at thermal
pressures less than
$P_{\rm min}$ only the warm phase (WNM) may exist (see also further
discussions in reviews by Shull 1987 and Begelman 1990).

We see that a two-phase (WNM+CNM) equilibrium is
possible
in the Galactic midplane at all Galactic radii between 3 kpc and 18 kpc.
The pressure
range $P_{\rm min}$ and $P_{\rm max}$ are listed in 
Table~\ref{tbl:physicalcond} for
each radius along with the range
in gas temperature and density for WNM and CNM gas between $P_{\rm min}$
and $P_{\rm max}$.
Also listed is the average pressure, where we adopt the
geometric mean of $\pmin$ and $\pmax$ to represent the average,
\beq
P_{\rm th,\,ave}(R) \equiv \left[P_{\rm min}(R) P_{\rm max}(R)\right]^{1/2}.
\label{eq:pthave}
\eeq
We also give the temperature and density at $P_{\rm th,\,ave}$ for WNM and
CNM gas.
Compared to the results in WHMTB at
$R=8.5$ kpc, $P_{\rm min}$  is higher by a factor of $\sim 2.0$ and  
$P_{\rm max}$ is higher by a factor of 
$\sim 1.3$. The difference in $\pmin$ is mainly due to the revised
(lower) gas phase carbon and oxygen abundances from those in WHMTB. The
lower abundance of coolants results in a higher gas temperature. 
The higher $\pmax$ is partly due to the lower abundance of coolants,
but mitigated by the effects of the collision rate parameter
$\phi_{\rm PAH}$. The collision parameter affects the electron fraction,
the photoelectric heating rate, and the cooling rate due to electrons
recombining onto positively charged grains. Lower values of
$\phi_{\rm PAH}$ result in a higher electron fraction in WNM gas.
At $\pmax$,
higher electron fractions and enhanced recombination cooling plays 
a role in limiting the maximum pressure.

We next examine the effects of the PAH collision rate parameter
$\phi_{\rm PAH}$ and the PAH abundance on the pressure curves.
We show in Figure \ref{fig:phipah} and list in
Table~\ref{tbl:depmodelparam} results for $R= 8.5$ kpc and 
$\phi_{\rm PAH} = 0.25$, 0.5 and 1.0,
with the standard PAH abundance ($n_{\rm PAH}/n=6\times 10^{-7}$ or
an amount of C in PAHs of $22\times 10^{-6}$ relative to H). We
also show results for $\phi_{\rm PAH} = 0.5$ and a lower PAH abundance of
$n_{\rm PAH}/n = 4\times 10^{-7}$ as used by WHMTB.  
Results for $G_0=1.1$ as is appropriate
for the interstellar radiation field of \cite{mat83} are also given
in Table~\ref{tbl:depmodelparam} but not shown in the figure since the
resulting pressures are very similar to the case for  low PAH abundance. 
Over the range of
$\phi_{\rm PAH}$ from 0.25 to 1, $P_{\rm min}$ changes by a factor of only
1.5 while $P_{\rm max}$ changes by 1.9 mainly due to the increased
effects of recombination cooling. 
We conclude that the results for the phases of
the ISM are very robust against variations in the PAH physical 
and chemical characteristics. For example, the average temperature
hardly varies. The largest change occurs for $\phi_{\rm PAH} =0.25$
in which $\pmax$ decreases by $\sim 35$\%.

The phase diagrams presented in Figure \ref{whmPnRfig}
used a column of atomic gas $N_{\rm cl} = 10^{19}$
cm$^{-2}$.
We illustrate in Figure \ref{whmPnRNfig} the effects of higher
($N_{\rm cl} = 10^{20}$ cm$^{-2}$) and lower
($N_{\rm cl} = 3\times 10^{18}$ cm$^{-2}$)
column densities at $R=R_0=8.5$ kpc. As the column density decreases
at a given Galactic radius,
the electron abundance rises (due to hydrogen photoionization
by EUV/X-rays). The
increased electron abundance neutralizes the grains and enhances
the rate of grain photoelectric heating. In addition, the EUV/X-ray 
heating rate rises. Therefore, lower column densities
result in higher temperatures and pressures, $P_{\rm min}$ and
$P_{\rm max}$. At columns of $N_{\rm cl} = 3\times 10^{18}$ cm$^{-2}$,
$P_{\rm min}/k =2560 $ K cm$^{-3}$ and $P_{\rm max}/k=7830$ K cm$^{-3}$
while at the higher column of
$N_{\rm cl} = 1\times 10^{20}$, $P_{\rm min}/k = 1240$ K cm$^{-3}$ and
$P_{\rm max}/k = 2310$ K cm$^{-3}$.  The calculated range in two
phase (CNM+WNM) thermal pressures over the
column densities $N_{\rm cl} = 3\times 10^{18}$ cm$^{-2}$ to
$N_{\rm cl} = 1\times 10^{20}$ cm$^{-2}$
are in reasonably good agreement with the thermal pressures
of $P/k = 10^3 - 10^4$ K cm$^{-3}$ that are observed in the local ISM
\citep{jen83,jen01}.  For the fiducial case $N_{\rm cl}=10^{19}$~cm\ee, the
average pressure at the solar circle is predicted to be 3070~K~cm\eee,
consistent with the C~I population ratio observed by
\cite{jen01} (see \S~\ref{sub:local}).

\subsection{Thermal Processes}
\label{sub:thermal}

Figures~\ref{fig:whmhcR}$a$ through $d$ show the dominant
thermal  processes at Galactic radii of $R=5$, 8.5, 11, and 17 kpc.
We see that at all
radii, photoelectric heating dominates in both the CNM and WNM phases.
The [C~II] 158 $\mu$m line emission dominates the cooling in the CNM and is
generally a factor of $\sim 20$ weaker (per H) in the WNM.
The photoelectric heating rate per hydrogen
is nearly the same as the results presented in WHMTB with
the current rates approximately 20\% lower
at a density of 30 cm$^{-3}$. 
The higher PAH abundance used in this paper is offset by a lower
gas phase carbon abundance (and lower
electron abundance in CNM gas), higher temperatures, 
and a $\phi_{\rm PAH} = 0.5$ collision rate parameter,
which decreases the photoelectric heating efficiency. 
Due to the higher CNM temperatures  however, we arrive at an
average CNM density which is a factor  $\sim 2$ lower than
WHMTB (see Table~\ref{tbl:depmodelparam}), which results in an
average 
C~II cooling rate per hydrogen a factor of $\sim 1.5$ lower than our
previous result. 

It is instructive to consider the total cooling luminosity of
the ISM, which is dominated by the [C~II] 158 $\mu$m line.  The [C~II]
luminosity of the Milky Way has been measured by COBE to be $5\times
10^7$ L$_{\odot}$ \citep{wright91}.  \cite{tie95}  considered  the 
possible global sources of this emission. In the inner galaxy, much of 
it may originate in extended low density H~II regions \citep{hei94} while
most of this emission in the Galaxy as a whole has to stem from the
CNM (WHMTB).  The WNM can not contribute much to the total [C~II] 
luminosity due to the low emission rates in such tenuous gas. 
We consider the potential for mechanical heating 
by supernovae in Appendix~\ref{appen:turbheating} and conclude 
that neither the shock
heating nor the turbulence generated by SNRs can contribute
substantially to heating the H~I gas. (Of course, mechanical energy dominates
the energy balance of the hot ionized medium and coronal gas in the
Milky Way.)  Thus, the thermal structure of the ISM
is largely dominated through the coupling of the gas to the 
stellar, non-ionizing radiation field. The fraction of [C~II] emission
in the inner Galaxy which arises from CNM or warm ionized gas depends 
on the mass and density (or filling fractions) of these components.
Fitting the observed profile of [C~II] emission versus Galactic
longitude may provide a test of the models presented here, however  
a detailed model of the ionized gas (in pressure equilibrium and 
in overdense regions) is beyond the scope of this
paper.   We shall discuss in a future paper the implications of 
our [C~II] emission
rates for the filling fractions of the WNM and CNM gas, and for the origin
of the [C~II] emission in the Galaxy.

\subsection{Predicted Infrared Radiation Field}\label{sub:IRcheck}

In this subsection we discuss a check on the distribution and local values
of the model opacity and interstellar radiation field by comparing
the calculated infrared intensity emitted by dust with observations
from the COBE satellite.

We calculate the integrated infrared continuum intensity along a line of sight
assuming that the predominantly far-infrared emission is optically thin,
\begin{equation}
I_{\rm IR} = \int_{\rm IR}d\nu\, I_\nu =
               \int_0^{s}ds' \int_{\rm IR}d\nu\,\int da\, B_\nu[T(a)]
               \kappa_\nu(a) \, ,
\end{equation}
where $B_\nu[T(a)]$ is the Planck function
for the grain-size dependent temperature $T(a)$
and $\kappa_\nu(a)$ is the grain absorptive
opacity for grains of size $a$.
In thermal equilibrium, this emission just balances the heating by UV
photons,
\begin{equation}
I_{\rm IR}  = \int_0^{s}ds'
\int_{UV} d\nu\,\int da\, J_\nu^{\rm ISRF} \kappa_\nu(a) \, ,
\end{equation}
where $J_\nu^{\rm ISRF}$ is the interstellar radiation field.
We have ignored the small amount of energy that goes into
photoelectric heating of the gas.
We take the spectral energy distribution of the interstellar field to
be that of \cite{dra78}  between 912 \AA $ \le \lambda < 2000$ \AA ,
\cite{dish82} between 2000 \AA $\le \lambda <  3400$ \AA , and
\cite{mat83} for $\lambda \ge 3400$ \AA . The grain absorptive
opacity $\kappa_\nu$ is calculated as in \cite{wol86} and \cite{wol94}
using a grain abundance and optical constants from Draine \& Lee (1984, 1987)
with grains distributed in size as a power law
($n[a]\, da\propto a^{-3.5}\, da$; Mathis et al.\ 1977). The dust opacity is
scaled with $R$ in proportion to the mean H~I density and metallicity,
\begin{equation}
   \kappa_{\nu}(R) = \kappa_{\nu} (R_0)\frac{\langle n_{\rm H\,I}(R)\rangle }
                 {\langle n_{\rm H\,I}(R_0) \rangle }
          {e}^{-(R-R_0)/H_R^Z}
\end{equation}
while the interstellar radiation field is scaled with $R$ according to
results reported in \S~\ref{sec:FUVradfield}.

We show in Figure~\ref{fig:IRfig} the calculated IR emission compared with
the COBE longitudinal profile reported by \cite{sod94}. We obtain a reasonably
good fit.
We have not compared with the region within 3 kpc of the Galactic
Center, since we have not modeled the H~I there.
The largest deviations occur at the locations of the spiral arms,
where some of the FUV radiation is absorbed by molecular gas
associated with star-forming regions; recall that we have
not included this in our models, which are for the diffuse ISM.
Integrating over galactic longitude, but excluding the region
near the Galactic Center, the mean intensity from COBE is
$1.4\times 10^{-5}$ W m\ee\ sr\e, whereas our model
gives $1.1\times 10^{-5}$ W m\ee\ sr\e.  According to
Parravano et al. (2002), about 80\% of the FUV radiation is absorbed
in the diffuse ISM, and 20\% is absorbed by the nearby 
natal giant molecular cloud. Using the interstellar radiation field
and opacity given in the preceding paragraph we estimate 
that approximately half the dust heating is  produced by
the FUV 
radiation from OB stars and half by longer wavelength 
radiation from older stars generally located far from molecular
clouds. Correcting for the OB starlight
absorbed by nearby molecular clouds, 
our model
gives $1.2\times 10^{-5}$ W m\ee\ sr\e\ for the total IR emission,
in reasonably good agreement with observation.

\subsection{Multiphase ISM}\label{sub:ISM2phase}

We have presented phase diagrams which show $\pmin$ and $\pmax$, and 
have presented observational evidence that both CNM and WNM gas exists,
but we have not yet tried to estimate the thermal pressure 
in order to make comparison to $\pmin$ or $\pmax$. One way to 
determine the pressure is from the density and volume filling
factors
of the H~I gas.
Our results are in terms of the local density $n$, but
they can be recast in terms of the mean atomic hydrogen 
density $\langle n_{\rm H\, I} \rangle $
which is often better determined observationally.
Let $n_{\rm WNM}(P_{\rm min})$ be the equilibrium density
of the WNM at the pressure $P_{\rm min}$, etc.  We generalize the
treatment of \cite{kro81} (who used the notation $\bar{n}$ for the mean
density) by allowing for the possibility
that the two-phase medium fills only a fraction $f_{\rm H\,I}$ of the volume
and by allowing for pressure fluctuations in the gas.
We follow Krolik
et al. in assuming that the gas is in thermal equilibrium,
which is not strictly true in a turbulent medium.

      The ratio of the CNM mass to the WNM mass is \citep{kro81}
\beq
\frac{M_{\rm CNM}}{M_{\rm WNM}}=
             \frac{[\langle n_{\rm H\, I} \rangle 
               /(n_{\rm WNM}f_{\rm H\,I})]-1}
             {1-[\langle n_{\rm H\, I}\rangle /(n_{\rm CNM}f_{\rm H\,I})]}.
\label{eq:mass}
\eeq
Similarly, one can show that the ratio of the volume filling
factors is
\beq
\frac{f_{\rm CNM}}{f_{\rm WNM}}=\frac{1-(n_{\rm WNM}f_{\rm H\,I}/\langle
  n_{\rm H\, I} \rangle) }{(n_{\rm CNM}f_{\rm H\,I}/\langle n_{\rm H\, I}
       \rangle ) -1}.
\label{eq:fill}
\eeq
Let ${\cal R}\equiv n_{\rm CNM}(P_{\rm min})/n_{\rm WNM}(P_{\rm max})$;
if ${\cal R}$ is large, then there is an extensive range of density
in which the H~I is thermally unstable.  The cases summarized
in Table~\ref{tbl:physicalcond} all have ${\cal R}\ga 6$.

There are five regimes for two-phase media:

\begin{enumerate}
\item $\langle n_{\rm H\, I} \rangle /f_{\rm H\,I}<n_{\rm WNM}(P_{\rm min})$.
Since the  average
density of the H~I is less than the minimum required for CNM to
exist, it follows that most, if not all, of the volume of H~I
is filled with WNM.  If the gas is isobaric, the H~I is all WNM.

\item $n_{\rm WNM}(P_{\rm min})<\langle n_{\rm H\, I} 
\rangle /f_{\rm H\,I}<n_{\rm WNM}(P_{\rm max})$.
Here again, most of the volume of H~I must be filled with WNM
(one can show from eq.~[\ref{eq:fill}] above
that $f_{\rm CNM}/f_{\rm WNM}<1/[{\cal R}-1]$), but even for
the isobaric case some CNM can coexist with the WNM.  In the presence
of turbulent pressure fluctuations, some WNM will be driven
into CNM, and both phases will occur

\item $n_{\rm WNM}(P_{\rm max})<\langle n_{\rm H\, I}\rangle /f_{\rm H\,I}
<n_{\rm CNM}(P_{\rm min})$.
If the local density $n$ were in this regime, the gas would
be thermally unstable and two phases would form.  As a result,
in this case two phases {\it must} exist.

\item $n_{\rm CNM}(P_{\rm min})<\langle n_{\rm H\, I}\rangle /f_{\rm H\,I}
<n_{\rm CNM}(P_{\rm max})$.
In this regime, most of the mass must be CNM
(one can show from eq.~[\ref{eq:mass}] above
that $M_{\rm CNM}/M_{\rm WNM}>{\cal R}-1$), but even for the
isobaric case some WNM can occur.  In the presence of turbulent
pressure fluctuations (which are negative as
well as positive), some CNM will be driven into WNM, and both
phases will occur. However, in this regime, it is possible that 
all the neutral gas is CNM. 

\item $n_{\rm CNM}(P_{\rm max})<\langle n_{\rm H\, I}\rangle /f_{\rm H\,I}$.
Finally, in this case most, if not all, the H~I must be
in the form of CNM.  If the gas is isobaric, all the H~I must be CNM.

\end{enumerate}

According to Table~\ref{tbl:physicalcond}, 
the local ISM is characterized by
$n_{\rm WNM}(P_{\rm min})=0.21$ cm\eee,
$n_{\rm WNM}(P_{\rm max})=0.86$ cm\eee,
$n_{\rm CNM}(P_{\rm min})=6.9$ cm\eee, and
$n_{\rm CNM}(P_{\rm max})=71$ cm\eee.  The locally observed value of
the mean atomic hydrogen density 
$\langle n_{\rm H\,I}\rangle $ in the Galactic plane is 0.57 cm\eee.
Therefore,
if the H~I is pervasive, the local ISM would be in the regime in
which most of the volume of H~I is primarily WNM.  Observations
show that most of the mass of the H~I in the local plane is CNM
\citep{dic90}.
Recently, \cite{hei02}, utilizing H~I 21 cm measurements,
estimated a local value of $f_{\rm H\, I} = 0.5$ based on the assumption
that the local thermal pressure of WNM is 2240 K cm$^{-3}$ and 
$T_{\rm WNM} = 4000$ K so that $n_{\rm WNM} = 0.56$ cm$^{-3}$. In this
paper, we argue that the local WNM pressure is approximately 3100 K cm$^{-3}$
and that $T_{\rm WNM} \simeq 8000$ K so that $n_{\rm WNM} \simeq 0.35$
cm$^{-3}$. The Heiles \& Troland result then becomes $f_{\rm H\, I} \simeq
0.79$. These results lie close to the theoretical estimates which range
from $f_{\rm H\, I} = 0.4$ (MO) to 0.8 \citep{slav93}.
If a substantial fraction
of the volume of the ISM is hot then $f_{\rm H\,I}\al
0.5$ and the local ISM is marginally
in the regime in which there {\em must} be two
phases. 

We extend our analysis to the inner and outer Galaxy
by  estimating  the thermal pressure in the midplane
in the limiting case in which the H~I is entirely in the form of WNM.
We derive a WNM pressure and check to see if it is self consistent
(i.e., is the thermal pressure less than $\pmin$ or $\pmax$) to 
assume that all the neutral gas is WNM.
To calculate the thermal pressure
we rely on the assumption that the total pressure in the midplane is balanced
by the weight of the overlying gas layers (i.e., hydrostatic equilibrium).
We first consider the
case in which the Galactic H~I resides in a continuous, thermally
supported  WNM layer, and then modify this thermal pressure calculation
for the effects of non-thermal support, including cosmic rays, magnetic
fields,  and turbulence.
We also estimate the thermal pressure
by using the mean atomic hydrogen density $\langle n_{\rm H\, I} \rangle$ 
shown in Figure~\ref{fig:HImean} to derive a ``mean'' thermal pressure which
we compare to $\pmin$ and $\pmax$ to find the region over which a
two-phase medium must exist.

\subsubsection{Thermally Supported WNM}\label{subsub:thermalsupwnm}

For an isothermal
gas in hydrostatic equilibrium in the Galactic gravitational field,
the vertical density distribution is given by
\begin{equation}
n(R,z) = n_0(R)e^{[\Phi (R,0)-\Phi (R,z)]/\sigma_{\rm th}^2}
\label{eq:nrz}
\end{equation}
where $n(R,z)$ is the gas density at Galactic radius $R$ and height
above the plane $z$, $n_0(R) = n(R, z = 0)$ is the density in the
midplane,
 $\Phi (R,z)$ is the Galactic gravitational
potential, and $\sigma_{\rm th}$ is the isothermal
sound speed. We take $\Phi (R,z)$ from
a variant of model 2b of \cite{deh98},
which has a disk mass of $5.3\times 10^{10}\, M_\odot$ (by comparison,
the disk model used by Wolfire et al. 1995b had a mass of $1.0
\times 10^{11}\, M_\odot$).  The numerical code to calculate the
potential was kindly provided by W. Dehnen.
Each density component in the disk is assumed to be of the form
\beq
\rho_=\frac{\Sigma_{d}}{2z_{d}}\exp\left(-\frac{R_m}{R}
        -\frac{R}{R_d}-\frac{|z|}{z_{d}}\right),
\eeq
where $z_d$ is the vertical scale height and $R_d$ is the
radial scale length of the disk.  The parameter $R_m$ allows for
the depression in the gas density observed in the inner several
kiloparsecs of the Galaxy; for the stellar disks, they set $R_m=0$.
\cite{deh98} assumed that the gas in the disk could be described by
a single component with $z_d=40$ pc and, for Model 2b, $R_d=5.1$ kpc.
A limitation of their model (which is relatively unimportant
for their application) is that the vertical scale height $z_d$ for
each component is assumed to be independent of radius.
We have altered their model for the gas to make it more
consistent with the discussion in \S~\ref{sec:gasdust}, with one component
for the molecular gas, two components for the H~I in order
to capture the radial variation, and one component for the H~II.
(We have included only the diffuse H~II component
discussed in \S~\ref{sub:wim}, since
the ionized gas contributed by the envelopes of H~II regions
is always very small compared to the column density of stars plus
H~I.  We have approximated ${\rm sech^2}[R_k/20]$ as $\exp[-R_k/30]$).
We have also altered their model for the thin stellar disk to allow the
scale height to increase in the outer Galaxy. For Galactic radii $R > R_0$
we assume that $z_d \propto 1/\Sigma$ where $\Sigma$ includes the
total (gas plus stars) surface density and then recalculate the potential
for each radius $R$ using the appropriate $z_d$.
The parameters entering the fit are summarized in Table~\ref{tbl:potential}.
The total surface density at the solar circle is $10.1\, M_\odot$ pc$^{-2}$.
This is $2.5\, M_\odot$ pc$^{-2}$ less than that of \cite{deh98}, so we
added $2.5\, M_\odot$ pc$^{-2}$ to the thin stellar disk in order to maintain
the same value of the total surface density.

We wish to calculate the pressure that the WNM in
the Galactic plane would have if all the H~I in the disk were in
the form of WNM and if the pressure support of the WNM were entirely
thermal. We label this hypothetical pressure $P_{\rm WNM'}$.
It is given in terms of the midplane H~I density $n_{\rm H\,I,0}(R)$ by
$P_{\rm WNM'}(R)=1.1n_{\rm H\,I,0}(R)kT$.
To determine $n_{\rm H\,I,0}(R)$, we first calculate the
column density from equation
(\ref{eq:nrz}), noting that the
the velocity dispersion of the neutral WNM is
$\sigma_{\rm th} = 7.2$ km s$^{-1}$ ($T=8000$ K).
We then equate this theoretical column density
to the observed column density,
$N_{\rm H\,I}(R)=2\int n_{\rm H\,I}(R,z)\, dz$,
and solve for $n_{\rm H\,I,0}(R)$.
Figures \ref{fig:PWNM}$a$ through \ref{fig:PWNM}$c$  show the pressure
obtained in this manner
for $N_{\rm cl} = 1\times 10^{19}$ cm$^{-2}$,
$1\times 10^{20}$ cm$^{-2}$, and $3\times 10^{18}$ cm$^{-2}$,
respectively, compared to
$P_{\rm min}$, $P_{\rm max}$, and the average
thermal pressure $\pthave (R)$.
The kinks in $P_{\rm WNM'}$ are a reflection of the kinks
in our adopted H~I surface density (eqs.~[\ref{eq:h1surf1}]
through [\ref{eq:h1surf4}]). In Appendix \ref{appen:thermalWNM}
we provide an analytic solution for  $P_{\rm WNM'}$ at the solar 
circle which provides a value of $P_{\rm WNM'} = 7800$ ${\rm K\ cm^{-3}}$,
in good agreement with the numerical solution of 
$P_{\rm WNM'} = 8615$ ${\rm K\ cm^{-3}}$. 
These values exceed $P_{\rm max}$ for $N_{\rm cl}\ag 3.0\times 10^{18}$
cm\ee (see Table~\ref{tbl:physicalcond}), 
so we conclude that a thermally supported H~I layer cannot be
all in the form of WNM gas and some of the H~I {\em must} be forced into
the CNM phase, at least in the solar neighborhood.
The numerical results (Figs.\
\ref{fig:PWNM}$a$ through \ref{fig:PWNM}$c$)
show that this is true over most of the disk
of the Galaxy.

\subsubsection{Turbulently Supported WNM}\label{subsub:turbsupwnm}

The discussion in  the previous section is based on a highly idealized 
model of the
ISM, in which the gas is supported entirely by thermal pressure.
In fact, turbulence makes a substantial contribution to the
support of the gas; cosmic rays and magnetic fields appear to make less
of a contribution to support of H~I near the plane since 
their scale heights are much greater than
that of the CNM and the gradient therefore weak, though
their substantial pressures must somehow be anchored by the weight
of the ISM \citep{bou90}.  The increase in the scale height
reduces the mean density of the gas, and hence the inferred
thermal pressure.
The turbulence leads to large pressure fluctuations, and these
will drive some of the
gas into the cold phase (e.g., Hennebelle \& P\'erault
1999, 2000).  The condition for a
substantial amount of CNM in a turbulent ISM is therefore
that the thermal pressure exceed $\pmin$, which ensures that the gas that
is compressed into the CNM can remain there.  

We analyze first the solar circle and compare the thermal pressure to $\pmin$
{\em assuming} all H~I is WNM.
We redistribute the $2.75\times 10^{20}$
H~I atoms cm\ee\ of CNM into the WNM components found by
\cite{dic90}; the result is a gas with a central density
of 0.31 cm\eee.  
This procedure should provide a lower bound on the thermal
pressure of this hypothetical ISM in which the H~I is pure WNM, since we are
assuming that there is enough additional turbulent energy injection to lift
the mass in CNM up to the height of the WNM.  With this assumption,
the resulting thermal pressure 
is $P/k=2700$ K cm\eee.  Since this significantly exceeds 
$P_{\rm min}/k=1960$ K cm\eee\ and since the local ISM is turbulent,
we conclude that it must be in two phases.

We can make analogous arguments to the rest of the Galaxy to show
that, assuming all of the H~I were WNM, the thermal pressure
would exceed $\pmin$, violating our assumption, and that therefore
CNM must exist. We begin by determining a lower limit 
on the thermal pressure produced by
WNM gas at other positions in the Galaxy.
In \S~\ref{sub:HI}, we have presented the mean density 
$\langle n_{\rm H\, I} \rangle$ of 
H~I in the plane
rather directly derived from the observations.  Assuming that the volume
filling factor of WNM is much greater than that of the CNM,
then
\beq
    \langle n_{\rm H\, I} \rangle = n_{\rm WNM} 
         \left( 1 + \frac{M_{\rm CNM}}{M_{\rm WNM}} \right) 
          f_{\rm H\, I}\,\, ,
\eeq
where $M_{\rm CNM}/M_{\rm WNM}$ is evaluated in the midplane.
The thermal pressure in the medium is given by
\beq
    P_{\rm th,\, WNM} = 1.1 n_{\rm WNM} k T_{\rm WNM}\,\, .
\eeq
In terms of the ``observable'' $\langle n_{\rm H\, I} \rangle$, 
$P_{\rm th,\, WNM}$ can be written as

\beq
    P_{\rm th,\, WNM} =  
1.1 \frac{\langle n_{\rm H\, I} \rangle  k T_{\rm WNM}}
  {f_{\rm H\, I} (1 + M_{\rm CNM}/M_{\rm WNM} )} \equiv 
   \frac{ \langle P_{\rm WNM} \rangle}
    {f_{\rm H\, I} (1 + M_{\rm CNM}/M_{\rm WNM} )} \,\, .
\eeq

The ``mean'' thermal pressure $\langle P_{\rm WNM} \rangle$, the
thermal pressure assuming the 
denominator is unity (for example, the 
observed mean density is all WNM and $f_{\rm H\, I} = 1$), 
is shown in Figure~\ref{fig:PWNM}$d$ (assuming $T_{\rm WNM} = 8000$ K)
along with
$\pmin$, $\pmax$, and $\pthave$ for our standard $N_{\rm cl} = 1.0\times
10^{19}$ cm$^{-2}$ case.
Following the distribution
of H~I in the Galaxy, $\langle P_{\rm WNM} \rangle$
is flat with galactocentric
radius out to 13 kpc where it starts to drop exponentially. 

As we did in the non-turbulent case, we can make the ansatz that all
the H~I gas is WNM and ask whether the thermal pressure exceeds
$\pmin$ or $\pmax$. 
In contrast to the argument above for the local ISM, we assume that
there is no additional turbulent energy injection to raise the height of the 
converted CNM.
In this case $\langle P_{\rm WNM} \rangle$ is 
a lower limit
to the thermal pressure since $f_{\rm H\, I}$ may be less than unity
(although it must be borne in mind that a hypothetical ISM in which
the H~I is pure WNM is likely to have $f_{\rm H\, I}$ closer to unity 
than the actual ISM).
Comparing $\langle P_{\rm WNM} \rangle$ to the model thermal pressure
curves, we conclude that in this case the thermal pressure would exceed
$\pmax$ for Galactic distances $8\,\,{\rm kpc}\al R \al 16$ kpc; our ansatz
is violated; and a two-phase medium {\em must} exist in the outer Galaxy.
Moreover, over much of the Galaxy 
$\langle P_{\rm WNM} \rangle$ exceeds $\pmin$ and hence 
pressure fluctuations in a turbulent ISM will produce a two-phase
medium. 

We also note in Figure $12d$ that $\langle P_{\rm WNM} \rangle$
falls
everywhere below the non-turbulent pressure $P_{\rm WNM'}$ derived from the 
weight of the H~I in the
galactic potential.  
Essentially, this high value of $P_{\rm WNM'}$ 
derives from the assumption of no turbulence and therefore
has a thermal scale hight of $\sim 80$ pc 
(see Appendix \ref{appen:thermalWNM}), whereas
$\langle P_{\rm WNM} \rangle$ takes into account 
the observed half height 
($\sim 265$ pc) of the H~I 
and reflects
the importance of turbulence for the
dynamics of the interstellar gas.  

\section{An Analytic Model for Two-Phase Equilibria}
\label{sec:toy}
In order to understand the physical reasons for the results just
presented, and to obtain scaling laws,
it is convenient to have an approximate analytic
model. The details of this model are presented in 
Appendix~\ref{appen:toymodel};
we summarize here our procedure and the results.

The thermal balance pressure $P$ is found as a function of $T$
by equating a simple analytic equation for the total ([CII]
+[OI]) cooling to the analytic expression for grain/PAH
photoelectric heating. The simple  cooling equation holds for
$100\,\, {\rm K} < T < 1000\,\, {\rm K}$, which marks the temperature
range of validity of our analytic solutions. Several other parameters
enter the pressure equation, including the FUV field $\gs$, the total
gas ionization rate by cosmic rays and EUV/X rays $\zio$, the
dust/PAH abundance $\zd$, and the gas phase metal abundance
(especially of the coolants C and O) $\zg$.  The primes denote 
normalization to the local values given in Table~\ref{tbl:modelparam1},
so that all these parameters have value unity at the solar circle.
In other words, $\gs=G_0/1.7$, and $\zio = \zeta _t /10^{-16}
$ s$^{-1}$, where $\zio$ includes primary and secondary ionizations.
The grain
photoelectric heating depends on the charge state of the PAHs/grains,
which in turn depends on the electron density $n_e$ in the gas.
At a typical column $N_{\rm cl}=10^{19}$ cm$^{-2}$, H$^+$ (not C$^+$)
supplies the electrons. Therefore, $n_e \simeq n({\rm H^+})$ and we solve
for $n({\rm H^+})$. The source of H$^+$ is the photoionization of H
by EUV/X-rays, and the sink is recombination with negatively charged
PAHs, or PAH$^-$.  PAH$^-$ is produced by electron attachment on
neutral PAHs (neutral PAHs dominate the PAH population in the
parameter space valid for our analytic $P$ expression), and is
destroyed by photodetachment in the FUV field ($\gs$).

With an analytical expression for $P$, we then take 
$d P/ d T =0$
to find the temperature $\tmin$ at the pressure minimum, which 
substituted into our expression for $P$
gives $\pmin$.  We put the parenthesis in the subscript for $\tmin$ 
to emphasize that this temperature is not a minimum temperature;
in fact it is the maximum temperature of the CNM!
The hydrogen density at $\pmin$ is then 
$\nmin =\pmin/[1.1k\tmin]$.  Appendix~\ref{appen:toymodel} describes 
all the essential
and competing reactions, and gives analytic expressions for the cooling,
 heating, PAH$^-$ abundance, $n_e$, and $P$. We summarize here
the results for $\pmin$, $\tmin$, and $\nmin$, which we emphasize
are valid over a range of parameter space centered on solar circle values,
as discussed below. 
We find
\beq
\pmin/k \equiv 1.1\nmin \tmin \simeq 8500 {{\gs \left({\zd \over \zg} \right)}
\over {1+3.1\left({{\gs \zd} \over {\zio}}\right)^{0.365}}} \ {\rm cm^{-3}\ K}
   \, ,
\label{eq:pmin}
\eeq
\beq
\tmin \simeq 243\ {\rm K}\, ,
\label{eq:tmin}
\eeq
and
\beq
 \nmin \simeq 31 {{\gs \left({\zd \over \zg} \right)}
\over {1+3.1\left({{\gs \zd} \over {\zio}}\right)^{0.365}}} \ {\rm cm^{-3}}
   \, .
\label{eq:nmin}
\eeq
These equations are valid over the following range of parameters.  The
first condition is that

\beq
4.6\times 10^{-2} \al {{ \gs \zd}\over {\zio}} \al 11 \, .
\label{eq:con1}
\eeq
The lower limit is set so that the grain/PAH photoelectric
heating is significantly affected by the positive charging of the
grain/PAHs, which simplifies the analytic expression for the gas heating.
It also ensures
that most PAHs be neutral and not PAH$^-$, and that
FUV photodetachment
destroys PAH$^-$ and not reaction with H$^+$; these assumptions were made
in deriving the analytic expression. The upper limit
is set so that H$^+$ is destroyed by PAH$^-$, and not neutral PAHs.
It also ensures the less restrictive condition that neutral PAHs,
and not ${\rm PAH^+}$ dominate the PAH population. The second condition
is that
\beq
6.5\times 10^{-3}Z_{\rm g}^{\prime 2} \al \left({\gs \over \zio}\right)^{0.27}
Z_{\rm d}^{\prime 2.27}\phi _{PAH}^2 \al 4.1 \, .
\label{eq:con2}
\eeq
The upper limit ensures that the gas-phase abundance of H$^+$ exceeds that
of C$^+$ (H$^+$ supplies the electrons in the gas). As an interesting
sidenote, we find that there is no $\pmin$ in the temperature
range where [C~II] 158 $\mu$m and [O~I] 63$\mu$m dominate the cooling,
{\it if the electrons are supplied by C$^+$ and if the grain heating
is significantly affected by positive charge}. The ionization of
H and He is therefore generally critical to the two-phase phenomenon.
The lower limit ensures that atomic hydrogen collisions, and not
electrons, dominate the excitation of the [C~II] line.  
The final condition is that
\beq
\zg  {Z_{\rm d}^{\prime -1.635}} \left({\gs \over \zio}\right)^{0.365}
    \al 38\phi _{PAH}^2 \, .
\label{eq:con3}
\eeq
This condition assures that H$^+$ is destroyed by PAH$^-$, and not
by recombining with gas phase electrons. We found it interesting
that this is generally the case, for the wide range of conditions
centered on local values.

Given that all the above conditions are met, equations
(\ref{eq:pmin}) through (\ref{eq:tmin}) give
not only the absolute values of $\pmin$ and $\tmin$, but the scaling
of $\pmin$ with
the parameters  $\gs$, $\zio$, $\zd$, and $\zg$.
It might be
noted that $\zd$ and $\zg$ may often scale linearly with respect
to each other, and with the total elemental metallicity $Z$. It
is also noteworthy that
in this simple analytic solution, $\tmin$ is independent
of all the parameters $\zg$, $\zd$, $\gs$, and $\zio$. 
It depends primarily on the
gas temperature dependence of the cooling rates and the grain photoelectric
heating rate.
In much of parameter space, the numerical solution gives
$\tmin \sim 180-350$ K, which lie within about $\pm 40$\% of the
constant 243 K analytic result.
The numerical 
calculation of $\tmin$ is sensitive to small contributions from
processes we have ignored or simplified in the analytic treatment
(e.g., electrons supplied
by C$^+$, our approximate cooling function, a simplification of
the grain photoelectric heating).  However, our analytic results
for $\pmin$ and the scaling of $\pmin$ are much more robust;
$\pmin$ is not sensitive to the exact value of $\tmin$, since
it lies at a minimum with respect to changes in $T$.

Figure~(\ref{fig:toy}) illustrates the region of validity for the analytic
solution for $\pmin$ (eq.\ [\ref{eq:pmin}]). We plot the
conditions given in equations
(\ref{eq:con1}) through (\ref{eq:con3}) as functions of $Z^\prime = \zg = \zd$
and $\gs/\zio$. The constraint that hydrogen collisions dominate the
excitation
of [C~II] (lower limit in eq.\ [\ref{eq:con2}]) is readily met and not
a factor as long as the other conditions are satisfied.
We have performed several checks of
the analytic solution for $\pmin$ and $\tmin$
against the detailed results of our numerical code. We held
three of the parameters fixed at the local values ($=1$ in the
notation of the analytic equations), and varied the fourth
over a factor of 100 from 0.1 to 10.
This test showed that over this range the absolute value
of the analytic $\pmin$ was within $\pm 50$\%  of the numerical value
as long as $\zg$ was less than 3 and $\zd$ was between 0.3 and 5.0.
The scaling
of $\pmin$ was accurate to about $\pm 45$\%, and $\tmin$ was
accurate to about a factor of 2.5. In the numerical runs, $\tmin$ 
varied from
180 to 630 K, with a value of 258 K at the solar circle.  
This test,
however, occasionally violated the conditions of validity.
If we restrict the test strictly to  the regime of validity,
the agreement for $\pmin$ is unchanged, but the
range of $\tmin$ in our numerical calculation is reduced to 180-465 K,
with much of parameter space $180-350$ K.

As another test of our analytic solution we compared $\pmin$
to our numerical results in which we varied
$\zg =\zd$ from 0.01-10 and $\gs/\zio$ from 0.1 to 300. The
shaded region in Figure~(\ref{fig:toy}) shows the range in
which the analytic solution agrees with the numerical results to within
$\pm 50$\%. It is clear that the applicable range of
validity extends well beyond that given by the restrictive
conditions expressed in equations (\ref{eq:con1}) through (\ref{eq:con3}).
As a final test we compared the analytic solution for $\pmin$ with the
calculated $\pmin$ as a function of Galactic radius (\S~\ref{sub:PnR}
and Table~\ref{tbl:physicalcond}). We find that the calculated $\pmin$ can be fit with an 
exponential in Galactic radius as
\beq
 P_{\rm min} = 1.1\times 10^{4} \exp(-R_k/4.9)\,\,\,\,\, {\rm K\,\,\, cm^{-3}}.
\eeq
This fit is good to within $\pm 17$\% between $R_k = 3$ and $R_k = 18$.
Substituting values for $\gs$, $\zio$, $\zd$, and $\zg$ from 
Table~\ref{tbl:modelparam2}
into equation~(\ref{eq:pmin}) yields analytical results which are good 
to within $\pm 15$\% of the numerical results.

\section{Comparison with Observations and Previous Work}
\label{sec:comp}

\subsection{Comparison with Local Observations}
\label{sub:local}

In this section we compare our model results with three critical
observations of the CNM: the thermal pressure, the gas temperature, and
the ${\rm C^+}$ cooling rate per hydrogen atom. We also briefly
examine the constraints imposed by the observed C~I/C~II ratio.
Our model results for the local Galaxy and the effects of various
model parameters are summarized in Tables~\ref{tbl:physicalcond} 
and \ref{tbl:depmodelparam}. The results presented in this paper
so far have been for the WNM/CNM interface at a cloud column of $N_{\rm cl}$. 
We also list in Table~\ref{tbl:depmodelparam} results for a CNM cloud 
interior at a depth of $1\times 10^{20}$ cm$^{-2}$.  
The mean CNM column density measured
by \cite{hei02} is 
$5\times 10^{20}$ cm$^{-2}$. For a slab cloud,
a typical point is at a depth of 1/4 the observed column density
($\sim 1.25\times 10^{20}$ cm$^{-2}$),
where half the mass 
is at greater column and half at lower column.
For a spherical cloud, the average line of sight passes through a
column of $2N/3$, where $N$ is the column density through the cloud
center, and the  half mass depth is at 0.1 $N$ from the
surface. This half mass depth corresponds to $\sim 0.75\times 10^{20}$
cm$^{-2}$. Our fiducial column is between the two limits for slab
and spherical clouds.

The thermal pressure in the ISM has been measured through studies of the
population distribution of the C~I fine-structure levels based upon FUV
absorption lines originating from these levels \citep{jen83,jen01}.
By necessity, such studies are largely limited to the local solar
neighborhood. \cite{jen01} find a mean observed 
${\rm C~I^*/C~I_{tot}}$ ratio of 0.196 and derive a 
 mean pressure of $P_{\rm th} = 2240$
K cm$^{-3}$ from this ratio. This pressure is somewhat lower than the 
previous result of
\cite{jen83} who found $P_{\rm th} \approx
4000$ K cm$^{-3}$. Our standard parameter set yields $P_{\rm min} = 1960$,
K cm$^{-3}$ $P_{\rm max} = 4810$ K cm$^{-3}$ and 
$\pthave = 3070$ K cm$^{-3}$ and our average
pressure would appear to be higher than the observed value. However,
\cite{jen01} note that the derived pressure is sensitive to the
CNM temperature and atomic constants where they used a CNM temperature
of T = 40 K and radiative decay rates from \cite{fro85}.
The more recent radiative rates of \cite{gal97} are approximately
10\% higher than those used by \cite{jen01} and our 
model temperature of 71 K in the CNM cloud interior 
(see Table~\ref{tbl:depmodelparam}) is
considerably higher than 40 K. Both of the these differences tend
to increase the derived thermal pressure for the same 
observed population ratio. (We note that the excitation
rates of C~I by collisions with H~I from Launay \& Roueff 1977 
are expected to be accurate to within 10\%; E. Roueff 2002
private communication). Including the updated radiative rates, 
plus the optical/UV
pumping of the C~I lines as discussed in \cite{jen01} and \cite{jen79}
we find a ${\rm C~I^*/C~I_{tot}}$ 
population ratio of 0.190 at the CNM
cloud surface, and a ${\rm C~I^*/C~I_{tot}}$ ratio of 0.201 in the
 cloud interior
(Table~\ref{tbl:depmodelparam}). These values are within 
3 percent of the 
observed ratio of 0.196. The ``Low $\phi_{\rm PAH}$'' 
($\phi_{\rm PAH} = 0.25$)
model compares least favorably to the observations
with a population ratio of between $0.145$ and $0.162$.
Although the differences between models
are not large, 
we conclude that our standard model with an average thermal pressure of 
$\pthave = 3070$ K cm$^{-3}$ provides the best  agreement 
with the observed ${\rm C~I^*/C~I_{tot}}$ ratio.

In addition to the thermal pressure, observational tests are provided by the
CNM temperature and ${\rm C^+}$ cooling rate. \cite{hei01},
using H~I emission/absorption experiments along 19 lines of sight,
found that of the total CNM column detected, most ($\sim 61$\%)
lies in a narrow temperature range between
25 K and 75 K with a peak near 50 K.  This value was recently revised
by \cite{hei02} when an additional 60 lines of sight became available 
and it was possible to derive temperatures for separate low latitude 
($|b| < 10^\circ$)
and high  latitude ($|b| > 10^\circ$)
directions. They find a mean mass weighted 
(low latitude) CNM temperature of 99 K and a median temperature of
63 K with half the column density above and half below this temperature.

Additional measures
of the gas temperature of diffuse clouds come from UV absorption 
line observations of the $J = 0$ and $J = 1$ level populations of 
${\rm H_2}$ in the ground vibrational state. The excitation temperature
of these levels (sometimes denoted $T_{01}$) will equal the
gas kinetic temperature as long as collisions with ${\rm H^+}$ are
sufficiently rapid to thermalize the ${\rm H_2}$ ortho/para ratio,
and that other processes (e.g., FUV pumping or ${\rm H_2}$ formation)
do not drive the ratio away from thermalization 
\citep[see e.g.,][]{dal77,bla87,bur92,ste99}. 
We estimate that the ionization provided by the X-ray/EUV
field provides sufficient ${\rm H^+}$ so that
$T_{01}$ is a good measure of the gas kinetic temperature in CNM clouds. 
Early
estimates of the gas temperature from Copernicus observations found
$T_{01} = 77 \pm 17$  K \citep{sav77} while recent estimates using FUSE 
report $T_{01} = 68 \pm 15$ K \citep{rac02}. Of the models shown
in Table~\ref{tbl:depmodelparam} the ``High $\phi_{\rm PAH}$'' 
($\phi_{\rm PAH} = 1$) 
calculation has  temperatures in the cloud interior
more than 1 sigma  higher than the ${\rm H_2}$ temperatures.
Our CNM temperature of 71 K for the standard model  
lies between the median and mean H~I temperatures of \cite{hei02}
and is in good agreement with the ${\rm H_2}$ measurements.  

We note that \cite{hei02} found $\sim 4$\% of the CNM column 
to lie at temperatures less than 25 K. Such extremely cold H~I
has been also reported, for example,  by \cite{gib00} who find local 
H~I gas
at $T=16-32$ K, and by \cite{knee01} who find an H~I supershell at 
$R= 16$ kpc with $T=10$ K. Our standard model is unable to produce these
low CNM temperatures.  
Such low temperatures might be produced in  
gas depleted in PAHs, which reduces 
the photoelectric heating (as suggested by Heiles \& Troland 2002), or
in  gas that contains a sufficient molecular ${\rm H_2}$ abundance  
to cool to these temperatures.
  
The reported [C~II] 
158 $\mu$m cooling rate per hydrogen atom  
varies widely depending on the method 
and direction of observation. The two main methods are derived 
from (1) UV absorption line measurements of 
the column density of ${\rm C~II^*}$,
$N({\rm C~II^*})$, where ${\rm C~II^*}$ means 
${\rm C^+\,\, ^3P_{3/2}}$, and (2)
IR observations of the [C~II] 158 $\mu$m line intensity $I([{\rm C~II}])$.
From $N({\rm C~II^*})$ along a line of sight,
and the total (${\rm H~I + H~II + 2H_2}$) column of 
hydrogen nuclei, $N({\rm H})$, the cooling rate 
is found from $n\Lambda = E_{21}
A_{21}N({\rm C~II^*})/N({\rm H)}$ where $E_{21}$ and $A_{21}$ are the
energy and radiative decay rate for the [C~II] 158 $\mu$m 
transition. The total hydrogen column is usually inferred from the 
column of ${\rm S^+}$. For the IR method, the results are 
usually reported using H~I 21 cm emission to obtain the column 
of neutral hydrogen $N({\rm H~I})$.  
From the intensity of the [C~II] line and $N({\rm H~I})$,
the average cooling rate is given by 
$n\Lambda = 4 \pi I([{\rm C~II}])/N({\rm H~I})$. 
Note that the UV and IR measures are not directly
comparable since the former is per H nucleus and 
latter is per H~I atom.
One source of uncertainty in both methods 
is the amount of WNM gas along a given line of sight. 
This is because the total column of hydrogen
arises from both WNM and CNM gas whereas the ${\rm C~II^*}$ 
resides in mainly the CNM gas alone. Thus the effect of WNM
gas is to lead to an underestimate of the derived cooling 
rates in the CNM.
This problem is especially severe for the high latitude 
IR observations in which the H~I column extends several
hundred parsecs above the plane where there is not much
cold gas. There are also processes which may lead  to overestimates
of the derived cooling rate.  The UV absorption method can be influenced by 
radiation from the background stars
which provides a UV radiation field, and  subsequent gas heating rate,
greater than
the average interstellar field. 
The IR line and UV absorption methods can be contaminated  by WIM gas,
which has a cooling rate per hydrogen in the [C~II] transition 
similar to the CNM so that the WIM contributes to both IR emission and
UV absorption. (However, generally along a line of sight the column
of WIM is much less that that of CNM).

The UV absorption method was used in the Galactic plane 
by \cite{pot79}, 
who found an average value of
$n\Lambda \approx 1 \times 10^{-25}$ erg s$^{-1}$ H$^{-1}$, 
and  by \cite{gry92} who reported $n\Lambda = 3.5^{+5.4}_{-2.1}\times 10^{-26}$
erg s$^{-1}$ H$^{-1}$. The Pottasch et al.\ (1979) value is
probably biased towards higher rates 
due to UV illumination  from background stars. 
On the other hand, the
Gry et al.\ (1992) results are probably biased towards
lower rates.  Their lines of sight were  
selected to have low column density and are thus expected
to contain more WNM gas on average then a typical line of sight.
The ${\rm C~II^*}$ columns have been determined for a few 
high latitude lines of sight from which the cooling rate can be
derived. For example, from \cite{fitz97},
we find $n\Lambda = 0.8-3.5\times 10^{-26}$ erg s$^{-1}$
H$^{-1}$ in the cool components toward HD 215733,
and \cite{sav93} find $n\Lambda = 1.4\times 10^{-26}$
erg s$^{-1}$ H$^{-1}$ toward 3C 273.
The Savage et al.\ (1993) 
observation is probably a lower limit to the CNM cooling rate. 
The direction towards 3C 273  is a well studied line of sight
with low mean density and much of the ${\rm C~II^*}$ may reside
in WNM gas. In addition, the 
hydrogen column (inferred from ${\rm S^+}$) samples the entire column
throughout the disk and halo and thus overestimates the column in 
CNM alone.

The IR line emission method has been applied to high latitude
directions. \cite{bock93} and \cite{mats97}  using a balloon borne experiment
measured the 158 $\mu$m line emission in a path covering Galactic
latitudes $b = 33^\circ$ to $50^\circ$. 
Matsuhara et al.\ (1997) obtain a cooling rate of 
$1.6 \pm 0.4 \times 10^{-26}$ erg s$^{-1}$ ${\rm (H~I)}^{-1}$
(i.e., per neutral hydrogen)
for  columns $N({\rm H~I})< 2\times 10^{20}$ cm$^{-2}$. 
Bock et al (1993) using the same C~II observations, but for
$N({\rm H~I}) > 10^{20}$ cm$^{-2}$  and excluding anomalously low
C~II/continuum ratios, found $2.6 \pm 0.6 \times 10^{-26}$  
erg s$^{-1}$ ${\rm (H~I)}^{-1}$. These rates are slightly lower
than the COBE observations of
$2.65\times 10^{-26}$ erg s$^{-1}$ ${\rm (H~I)}^{-1}$ \citep{benn94}.

We can estimate the correction to these high latitude IR observations 
due to WNM and WIM gas (see discussion in WHMTB). The ${\rm C^+}$
line intensity arising in the ionized gas can be estimated from 
the ${\rm H}\alpha$ line intensity available from the WHAM data set
(Reynolds et al.\ 1999)
\footnote{\url{ http://www.astro.wisc.edu/wham/index.html}}. 
We find that the ${\rm H}\alpha$ line intensity
varies from about 0.5 to 1 R over the path of the Bock et al (1993) 
experiment. Using $I({\rm H}\alpha) = 1$ R  and the conversion between
$I({\rm H}\alpha)$ and $I([{\rm C~II]})$ \cite[e.g.,][]{rey92} we find
$I({\rm [C~II]}) = 1.5\times 10^{-7}$  
erg s$^{-1}$ cm$^{-2}$ sr$^{-1}$ from the
ionized gas. Subtracting this from the Matsuhara et al.\ (1997)
cooling rate we get roughly $1.1\times 10^{-26}$ erg s$^{-1}$ 
$({\rm H~I})^{-1}$. Following WHMTB we use the average WNM and 
CNM columns derived from the observations of
\cite{kul85} yielding
$N_{\rm CNM} = 0.4\times 10^{20}$ csc $|b|$ cm$^{-2}$ 
and
$N_{\rm WNM} = 1.0\times 10^{20}$ csc $|b|$ cm$^{-2}$. From
WHMTB equation (12) and
neglecting the neutral component of the WIM we find a cooling rate
in CNM of $3.9\times 10^{-26}$ erg s$^{-1}$ H$^{-1}$.
We conclude that although the rates are quite uncertain
 due to contributions from  WNM and WIM gas, a conservative lower limit is 
that the [C~II] cooling rate in the CNM is 
greater than $\ag 3.0\times 10^{-26}$ erg s$^{-1}$ H$^{-1}$.
All of the models presented in Table~\ref{tbl:depmodelparam} 
satisfy this constraint although the low $\phi_{\rm PAH}$,
low PAH abundance, and low $G_0$ cases barely do so. 

We finally consider the model constraints imposed by 
the C~I/C~II ratio. From the C~I and H~I data compiled 
in \cite{wel01}, and assuming a constant C~II/H~I ratio of 
$1.4\times 10^{-4}$, we find that C~I/C~II $\al 3\times 10^{-3}$. 
The abundance of neutral metals, as typified by the C~I/C~II ratio, 
strongly rises with increasing  $\phi_{\rm PAH}$ through the 
direct recombination
of ${\rm C^+}$ on ${\rm PAH^-}$, and the C~I/C~II ratio increases by a factor
8.2 as $\phi_{\rm PAH}$ increases from 0.25 to 1. The effect of a 
lower PAH abundance is to diminish both the ion recombination on 
grains and the 
photoelectric heating rate, resulting in both lower thermal 
pressures and lower
C~I/C~II ratios. Decreasing the FUV field from 1.7 to 1.1 decreases the
rate of photoionization of C~I and the C~I/C~II ratio rises
by a factor of 1.6. We find that the ``High $\phi_{\rm PAH}$ and 
``Low $G_0$'' models can be safely ruled out by the required C~I/C~II
ratio. We shall present in a future paper a more detailed analysis
of the implications of the C~I/C~II ratio for translucent cloud models.
In summary, we find that the average thermal pressures are not dramatically
affected by the range of parameters shown in Table~\ref{tbl:depmodelparam},
however in a detailed comparison with observations we find 
(1) ``Low $\phi_{\rm PAH}$'' models poorly match the ${\rm C~I^*/C_{tot}}$
ratio and ${\rm C^+}$ cooling rate, with values lower than observed
(2) ``High $\phi_{\rm PAH}$''  models produce temperatures and 
C~I/C~II ratios higher than observed,
(3) ``Low PAH'' models produce low values for the ${\rm C~I^*/C_{tot}}$
ratios and ${\rm C^+}$ cooling rate,  and
(4) ``Low $G_0$'' models produce low vaules for the ${\rm C~I^*/C_{tot}}$
ratios and ${\rm C^+}$ cooling rate,  and high values for the 
C~I/C~II ratio. 

The discussion in this section highlights the need for more observational
and laboratory study in the areas of PAH chemistry and thermal processes
in the diffuse interstellar medium.
In the near future several ground-based,
airborne, and space based telescopes (ALMA, SOFIA, Herschel) may enable
us to measure C~I fine-structure level populations throughout the Milky Way
as well as in other galaxies.  These observations  will provide
stringent tests for our
model and for the general importance of the thermal instability for the
two-phase medium of interstellar gas.  We note that SOFIA and Herschel
will also allow us to study the [C~II] 158 $\mu$m emission along specific
sight lines in the Milky Way at high spatial and spectral resolution.
Together, this will provide a powerful way to study the physical
conditions in the ISM and the processes that dominate its structure.

\subsection{Comparison with Extragalactic Observations}
\label{sub:extragalobs}

In comparison to the \cite{braun97} observations, we find that based
on a phase diagram analysis alone we do not expect a strict cut-off
in the cold gas at the Galactic $R_{25}$ radius
($R_{25} = 12.25$ kpc). We also find a
CNM temperature that decreases in the outer Galaxy,
where as \cite{braun97} finds peak brightness temperatures to increase.
To reconcile these differences, we suggest
that \cite{braun97}
is mainly detecting the photodissociated outer
envelopes of molecular clouds. The inferred high column densities
($N\ag 3\times 10^{21}$ cm$^{-2}$)
and temperatures ($T\sim 80-150$ K) reported by \cite{braun97}
are more typical of photodissociated
gas near star forming regions
than of the diffuse CNM clouds illuminated by the general interstellar
radiation field on which we have been
concentrating. A similar interpretation of the H~I emission in M101 has
been proposed by \cite{smith00}.

Using the photodissociation region models of
\cite{kauf99}, modified by the chemistry and
thermal processes discussed in this paper,  we find that if the gas density and
UV field illumination are fixed, the H~I brightness temperature increases
if the metallicity (and therefore the gas phase coolants and grain abundances)
decreases
as expected in the outer portions of galaxies.
Near regions of star formation, the UV radiation field does not drop
with the disk radial scale length, but remains relatively constant
and depends mainly on the
illumination from the nearest star-forming region.
For local ISM  metallicities,
the H~I 21 cm line is only marginally opaque,
so that the maximum brightness temperature is set by the cool interior
regions of H~I clouds.  However,
at lower
metallicities (and lower dust-to-gas ratios) the same visual optical
depth is achieved at larger H~I columns and thus the H~I to ${\rm H_2}$
transition is pushed deeper into the cloud.
The larger H~I column
forces the 21 cm transition to become optically thick while still
within the warm regions, thereby increasing the H~I brightness
temperature. For metallicities as low as 0.2-0.3 times cosmic, typical 
of those
at the $R_{25}$ radius in M101 and other spirals \citep{gar97}
we find  that the temperature
increases by $40-60$ K, comparable to that seen by \cite{braun97}.

As discussed by \cite{sel99}, a fairly uniform H~I 21 cm line width
of $\sim 6$ km s$^{-1}$ is seen in the outer disk of spiral galaxies.
Instead of indicating WNM gas, \cite{sel99} suggest that the
dispersion arises from a cold, MHD-driven turbulent gas. Based on the
results of WHMTB, they argue that a 5000 K gas inferred from the
line width, would be thermally unstable. Our temperature
results for WNM in the the outer Galaxy shown in Table~\ref{tbl:physicalcond}
confirm that 5000 K gas is indeed thermally unstable there.
Furthermore, our
thermal pressure arguments for cold gas in the outer Galaxy,
are also consistent with their turbulence hypothesis.
However, we also anticipate that the H~I in the outer Galaxy
is a two-phase medium with a substantial amount of WNM.

\subsection{Comparison with Previous Work}
\label{sub:prev}

Our results differ from those reported in \cite{elm94}, who found that
CNM gas extends to only $\sim 12 - 14$ kpc in the Galactic plane.
Using the thermal processes and chemistry from \cite{par87},
they estimated the range of thermal pressures ($P_{\rm min}<P
<P_{\rm max}$) over which two phases are possible, as a function
of FUV radiation field and metallicity. They also estimated the
midplane thermal pressure and concluded that the thermal pressure
would drop below $P_{\rm min}$ by a Galactic radius of $\sim 12 - 14$ kpc.

The difference between our results and theirs
arises in large part due to the approximations they
made in estimating the pressure.
Following \cite{elm89}, they wrote the total pressure, $P_{\rm tot}$,
as being proportional to the product of the gas 
surface density $\Sigma_{\rm gas}$ and the total
(gas plus stars) surface density, $\Sigma_{\rm gas+stars}$, or
$P_{\rm tot} \propto \Sigma_{\rm gas} \Sigma_{\rm gas+stars}$.
They then assumed
that the total surface density is proportional
to the gas surface density, $\Sigma_{\rm gas+stars}\propto \Sigma_{\rm gas}$,
so that $P_{\rm tot}\propto \Sigma_{\rm gas}^2$,
and that the ratio of thermal to total
pressure is constant so that
$P_{\rm th} \propto \Sigma_{\rm gas}^2$.
Finally, and most important, they assumed that
$\Sigma_{\rm gas}\propto \exp(-R/H_R)$ with $H_R\sim 4$ kpc.
As a result, they obtained
$\pth/k\approx 10^{5.0}\exp(-2R/H_R)$, which
results in a thermal pressure of only
$P/k \sim 30$ K cm$^{-3}$ at 16 kpc.  In fact, as discussed in
\S~\ref{sec:gasdust}, the H~I surface density remains above
$5\,\, M_\odot$~pc\ee\ out to $R\simeq 15$ kpc.
Our value for the pressure of a pure WNM,
$P_{\rm WNM'}$, which is based on the
observed H~I surface density, is a factor 67 higher
than theirs at 16 kpc. This higher
pressure extends the region of cold gas to much greater radial distances
than found by \cite{elm94}.

\section{Summary and Discussion}
\label{sec:disc}

\subsection{Model Assumptions}

Our model for the gas heating and ionization in the Galactic disk
is based on a cosmic-ray rate that is constrained by comparisons of
observations to  chemical models
\cite[e.g.,][]{dish86,fed96,tak00}, direct observations of the local
FUV field and soft X-ray field,  and
a theoretical estimate of the EUV intensity.
The soft X-ray and EUV intensity is partly derived from the calculations of
Slavin et al.\ (2000),
who find that radiation from cooling supernova remnants can
produce the fractional ionization seen in clouds at high latitude.
The intensity used in this paper is an extension of their calculation to
the Galactic plane.  In addition,
a stellar EUV component from Slavin et al.\ (1998) is added so that the total
ionizing photon flux from the Galactic disk matches the recombination
rate derived from H$\alpha$ observations. An important component of the EUV
and soft X-ray radiation transfer is the opacity produced by the
WNM gas component.
We have assumed that the opacity is provided by an ensemble of WNM clouds
each of column density $N_{\rm cl}({\rm H~I})$.  The radiation
fields at the WNM/CNM interface and cloud interior
are found by passing the incident radiation through
additional columns of 
$N_{\rm cl}({\rm H~I})$, and $1\times 10^{20}$ cm$^{-2}$,
respectively.

Our confidence in the adopted local parameters is strengthened
by our successful modeling of the densities and temperatures in the
local WNM and CNM gas. Furthermore, we obtain a good fit to the
[C II] cooling rate per hydrogen as derived by UV line absorption and IR
line emission studies.
The good match between theory and observations further indicates that we
have included all the relevant
heating and cooling terms in our model, an argument that is especially
strong for the CNM phase in which [C~II] dominates the cooling; any additional
heating sources in our model would emerge as excess [C~II] emission.
This case is not as strong for the WNM in which the [C~II] emission
contributes only  $\sim 14$\% of the cooling and
is a factor 16 weaker per hydrogen than in CNM.
Additional heating terms do not strongly affect
the fine-structure
line emission due to the strong temperature regulation by Ly$\alpha$
cooling.

To obtain the FUV field in other regions of the Galaxy we have relied
on observations
of the gas surface density, metallicity, and OB star distribution, along with
a numerical integration for the mean intensity. To check our assumptions,
in \S~\ref{sub:IRcheck}, we compared the infrared emission produced
by dust grains heated by the interstellar radiation field
with observations taken by the COBE satellite.
The processes that determine the X-ray and cosmic ray distributions
are certainly more complicated than our models allow.
For example, the soft X-ray flux depends on the temperature and emission
measure in the  hot ionized gas component and the optical depth to
the emission regions.
The temperatures and emission measures depend on the metallicity,
and the optical depth depends on the structure of the interstellar
medium. We note, however, that Slavin et al.\ (2000) find that
the intensity of
X-rays produced by SNR emission does not
depend sensitively on the ambient density of the preshock gas.
We have carried through our analysis by adopting
a simple approach in which we
use plausible arguments for scaling the soft X-ray
and cosmic ray rates to other regions of the Galaxy based on the
distribution of production sources
(OB stars) and destruction sinks (various gas components).

Note that we have chosen to extend the OB star distribution with
a constant scale length $H_R^{\rm OB} = 3.5$ kpc out to
$R=18$ kpc.
Had we adopted the OB star distribution of \cite{bron00},
$P_{\rm min}$ and $P_{\rm max}$ would have been lower and it
would have been easier to form CNM in the outer Galaxy.
Since the actual OB star distribution in the outer Galaxy is
if anything below the one we have adopted, our conclusion
that CNM must exist in the outer Galaxy is strengthened.

Although the distribution of total H~I surface density is constrained by
observations, the separate column densities of CNM and WNM gas are not well
determined away from the solar neighborhood.
In calculating the opacity for EUV and soft X-ray photons,
we have assumed that the ratio of WNM to CNM surface densities and
scale heights are the  solar neighborhood
values \citep{dic90} throughout the Galaxy.
This in turn implies that the ratio of CNM to WNM volume filling
factors in the midplane is held constant with Galactic radius.
A somewhat different  prescription is given by \cite{fer98}
in which the volume fraction of WNM increases in the outer Galaxy.
Note that if we allowed the WNM fraction to increase in the outer
Galaxy, then the opacity to soft X-ray and EUV radiation would
increase as well, thereby reducing $P_{\rm min}$ and $P_{\rm max}$
and making conditions less favorable for the existence of WNM.
Another difference in our H~I distributions is that 
\cite{fer98} used a constant surface density of 
$\Sigma_{\rm H\,I} = 5$ $M_\odot$ pc$^{-2}$ in the outer Galaxy to
$R=20$ kpc where our distribution drops below 
5 $M_\odot$ pc$^{-2}$ beyond $R >  15$ kpc.  For constant
surface density, at $R=18$ kpc the 
pressure $P_{\rm WNM'}$ is a factor $\sim 2$ higher than for our H~I
distribution and would make it more likely 
that CNM gas can exist.
In a future paper we shall compare the calculated
[C II] emission with the observational data in an effort to independently
derive the volume fractions of CNM and WNM gas.
This will be particularly telling for the outer Galaxy, where
the relative absence of photodissociation regions and
H~II regions should permit
a clean distinction between WNM and CNM without the confusion
of predominantly molecular or ionized gas.

\subsection{Galactic Distribution of Two-Phase ISM}
\label{sub:disc-distb}

As discussed in \S~\ref{sub:ISM2phase},
in Figures~\ref{fig:PWNM}$a$ through \ref{fig:PWNM}$c$ we plot
$P_{\rm min}$ and
$P_{\rm max}$ as a function of position in the Galactic midplane
for 3 values of the WNM cloud column $N_{\rm cl}$. We also plot the
thermal pressure $P_{\rm WNM'}$ in the Galactic midplane
that would result if all of the H~I layer were WNM gas supported by thermal
pressure. For the case
in which there is no turbulence
and the thermal pressure dominates the pressure,
regions in which
$P_{\rm WNM'} > P_{\rm max}$
{\em must} have CNM gas. This is because
only CNM gas can exist at these pressures, and mass will be converted
from the WNM phase to the CNM until the pressure drops below
$P_{\rm max}$, where a two-phase medium can exist. Our figures show that
this condition is satisfied over much of the Galactic disk.

$P_{\rm WNM'}$ is calculated assuming all the diffuse gas is WNM and that 
thermal pressure dominates and determines the vertical scale height,
which we calculate in this case locally to be $\sim 80$ pc. However,
the observed half-height of the ``WNM component'' seen by
\cite{dic90} is $\sim 265$ pc. This result demonstrates that nonthermal
pressure (due to turbulent motions, magnetic fields, and cosmic rays)
dominates. 
We take an analogous approach to
analyze the turbulent case. Assuming again that all the H~I gas
is WNM, we compare the thermal pressure to $\pmin$ and $\pmax$. We
use the {\em observed} $\langle n_{\rm H\, I} \rangle$  (which includes
the effects of turbulence and the greater scale heights which lowers
$\langle n_{\rm H\, I} \rangle$) to estimate a lower limit 
$\langle P_{\rm WNM} \rangle$ on the thermal pressure
$P_{\rm th,\, WNM} = \langle P_{\rm WNM} \rangle/f_{\rm H\, I}$ 
where $f_{\rm H\, I}$ is the volume filling factor of the WNM
(the rest is HIM). Since $\langle P_{\rm WNM} \rangle$ exceeds 
$\pmax$ in the outer ($8\,\, {\rm kpc} \al R \al 16$ kpc) Galaxy, 
CNM {\em must}
exist in these regions. Since $\langle P_{\rm WNM} \rangle$ exceeds
$\pmin$ (and turbulence likely drives the local pressures above
$\pmax$ occasionally), and since $f_{\rm H\, I} < 1$, we conclude CNM 
very likely exists at $3\,\, {\rm kpc} \al R \al 18$ kpc.

It is difficult, however, from our theoretical models, to
rule out an interstellar medium with only HIM and CNM  (and no WNM)
in which the intercloud
medium is filled with HIM that maintains a pressure  $P > \pmax$
on the CNM clouds. However, in such a scenario, the volume filling factor
of the HIM must be  nearly unity. If the HIM does not fill the
intercloud medium, CNM would partially convert to WNM to fill the vacuum,
and the pressure in the pervasive WNM would adjust such that
$\pmax > P > \pmin$ (see Parravano et al.\ 2002 for further
discussion). We also note that the observation of H~I 21 cm emission 
and absorption throughout the Galaxy strongly suggests the presence
of a pervasive WNM.

Assuming that $P$ lies between $\pmin$ and $\pmax$,
we can use our models to
predict the average thermal
pressure from equation (\ref{eq:pthave}), and the results are
given in Table~\ref{tbl:physicalcond}.
An approximate analytic fit to these results for our
fiducial column density of $N_{\rm cl}=10^{19}$ cm\ee\ is
\beq
P_{\rm th,\, ave}/k =1.4\times 10^4\exp(-R_k/5.5)~~~~~{\rm K~cm^{-3}}.
\eeq
We find that at fixed
Galactic radius, the pressure
does not change by more that a factor $\sim 3$ over our range
of cloud columns.
For our fiducial column, the
pressure drops from about 8,200 K cm\eee\ at 3 kpc to
3100 K cm\eee\ at the solar circle and to 600 K cm\eee\ at
18 kpc.  The drop in the thermal pressure from 3 kpc
to the solar circle (a factor 2.7) closely matches the drop in the
magnetic pressure inferred from radio observations:
Beck (2001) estimates that $B$ drops from about 10 $\mu$G
at 3 kpc to 6 $\mu$G locally, corresponding to a pressure
drop by a factor 2.8. (Note that these values for
the field are larger than the rms field; as Beck points out, his values of
$B$ are $\langle B^{3.9}\rangle^{1/3.9}$.)

We can now test the validity of our assumption
that the turbulence parameter $\Upsilon\equiv t_{\rm cool}/
t_{\rm shk}\al 1$, so that it
is meaningful to discuss a two-phase medium.
Our results show that locally $\Upsilon \approx 0.1$ for CNM and
$\Upsilon \approx 0.3$ for WNM at a pressure of $P_{\rm th}/k = 3000$
K cm$^{-3}$ (see eq.~[\ref{eq:upsilon}]). As a function of Galactic 
radius we find that
$t_{\rm cool}\propto T/(n\Lambda)\propto\exp(R_k/2.94)$.
Ignoring the
weak variation of $t_{\rm shk}$ due to the variation
in $\Sigma_{\rm WNM}$ and in $n_{\rm WNM}^{-0.1}$, we
have $t_{\rm shk}\propto\dot\varsigma_{\rm SN}^{-1}
\propto \exp(R_k/3.5)$.  As a result, we
find $\Upsilon \propto \exp(R_k/18.4)$, and we conclude
that the turbulence parameter is $\al 1$  and weakly 
dependent on  radius throughout the Galactic disk.

In Appendix~\ref{appen:turbheating} we discuss the potential role
of turbulence in heating the WNM and CNM phases. Using the
admittedly uncertain turbulent heating rate as a function of Galactic radius
given by equation~(\ref{eq:gturb1}), at $R=17$ kpc  we find that
 $P_{\rm min}$, $P_{\rm th,\, ave}$, and $P_{\rm max}$ are about a factor
2 higher than for the non-turbulent heating case.
However,
since $\langle P_{\rm WNM} \rangle$ remains above $P_{\rm min}$ to
$R = 18$ kpc, turbulent heating does not change our conclusion
that turbulent fluctuations will produce
cold gas that is thermally stable in the outer Galaxy. The rate of
turbulent heating does not exceed the rate of
photoelectric heating out to $R\sim 18$ kpc, and from equation 
(\ref{eq:upsilonturb})
we conclude that our assumption of thermal balance remains
approximately valid.

We conclude by summarizing our most important results. We have shown
that both observational evidence and our theoretical models
presented here indicate that the thermal pressure in the ISM of
the Galaxy lies in the relatively narrow range between
$\pmin$ and $\pmax$ for 3 kpc $< R < 18$ kpc. We have calculated
$\pmin (R)$, $\pmax (R)$ and an estimate of the thermal pressure
$P_{\rm th,\,ave}(R)$ in the Galaxy. We have shown that CNM gas must
exist out to 18 kpc. We present phase diagrams for several galactocentric
radii and for several cases of varying opacity to EUV and soft X-ray
flux. Understanding the neutral phases of the ISM and their dependence
on the radiation field is an important step in understanding the formation
of molecular clouds and the global star formation rates in a galaxy.

Acknowledgments.  We thank L. Blitz, J. Dickey, C. Heiles,
A. Lazarian, H. Liszt, E. Ostriker, and R. Snell for helpful comments,
W. Dehnen for providing his code to calculate the Galactic
potential, J. Slavin for providing the stellar EUV and
SNR X-ray spectra, and T. Sodroski, and N. Odegard for the COBE Galactic
longitude profile. We also thank the referee Don Cox for his insightful
comments.
MGW is supported in part by a NASA LTSA grant
NAG5-9271.
The research of CFM is supported in part by NSF
grant AST-0098365. The research of DJH is supported by NASA RTOP
344-04-10-02, which funds the Center for Star Formation Studies,
a consortium of researchers from NASA Ames, University of California
at Berkeley, and University of California at Santa Cruz.

\newpage

\appendix

\section{Analytic Solution for Thermally Supported WNM at Solar 
                                     Circle}\label{appen:thermalWNM}

  We can treat analytically the problem discussed 
in \S~\ref{subsub:thermalsupwnm} for the thermal support of the WNM 
if we assume that the
mass in the disk is distributed exponentially with height above
the plane,
\beq
\rho_t=\rho_{t0} e^{-z/z_d},
\eeq
where $z_d$ is the scale height and $\rho_t$ is the mass density of
all the matter---stars, gas, and dark matter.
The dominant contribution to the mass in the disk at $R=R_0$ is the 
``thin disk"
of stars with $z_d=180$ pc \citep{deh98}.
Solution of the
equation of hydrostatic equilibrium for the gas, which is
assumed to be isothermal with sound speed $\sigma_{\rm th}$, gives
\beq
n=n_0\exp\left[-\beta x+\beta(1-e^{-x})\right],
\eeq
where $x\equiv z/z_d$ and
\beq
\beta\equiv\frac{4\pi G\rho_{t0}z_d^2}{\sigma_{\rm th}^2}
        \equiv\frac{z_d^2}{h^2}.
\label{eq:beta}
\eeq
In the limit of large
$x$ ($z\gg z_d\sim 180$ pc), the distribution becomes 
an exponential with a scale height
$z_d/\beta=h^2/z_d$; in the limit
of small $x$ (which is more relevant for
the gas distribution in the disk), the gas density approaches a Gaussian,
\beq
n\rightarrow n_0\exp\left(-\frac{z^2}{2h^2}\right).
\eeq
Numerically, we have $h=0.89 T^{1/2}$ pc at the solar
circle, based on a stellar density in
the midplane of $0.115\,\, M_\odot$~pc\eee\ from \cite{deh98}
(with the $2.5\,\, M_\odot$ pc$^{-2}$ augmentation to the thin stellar
disk discussed above)
and a gas density in the
midplane of $0.034\,\, M_\odot$~pc\eee\ from \cite{dic90} and
\cite{mck90}.  For an adopted temperature of 8000 K, this
gives $h=80$ pc for the WNM.

      As remarked above, the value of the midplane density is determined
by requiring that the column density agree with the observed value,
\beq
N_0=2n_0 h \beta^{1/2}\int_0^\infty
      \exp\left[-\beta x+\beta(1-e^{-x})\right] dx \equiv
      (2\pi)^{1/2}n_0 h\phi_\beta.
\label{eq:s0}
\eeq
The fact that the mass is distributed exponentially rather than
uniformly increases
the Gaussian scale height of the gas from
$h$ to $\phi_\beta h$.
A little algebra shows that the factor $\phi_\beta$ is approximately
\beq
\phi_\beta\simeq 1+\frac 13 \left(\frac{2}{\pi\beta}\right)^{1/2}
    +\frac{1}{12\beta}.
\eeq
This approximation is accurate to within 2\% for $\beta>1$.
For $\beta>1$ we have $1<\phi_\beta<1.35$, so the deviation
from Gaussian behavior is not large.
With these results, we then find that the pressure the H~I would
exert in the midplane if it were all WNM is
\beq
\frac{P_{\rm WNM'}}{k}=\frac{1.1 N_{\rm H\,I} T}{(2\pi)^{1/2}
        \phi_\beta h}.
\eeq
For our adopted parameters ($N_{\rm H\,I}=6.2\times 10^{20}$ cm\eee,
$z_d=180$ pc, $T=8000$ K and $h=80$ pc), we find
$P_{\rm WNM'}/k=7800$ K cm\eee\ at the solar circle.  By comparison,
the exact numerical
solution using the \cite{deh98} potential gives
$P_{\rm WNM'}/k=8615$ K cm\eee, which is satisfactory agreement.

\section{Turbulent Heating}\label{appen:turbheating}

In addition to the heating processes we have
considered---photoelectric, X-ray, and cosmic-ray heating---
turbulent, or mechanical, heating may also be important.
\cite{cox79}  estimated that about 30\% of the energy of a
supernova would go into the compression of interstellar
clouds followed by radiative losses.
Numerical simulations \citep{cowie81}
confirmed this estimate. \cite{spi82}  estimated
that about 4\% of the energy of a supernova would go into
acoustic waves, and suggested that the absorption of these
sound waves could be an important source of heating for
the warm phase of the ISM.  \cite{fer88}  generalized
this discussion to consider the generation and damping
of hydromagnetic waves produced by supernova remnants (SNRs).
\cite{mint97}  used observations of interstellar
scintillation to infer the amplitude of fluctuations in the
ISM, and then set constraints on how these fluctuations
damp.  \cite{mintbal97} and \cite{mathis00} studied
the effect of a turbulent heating rate of $\Gamma\sim 10^{-25}$
erg s\e\ H\e\ on the WIM.  \cite{sel99}  have
estimated the heating rate due to the dissipation of turbulence
generated by Galactic differential rotation.
Turbulent heating has also
been investigated in molecular clouds \citep[e.g.,][]{sto98,mac99}.

  Despite more than two decades of work on turbulent heating
of the ISM, the rate remains very uncertain.  This uncertainty
stems directly from our lack of understanding of interstellar
turbulence---how it is generated, how it propagates, and how
it dissipates.  We have already encountered this uncertainty
when we tried to estimate $\Upsilon$, which measures
the degree to which non-turbulent heating is in balance
with radiative cooling (see \S\ 2).
Most of the heating associated with SNRs is very
intermittent, with gas being heated by shocks
that are separated by long time intervals.  As
a result, even though about 30\% of the energy of
a SNR may go into cloud heating, most of this
heat may be radiated away in a short time while
the gas is substantially hotter than average.
This is particularly true for the WNM, which has a
rapidly rising cooling rate above $10^4$ K due to Ly$\alpha$
cooling.
The fact that \cite{hei02} did not
find many WNM features with line widths 
above that 
corresponding
to the temperature we have calculated
for the WNM suggests that
impulsive heating is not a dominant process in
determining the temperature of the interstellar H~I.
\cite{cox79} reached the same conclusion based on
the fact that the observed level of turbulence
in clouds is relatively small.

         In view of the uncertainties associated
with turbulent heating in the ISM, we have
not included it in our basic models.
Here we shall estimate the rate of turbulent heating
and determine how it would affect our results.

\subsection{Dissipation of Turbulent Energy}

   Let $\dot \epsilon$ be the rate of dissipation of
turbulent energy per unit mass, which is equivalent to
the turbulent heating rate per unit mass.  On dimensional
grounds, we expect $\dot\epsilon\sim \delta v^3/\ell$,
where $\delta v$ is the rms velocity in a region
of size $\ell$
(Landau \& Lifshitz 1987; Stone et al.\ 1998; Mac Low 1999).
In other words, the kinetic energy per unit mass is dissipated
in a time of order $\ell/\delta v$. A simple global estimate
shows that turbulent heating is in fact unimportant in
the ISM: The turbulent energy in the ISM is about
$0.5M\delta v^2\sim 10^{54}$ erg, where we set $M=10^9\; M_\odot$
and $\delta v=10$ km s\e.
The scale on which the turbulence in the WNM has an amplitude
of 10 km s\e\ is about 200 pc (\S~\ref{sec:tur}), so the turbulent dissipation
time is about 20 Myr.\footnote{At first glance, it is puzzling
that this estimate for the dissipation time exceeds the
estimated time interval between shocks in the WNM of about 5 Myr
in \S2.  However, as discussed above, much of the
shock energy is radiated
promptly and does not contribute to the level of subsonic
turbulence.}
The resulting heating rate is about
$1.6\times 10^{39}$ erg s\e, or $4\times 10^5\;L_\odot$.
By comparison, the luminosity of the Galaxy in the C~II
158 $\mu$m line is $5\times 10^7\;L_\odot$ \citep{wright91}, 
so turbulent
heating is negligible on a galactic scale.  However, as
we shall see, it is relatively more important in the outer
Galaxy.

  We make the dimensional argument for the heating
rate 
exact by introducing the constant $\phied$,
\beq
\dot\epsilon\equiv\phied\;\frac{\delta v^3}{\ell}
        =3^{3/2}\phied\;\frac{\sigma^3}{\ell}\;,
\label{eq:epsdot}
\eeq
where $\sigma$ is the 1D turbulent velocity dispersion.
The turbulent heating rate per hydrogen, $\gturb$,
is then given by
\beq
\gturb  =  \frac{\rho\dot\epsilon}{n}=\muh\dot\epsilon
        = 3^{3/2}\muh \phied \;\frac{\sigma^3}{\ell}\;,
\label{eq:gturb}
\eeq
where $\muh=2.34\times 10^{-24}$~g is the mass per hydrogen.

      The numerical simulations of Stone et al.\ (1998) and of \cite{mac99}
 show that equation (\ref{eq:epsdot}) applies to both subsonic
and supersonic turbulence.  In supersonic turbulence, the energy is
dissipated primarily in shocks, so the dissipation is highly localized
in space and time.  As discussed above, for this reason
shocks do not contribute
effectively to the general heating of the CNM and WNM, so we are more
interested in the subsonic case.  In the subsonic,
non-magnetic case, we expect the
turbulence to follow the Kolmogorov scaling, in which $\dot\epsilon$
is independent of scale ($\sigma\propto \ell^{1/3}$).
When magnetic fields are included, \cite{gold95}  found
the same scaling, so this should be generally valid in the
ISM for scales such that the flow is subsonic.
Since $\dot\epsilon$ is independent of scale for subsonic turbulence,
the turbulent heating in this case should be widely distributed,
albeit with substantial fluctuations associated with intermittency.
On the other hand,
for supersonic turbulence, we expect 
$\sigma\propto \ell^{q}$ with $q > 1/3$
\citep{lar79,bold02}. 
This scaling is
observed in molecular clouds \citep{hey97},
which are highly supersonic.
In this case we have
$\dot\epsilon\propto\sigma\propto\calm^{3-1/q}$, where $\calm$ is the
Mach number of the flow.  The fraction of the total turbulent heating
that is subsonic and thus widely distributed is thus about $1/\calm^{3-1/q}$
for $\calm\geq 1$, corresponding to $1/\calm$ for $q=\frac 12$ \citep{lar79}
and $1/\calm^{1/3}$ for $q=0.375$ \citep{bold02}.
The length scale that separates subsonic from supersonic
turbulent motions in the WNM, and of CNM clouds in the WNM,
is $\ellp\sim 215$~pc (\S\ 2).

The parameter $\Upsilon\equiv t_{\rm cool}/t_{\rm shk}$
that describes the strength
of the turbulence is directly proportional to $\gturb$.
Recall from \S\ 2 that $t_{\rm cool}=\frac 52 (1.1 nkT)/n^2\Lambda$
for a neutral gas with 10\% He; in
terms of the isothermal sound speed $\sth$,
this is $\frac 52 (\muh\sth^2
/n\Lambda)$. We estimated the shock time to be $t_{\rm shk}
=\ell_P/(\surd 2\sth)$.  For a Kolmogorov-type spectrum
($\sigma^3\propto \ell$), this becomes $t_{\rm shk}=(\sth^2/\surd
2)(\ell/\sigma^3)$.  We then find
\beq
\Upsilon=\frac{t_{\rm cool}}{t_{\rm shk}}
        =0.68\;\frac{(\gturb/\phied)}{n\Lambda}.
\label{eq:upsilonturb}
\eeq
Thus, shocks are important in driving the gas away from thermal
balance if and only if turbulent heating (evaluated with
$\phied=1$) is important.

        What do numerical simulations say about the value
of the parameter $\phied$?
Stone et al.\ (1998) considered MHD turbulence driven by a range of
wavelengths, with the power peaking at a length scale
$\ell_d$.\footnote{In comparing with their work, it must be kept
in mind that they expressed their results in terms of the
size of the simulation box $L$, not $\ell_d=L/8$, and they
left open the question as to whether the dissipation rate
is $\sim\delta v^3/L$ or $\sim\delta v^3/\ell_d$. We assume the
latter, which is consistent with the subsequent work of \cite{mac99}.}
Their results imply
$\phied=0.94$ for the case in which the
initial field strength was such that
the Alfven velocity $v_A$ and the isothermal sound speed
$\sth$ were equal,
as is approximately true in the diffuse ISM.
In his study of turbulent dissipation,
\cite{mac99}  calculated three models for the case $v_A=\sth$.
He used
two slightly different approaches for measuring the dissipation
rate, one in terms of the volume-averaged rms velocity
and one in terms of the mass-averaged rms velocity. Like
Stone et al.\ (1998), we have used the latter approach.
In our notation, Mac Low found $\phied=0.9\pm0.13$~dex for this case, in
fortuitously good agreement with Stone et al.
(this agreement between the two calculations was not as good for other values
of $\sth /v_A$).

          In both these simulations,
the turbulence is ``balanced," in that the average wave power
is the same in both directions along the field.  Since the
sources of interstellar turbulence are intermittent in space
and time, actual interstellar turbulence is likely to
be imbalanced, and \cite{cho02}
have shown that this can substantially
reduce the decay rate of the turbulence.  The value $\phied\sim 1$
found in these simulations is thus an upper limit to the value
expected in the ISM.  In our work, we shall somewhat arbitrarily
adopt $\phied=0.5$ as representative of IS turbulence.

      Recall that we parameterized the strength of the turbulence
in terms of the 1D velocity dispersion at a scale of
1 pc, $\sigma(1)$.  We estimated this from observation
by assuming Kolmogorov scaling, which is reasonable
since the motions of the CNM clouds observed by
\cite{hei02}  have a velocity dispersion
that is very nearly the same as the thermal velocity
in the WNM (for $T=8000$ K, both are $\sim 7.1$ km s\e).
In order to extend the estimate of the heating rate to
other parts of the Galaxy, we assume that $\sigma$ is approximately
constant and that $\ell$ scales as the thickness of the H~I disk,
so that $\sigma^3/\ell=\sigma(1;R_0)^3\propto 1/H_z^{\rm HI}$,
where $\sigma(1;R_0)$ is the value of $\sigma(1)$ at the solar
circle.
From equation (\ref{eq:gturb}) we find that
the turbulent heating rate is then
\begin{eqnarray}
\gturb & = & 3.94\times 10^{-27}\phied\left[\frac{\sigma(1;R_0)}{1\
       {\rm km\ s^{-1}}}\right]^3\frac{H_z^{\rm HI}(R_0)}{H_z^{\rm HI}(R)}
       ~~~{\rm erg~s^{-1}~H^{-1}}\; ,\\
       & = & 3.40\times 10^{-27}\left(\frac{\phied}{0.5}\right)
        \left[\frac{\sigma(1;R_0)}{1.2~{\rm km\ s^{-1}}}\right]^3
        {\rm min}\left[1,\;\exp\left(\frac{8.5-R_k}{6.7}\right)\right]
        {\rm erg~s^{-1}~H^{-1}}.
\label{eq:gturb1}
\end{eqnarray}
Note that the estimated turbulent heating rate inside $R_0$
is constant, since
the scale height of the H~I does not change there.

\subsection{Turbulent Energy from Differential Rotation}

    Next, consider the extraction of turbulent energy from
differential rotation.  As pointed out by \cite{sel99},
this process occurs
at a rate
\beq
\rho\dot\epsilon=-\left(-\frac{B_rB_\phi}{4\pi}+\rho v_r\delta v_\phi
        \right)\frac{d\Omega}{d\ln R}\; ,
\eeq
where the rotation velocity is $\Omega R \hat{\bf\phi}$
and the shear is in the radial direction.
In a steady state, this rate of extraction of energy from
differential rotation will be balanced by dissipation of
energy, and so long as the velocities induced by the
differential rotation are subsonic, much of this energy
should be dissipated in a turbulent cascade. \cite{haw95}
have carried out MHD simulations of the generation of
turbulent velocities and magnetic fields in a shearing box.
They present
detailed results for one model of a shearing box simulation;
using these results, we find that the excitation
rate simplifies to
\beq
\dot\epsilon =  0.75\delta v^2\left(-\frac{d\Omega}{d\ln R}\right)\; .
\eeq
Note that this expression is based on the assumption that
the velocities are generated by the differential rotation,
as may be the case in the outer Galaxy; it may not apply if
the velocities are generated by other mechanisms, such
as supernovae.  Even in regions where the turbulence is
primarily due to differential rotation, this expression must
be regarded as higly approximate since it does not take into
account either the vertical structure or the multi-phase nature
of the ISM.

   The dynamical model of the Galaxy that we have adopted from
\cite{deh98} has a rotational velocity that declines
slowly beyond the solar circle,
$d\Omega/d\ln R=(-244$~km~s\e)$/R$ for
$R\ga 8.5$~kpc.
We adopt 6 km s\e\  as a typical
velocity dispersion for gas in the outer parts of disk galaxies
\citep[e.g.,][]{mar01},
somewhat less than the
7 km s\e\ for the CNM and the 11 km s\e\ for the WNM in
the solar neighborhood \citep{hei02}.
The turbulent heating rate due to differential rotation is then
\beq
\gturb=1.50\times 10^{-26}\left(\frac{\sigma}{6~{\rm km\ s^{-1}}}
                \right)^2\frac{1}{R_k}~~~{\rm erg\ s^{-1}\ H^{-1}}
                ~~~~~(R_k>8.5~{\rm kpc}).
\label{eq:gturb2}
\eeq
This heating rate is within a factor 2 of that in
equation (\ref{eq:gturb1}) for $8.5<R_k\la 25$~kpc.  Given
the uncertainties, these two estimates are in satisfactory
agreement.  The generic estimate in equation (\ref{eq:gturb1})
is about twice that for differential rotation in equation
(\ref{eq:gturb2}) at the solar circle, consistent with the
idea that supernovae are an important source of turbulent
motions in the local ISM.

\subsection{Results}

To estimate the effects of turbulent heating in the Galaxy,
we adopt equation (\ref{eq:gturb1}), since it applies to both the
case in which energy is injected by differential rotation
and that in which it is injected by explosive events such as supernovae.
It should be borne in mind that this estimate for the
turbulent heating is quite uncertain, since it is based on
highly idealized numerical simulations and an uncertain correction
for an imbalanced turbulent cascade; on the other hand, it is
reassuring that it is in accord with simple dimensional analysis
(eq.\ \ref{eq:gturb} with $\phied$ of order unity).
The estimate also depends on the amplitude of the turbulence
in the subsonic regime, which is uncertain at present.
Our estimate for the turbulent heating in the diffuse H~I is about
30 times less than that invoked by \cite{mintbal97} and
\cite{mathis00} in their studies of heating of the diffuse H~II.

       Turbulent heating, unlike heating by cosmic rays,
 EUV/soft X-rays, and photoelectric heating, is
independent of the depth
into the cloud, and it therefore becomes more important at high
column densities.
If turbulent heating at the rate given by
equation (\ref{eq:gturb1}) occurs throughout the Galactic
disk, then
for our fiducial column density of
$10^{19}$ cm\ee, at $P_{\rm min}$, our estimate of the
turbulent heating rate
exceeds the cosmic ray rate for Galactic radii between 3 kpc and
18 kpc and amounts to 70\% of the photoelectric
heating rate at $R_k = 17$ kpc.
For $N=10^{20}$~cm\ee, the turbulent rate is always greater
than the cosmic ray rate and is equal to 93\% of the photoelectric
rate at $R_k = 17$ pc. The average thermal pressure in a two-phase ISM 
including turbulent heating is
given by
\beq
P_{\rm th,\, ave}/k =1.2\times 10^4\exp(-R_k/7.5)~~~~~{\rm K~cm^{-3}},
\eeq
where the fit is good to $\pm 10$\% between $ 3 < R_k < 18$ except at
our $R_k= 11$ model point where the fit overestimates the thermal
pressure by  $25$\% .

The turbulent heating has the greatest effect in the outer Galaxy where
heating rates based on stellar photons or supernovae are small.
With turbulent heating, at $R_k = 17$ kpc, we find
$P_{\rm min}/k = 650$ K ${\rm cm^{-3}}$,
$P_{\rm th,\, ave}/k = 1370$ K ${\rm cm^{-3}}$, and
$P_{\rm max}/k = 2900$ K ${\rm cm^{-3}}$ which are factors of
1.7, 1.9, and 2.1 respectively times the non-turbulent case.
We find that $P_{\rm WNM'}$ (the thermal pressure in
the midplane based on the simple---and incorrect---assumption
that the WNM is supported only by thermal pressure) 
falls below $P_{\rm max}$ at
$R_k = 15$ kpc.
The ``mean'' thermal pressure
$\langle P_{\rm WNM} \rangle$ 
(an estimate of the 
thermal pressure in the midplane based on the assumptions that
all the H~I is WNM and that the WNM fills space)
falls below $P_{\rm max}$ at approximately
$R_k = 13.5$. It follows that the H~I cannot be all WNM
out to $R_k=13.5$, and a two-phase medium {\it must}
exist out to that point.   
Since $\langle P_{\rm WNM} \rangle$ 
remains above $P_{\rm min}$ out to $R = 18$ kpc,
we conclude that pressure
fluctuations will in fact produce a two-phase medium to these distances.

\section{Analytic Thermal Balance Model for Cool Gas}\label{appen:toymodel}

When gas is in thermal balance, the thermal pressure $P$ can be
expressed as a function of either density or temperature. We seek
$P(T)$ for the temperature range $T\al 1000$ K where [C~II]
158 $\mu$m and [O~I] 63 $\mu$m radiation dominates the gas cooling,
and for $n < n_{\rm cr}^{\rm H}({\rm [C~II]}) \simeq 3000$ cm$^{-3}$
and $n_e < n_{\rm cr}^{\rm e}({\rm [C~II]}) \simeq 30$ cm$^{-3}$ so
that the [C~II] + [O~I] cooling is proportional to $n^2$.

\subsection{Thermal Balance: Heating and Cooling}

Cooling rates per unit volume can be written as
$n_{\rm c} n \Lambda_{\rm s}^{\rm c}$, where $n_{\rm c}$
is the density of the collisional agent ($n_{\rm e}$ or $n_{\rm H}$),
$n$ is the hydrogen nucleus density, and $\Lambda_{\rm s}^{\rm c}$
is the cooling rate
coefficient, which takes into account the gas phase abundance of the
species at the solar circle (Table~\ref{tbl:modelparam1}). We find
\beq
\Lambda_{\rm C~II}^{\rm H} = 3.15\times 10^{-27}e^{-0.92/T_2}\zg
                        \,\,\,\,\,\,\,\, {\rm erg\,\, cm^{3}\,\, s^{-1}}\, ,
\eeq
\beq
\Lambda_{\rm C~II}^{\rm e} = 1.4\times 10^{-24}T_2^{-1/2}
                                      e^{-0.92/T_2}\zg
                        \,\,\,\,\,\,\,\, {\rm erg\,\, cm^{3}\,\, s^{-1}}\, ,
\eeq
and
\beq
\Lambda_{\rm O~I}^{\rm H} = 2.5\times 10^{-27}T_2^{0.4}
                                      e^{-2.28/T_2}\zg
                           \,\,\,\,\,\,\,\, {\rm erg\,\, cm^{3}\,\, s^{-1}}\, ,
\eeq
where $T_2 =T/(100\,\, {\rm K})$. We assume that H atom collisions
dominate electron collisions. Comparison of
$n_{\rm H}\Lambda_{\rm C~II}^{\rm H}$
with $n_{\rm e}\Lambda_{\rm C~II}^{\rm e}$ shows that this condition is
equivalent to $x_{\rm e}\equiv n_{\rm e}/n < 2.3\times 10^{-3}T_2^{1/2}$.

Assuming that H atoms dominate the excitation of [C~II] and [O~I],
that $n < n_{\rm cr} \simeq 3000$ cm$^{-3}$, and
$100\,\,{\rm K} < T < 1000\,\, {\rm K}$
(the relevant regime for $\tmin$), we use numerical
results using the above cooling coefficients  $\Lambda_{\rm C~II}^{\rm H}$
and  $\Lambda_{\rm O~I}^{\rm H}$ to obtain a simple form for
the total cooling coefficient
\beq
\Lambda_{\rm tot}^{\rm H} = 5.4\times 10^{-27}T_2^{0.2}
                                      e^{-1.5/T_2}\zg
                       \,\,\,\,\,\,\,\, {\rm erg\,\, cm^{3}\,\, s^{-1}}\, ,
\label{eq:cool}
\eeq
which is good to $\pm 20$\%. 
A simple form such as this is required,
rather than $\Lambda_{\rm C~II}^{\rm H} + \Lambda_{\rm O~I}^{\rm H}$,
in order to analytically determine $\tmin$.

We modify the grain photoelectric heating rate of \cite{bak94} by
multiplying by a factor of 1.3, which accounts for a higher PAH
abundance ($6\times 10^{-7}$ by number relative to hydrogen) compared
with that assumed in \cite{bak94}.
The heating rate per unit volume for $T \al 1000$ K
is given approximately [see \S~5, eq.\ (23)]
\beq
n\Gamma_{\rm pe} = \frac{1.1\times 10^{-25} \gs \zd n}
{1 + 3.2\times 10^{-2}\left(\frac{\gs T_2^{1/2}}{n_{\rm e}\phi _{PAH}}\right)^{0.73}}
\,\,\,\,\,\,\,\,\, {\rm erg\,\,cm^{-3}\,\, s^{-1}}\, .
\label{eq:gphe}
\eeq
Recall (see \S\ 5) that $\phi _{PAH}$ is a  parameter of order unity which scales the
PAH collision rates; $\phi _{PAH} = 0.5$ in our standard model.

Note that when the unity term dominates in the denominator, the photoelectric
heating is not significantly suppressed by positive charging, i.e., the
second term in the denominator represents effects of grain charging. To obtain
an analytic solution for $\tmin$ below, we will require the second term
to dominate, as it often does for a range of conditions centered on solar
neighborhood values.

\subsection{The Electron Density ${\rm n_e}$}

In order to obtain an analytic expression for $P$ under thermal
balance conditions, an analytic expression for $n_e$ is required
to substitute into the grain photoelectric heating equation~(\ref{eq:gphe}).
The chemistry leading to steady state electron abundances is quite
complex and interesting. We use the results of the numerical code
to determine the dominant reaction chains leading to $n_{\rm e}$, and
to determine the major competitors to these reactions. We then find
an analytic expression for $n_{\rm e}$, along with the conditions
required to ensure the assumed reaction chain.

For densities and temperatures near $n_{\rm min}$ and $\tmin$
at the solar circle, ${\rm H^+}$ is the dominant ion species and
therefore $n_{\rm e}\approx n_{\rm H^+}$. ${\rm H^+}$ is produced
mostly by the EUV/soft X-ray photoionization of H with a minor contribution
from cosmic rays. We find that $n_{\rm He^+}\approx 0.3 n_{\rm H^+}$
under a variety of conditions at a column of $N_{\rm cl}\simeq 10^{19}$
cm$^{-2}$, a result due to the higher photoionization cross section
of He at soft X-ray energies counterbalanced by the 0.1 abundance
of He relative to H. Since ${\rm H^+}$ is a surrogate to obtain
the electron abundances, we roughly account for ${\rm He^+}$
by increasing the photoionization rate of H by 30\% to a rate
$1.3\times 10^{-16}$ s$^{-1}$ at $N_{\rm cl}=10^{19}$ cm$^{-2}$
appropriate to the solar neighborhood. The rate
$\zio$ is expressed as
$\zio = \zeta_t/ 10^{-16}$ s$^{-1}$, where $\zeta_t$
is the total ionization rate (including primary and secondary ionizations)
 of H  by photons and cosmic rays.
The destruction of ${\rm H^+}$ is dominated by reactions with
${\rm PAH^-}$ or
\beq
{\rm H^+  + PAH^-}  \rightarrow  {\rm H + PAH^0}\,, \,\,\,\,
\kappa_1  = 8.3\times 10^{-7} \phi _{PAH}T_2^{-0.5}\,\,\,{\rm cm^3\,\, s^{-1} }\, ,
\eeq
where $\kappa_1$ is the rate coefficient calculated from \cite{draine87}
using disk PAHs with $N_{\rm C}= 25$ carbon atoms and the disk radius
$a= (N_{\rm C}/1.222)^{0.5}$. Competing reactions are
\beq
{\rm H^+  + PAH^0}  \rightarrow  {\rm PAH^+ + H}\,, \,\,\,\,
\kappa_2  = 3.1\times 10^{-8} \phi _{PAH} T_2^{-0.5}\,\,\,{\rm cm^3\,\, s^{-1} }\, ,
\eeq
and
\beq
{\rm H^+  + e}  \rightarrow  {\rm H} + h\nu \,, \,\,\,\,
\kappa_3  = 8.0\times 10^{-12} T_2^{-0.75}\,\,\,{\rm cm^3\,\, s^{-1} }\, ,
\eeq
where the reaction coefficients $\kappa_2$ and $\kappa_3$  are
calculated similarly to $\kappa_1$. Equating the formation of ${\rm H^+}$
to the destruction of ${\rm H^+}$, we obtain for the electron density
\beq
n_e  =  \frac{1.3\times 10^{-16} \zio \ n}
            {\kappa_1 n_{\rm PAH^-}}
     =  1.6\times 10^{-10} \frac{\zio T_2^{1/2} n}{\phi _{PAH}n_{\rm PAH^-}}
                                    \,\,\,\,\,\,\,{\rm cm^{-3}}\, .
\label{eq:ne1}
\eeq

The analytic solution for $n_{\rm e}$ therefore requires a solution for
$n_{\rm PAH^-}$. We simplify PAH ionization state calculations by assuming
single-sized PAHs with three ionization states, ${\rm PAH^-}$, ${\rm PAH^0}$,
and ${\rm PAH^+}$. We find numerically that, over a wide range of
relevant parameter space, neutral ${\rm PAH^0}$ dominates the population with
$n_{\rm PAH^0} \simeq 0.7 n_{\rm PAH}$ where
$n_{\rm PAH} = 6.0\times 10^{-7}n\zd $ is the total density of PAHs in
all states. ${\rm PAH^+}$ is formed by the FUV photoreaction
\beq
h\nu   + {\rm PAH^0}  \rightarrow  {\rm PAH^+ + e} \,, \,\,\,\,
\kappa_4  = 7.85\times 10^{-9}\gs \,\,\,{\rm s^{-1} }\, ,
\eeq
where the rate coefficient $\kappa_4$ is calculated from \cite{bak94}
using $N_{\rm C} = 35$ carbon atoms and assuming a disk geometry.
${\rm PAH^+}$ is destroyed primarily by recombination with electrons
\beq
{\rm PAH^+ + e}  \rightarrow  {\rm PAH^0}\,, \,\,\,\,
\kappa_5  = 3.5\times 10^{-5} \phi _{PAH} T_2^{-0.5}\,\,\,{\rm cm^3\,\, s^{-1} }\, .
\eeq
The condition that $n_{\rm PAH^+}/n_{\rm PAH^0} < 1 $ is then
$7.85\times 10^{-9} \gs/(\kappa_5 n_{\rm e}) < 1$ .
${\rm PAH^-}$ is formed by
\beq
{\rm PAH^0 + e}  \rightarrow  {\rm PAH^-}\,, \,\,\,\,
\kappa_6  = 1.34\times 10^{-6}\phi _{PAH} \,\,\,{\rm cm^3\,\, s^{-1} }\, ,
\eeq
and ${\rm PAH^-}$ is destroyed primarily by FUV photodetachment
\beq
{\rm PAH^-} + h\nu  \rightarrow  {\rm PAH^0 + e}\,, \,\,\,\,
\kappa_7  = 2.00\times 10^{-8}\gs \,\,\,{\rm s^{-1} }\, .
\eeq
The condition $n_{\rm PAH^-}/n_{\rm PAH^0} < 1$ is then
$\kappa_6 n_{\rm e}/2.0\times 10^{-8} \gs < 1$.

Assuming $n_{\rm PAH^0} = 0.7n_{\rm PAH}$, we find (with $n$ and $n_{\rm e}$
in units of cm$^{-3}$)
\beq
n_{\rm PAH^-} = 2.8\times 10^{-5} n_{\rm e} n \phi _{PAH} \zd G_0^{\prime -1}
                       \,\,\,\, {\rm cm^{-3}}\, ,
\eeq
or substituting into equation~\ref{eq:ne1}, we find
\beq
n_{\rm e} = 2.4\times 10^{-3} \zeta_t^{\prime 1/2}T_2^{1/4}
   G_{0}^{\prime {1/2}} Z_{\rm d}^{\prime -1/2}\phi _{PAH}^{-1} \,\,\,\, {\rm cm^{-3}}\, .
\label{eq:ne2}
\eeq

\subsection{Analytic Expression for P(T) and Conditions for Validity}

We substitute equation~(\ref{eq:ne2}) for $n_{\rm e}$ into equation~(\ref{eq:gphe}) for
the heating, equate the heating to the cooling given by
equation~(\ref{eq:cool}),
and obtain for neutral gas in thermal balance with
$100\,\,{\rm K} \al T \al 1000\,\,{\rm K}$
\beq
   P/k = 1.1nT =
       \frac{2240\, \gs \left( \frac{\zd}{\zg} \right) T_2^{0.8}e^{1.5/T_2}}
      {1 + 2.6 \left( \frac{\gs T_2^{1/2} \zd}{\zio}\right)^{0.365}}\, .
\label{eq:pt}
\eeq
This equation is valid if all of the following conditions are met:
\beq
(1)\hspace{1.0in}\frac{\gs T_2^{1/2} \zd}{\zio} < 17
\eeq
Condition (1) ensures that ${\rm H^+}$ is mostly destroyed by ${\rm PAH^-}$,
and not ${\rm PAH^0}$. This condition and the rest below are derived by
using the rate coefficients above, along with equation ~(\ref{eq:ne2}) for
$n_{\rm e}$.
\beq
(2)\hspace{1.0in}\frac{\gs T_2^{1/2} \zd}{\zio} < 112
\eeq
Condition (2) ensures that neutral ${\rm PAH^0}$ and not ${\rm PAH^+}$
dominates the PAH population. Note that if condition (1) is satisfied,
condition (2) is automatically satisfied.
\beq
(3)\hspace{1.0in}\frac{\gs \zio}{T_2^{1/2}\zd} < 0.9 n^2 \phi _{PAH}^2
\eeq
Condition (3) ensures that H atoms, and not electrons, dominate
collisional excitation of [C~II] 158 $\mu$m. In these equations,
the hydrogen nucleus number density $n$ is in ${\rm cm^{-3}}$.
\beq
(4)\hspace{1.0in}\frac{\gs}{\zd T_2^{1/4}} < 2.9 n\phi _{PAH}^2
\eeq
Condition (4) ensures that ${\rm H^+}$ is destroyed by ${\rm PAH^-}$, and
not by electron recombination.
\beq
(5)\hspace{1.0in}\frac{\gs\zd}{T_2^{1/2}\zio} > 2.5\times 10^{-2}
\eeq
Condition (5) ensures that neutral ${\rm PAH^0}$, and not ${\rm PAH^-}$
dominate the PAH population.
\beq
(6)\hspace{1.0in}\frac{\gs\zd T_2^{1/2}}{\zio} > 9.5\times 10^{-3}
\eeq
Condition (6) ensures that FUV dominates the destruction of ${\rm PAH^-}$,
and not reactions with ${\rm H^+}$. Note that for $T_2\simeq 1-10$,
condition (5) automatically insures condition (6).
\beq
(7)\hspace{1.0in}\frac{\gs\zio T_2^{1/2}}{\zd Z_{\rm g}^{\prime 2}} > 3.5\times 10^{-3}n^2
\phi _{PAH}^2
\eeq
Condition (7) ensures that the dominant ion is ${\rm H^+}$ and not
${\rm C^+}$, which is ignored in the analytic analysis. These conditions
ensure the validity of equation ~(\ref{eq:pt}) for $P(T)$.

Although the number of conditions is large, inspection of each
of them reveals that for $n\sim 10$ cm$^{-3}$ and $\phi _{PAH}=0.5$, all of the conditions
are met for a wide range of conditions centered on solar neighborhood
values. The motivation for deriving $P(T)$ is to find an analytic solution
for $\pmin$, $\tmin$, and $\nmin$. Thus, the range of validity is
centered on $T\simeq \tmin \simeq 250$ K and $n\simeq \nmin \simeq 10$
cm$^{-3}$, as we shall derive below. To calculate $\pmin$, we take
$d P/d T = 0$ and solve for $\tmin$. Substitution
of $\tmin$ into equation (\ref{eq:pt}) gives $\pmin$. No analytic
solution is possible unless we simplify the denominator of
$P(T)$ (eq.~[\ref{eq:pt}]). If conditions (1-7) are satisfied, then
the second term in the denominator is usually larger than unity.
The denominator originates from the grain photoelectric heating
equation, and the second term corresponds to the effects of
positive charging on the grains/PAHs. The condition that the
second term dominates is
\beq
(8)\hspace{1.0in}\frac{\gs \zd T_2^{1/2}}{\zio} > 0.072\, .
\eeq
Note that condition (8) and condition (5) are nearly identical,
for $T_2\sim 2.5$. With all eight conditions satisfied, we find the
solutions for $\tmin$, $\pmin$, and $\nmin$ given in the
main text. Substitution of $\nmin$ (with the factor of 
unity in the denominator removed) and $\tmin$ into the
eight conditions give the simplified set of conditions given in
the text.

\clearpage
{\small
\tablenum{1}
\begin{tabular}{llc}
\multicolumn{3}{c}{Table 1.\ \ Model Parameters 
              - Local Values\tablenotemark{a}\label{tbl:modelparam1}}\\ \hline\hline
 {Parameter} & {Value} &{Reference} \\ \hline
$J^{\rm FUV}(R_0)$\tablenotemark{b} & $2.2\times 10^{-4}$  erg cm$^{-2}$ 
   s$^{-1}$ sr$^{-1}$ & 1,2\\
$\zeta_{\rm CR}$\tablenotemark{c} & $1.8\times 10^{-17}$ ${\rm s^{-1}}$ & 3\\
$\zeta_{\rm XR}$\tablenotemark{d} & $1.6\times 10^{-17}$ ${\rm s^{-1}}$ & 4\\
$\cal{A}_{\rm C}$\tablenotemark{e} & $1.4\times 10^{-4}$ & 5,6\\
$\cal{A}_{\rm O}$\tablenotemark{f} & $3.2\times 10^{-4}$ & 7\\
$\cal{A}_{\rm PAH}$\tablenotemark{g} & $6\times 10^{-7}$ & 4,8\\
$A_{\rm V}$\tablenotemark{h} &$N/(2\times 10^{21}$ cm$^{-2}$) & 9\\
$\Sigma_{\rm H\, I}$\tablenotemark{i}  & 5 $M_{\odot}$ pc$^{-2}$ & 10\\
$\Sigma_{\rm H_2}$\tablenotemark{j}  & 1.4 $M_{\odot}$ pc$^{-2}$& 11\\
$\Sigma_{\rm WNM}$\tablenotemark{k}  & 2.75 $M_{\odot}$ pc$^{-2}$ & 10\\
$\langle n_{\rm H\, I} \rangle$\tablenotemark{l} &  0.57 cm$^{-3}$ & 10\\
$H_z^{\rm H~I}$\tablenotemark{m} & 115 pc & 10\\
$H_z^{\rm H_2}$\tablenotemark{n} & 59 pc & 11\\ \hline
\end{tabular}

\noindent
Refs.--{(1)~\cite{dra78}, (2)~\cite{hab68}; (3)~WHMTB;
(4)~This paper; (5)~\cite{car96}; (6)~\cite{sof97}; 
(7)~Meyer et al.\ (1998); (8)~\cite{tie99};
(9)~\cite{boh78}; (10)~\cite{dic90}; (11)~\cite{bron00}}\\
\noindent
$^{\rm {a}}${Values at solar circle  $R=R_0=8.5$ kpc.}\\
$^{\rm {b}}${Intensity of FUV (6 eV $< h\nu  <$ 13.6 eV)
   interstellar radiation field. This intensity is a factor 1.7 higher
    than the integrated field
    of Habing (1968) which has a value of
    $J^{\rm FUV}(R_0) = 1.3\times 10^{-4}$ erg cm$^{-2}$ s$^{-1}$
    sr$^{-1}$. The interstellar field has a value of
    $G_0 = 1.7$  in  units of the Habing field.}\\
$^{\rm {c}}${Primary cosmic-ray ionization rate of hydrogen.}\\
$^{\rm {d}}${Primary EUV plus soft X-ray ionization rate of hydrogen at
a cloud depth of $N_{\rm cl} = 10^{19}$ cm$^{-2}$.
The primary rate  depends on the adopted value of the WNM
cloud column, $N_{\rm cl}$. The total rate is higher than 
the primary rate due to secondary ionizations and increases with lower
electron fraction. For our standard model, over the 2 phase range,
the total rate is a factor $3-5$ higher than the primary rate.
For
$N_{\rm cl} = 1.0\times 10^{20}$ cm$^{-2}$,
$\zeta_{\rm XR}(R_0) = 8.9\times 10^{-19}$ s$^{-1}$;
$N_{\rm cl} = 3.0\times 10^{18}$ cm$^{-2}$,
$\zeta_{\rm XR}(R_0) = 7.6\times 10^{-17}$ s$^{-1}$.}\\
$^{\rm {e}}${Gas phase carbon abundance per H nucleus.}\\
$^{\rm {f}}${Gas phase oxygen abundance per H nucleus.}\\
$^{\rm {g}}${PAH abundance per H nucleus. This abundance gives a total
number of C atoms in PAHs of $22\times 10^{-6}$ relative to hydrogen.}\\
$^{\rm {h}}${Magnitudes of visual extinction per hydrogen column density.}\\
$^{\rm {i}}${Atomic hydrogen surface density through full disk,
           $N({\rm H~I})= 6.25\times 10^{20}$ cm$^{-2}$.}\\
$^{\rm {j}}${Molecular hydrogen surface density through full disk,
           $2N({\rm H_2})=1.75\times 10^{20}$ cm$^{-2}$.}\\
$^{\rm {k}}${WNM column density through full disk,
           $N_{\rm WNM,\, d}=3.45\times 10^{20}$ cm$^{-2}$.}\\
$^{\rm {l}}${Mean H~I density.}\\
$^{\rm {m}}${H~I half height to half intensity.}\\
$^{\rm {n}}${${\rm H_2}$ half height to half intensity.}\\
}

\tablenum{2}
\begin{tabular}{lcccc}
\multicolumn{5}{c}
{Table 2.\ \ Model Parameters 
      - Galactic Values\tablenotemark{a}\label{tbl:modelparam2}} \\\hline\hline
{$R$} & {$G_0^{\prime}$\tablenotemark{b}} & 
{$\zeta_{{\rm CR}}^{\prime}$\tablenotemark{c}} &
{$\zeta_{{\rm XR}}^{\prime}$\tablenotemark{d}} &
{$\zd = \zg$\tablenotemark{e}} \\
{(kpc)} & &  & & \\ \hline
3 & 2.95 & 3.64 & 5.02 & 2.43 \\
4 & 2.73 & 2.56 & 3.62 & 2.07 \\
5 & 2.21 & 1.83 & 2.72 & 1.76 \\
8.5 ($=R_0$)& 1.00 & 1.00 & 1.00 & 1.00 \\
11 & 0.509 & 0.424 & 0.359 & 0.668 \\
15 & 0.198 & 0.194 & 0.156 & 0.351 \\
17 & 0.116 & 0.182 & 0.145 & 0.254 \\
18 & 0.0634 & 0.176&  0.140 & 0.216 \\ \hline
\end{tabular}

\noindent
$^{\rm {a}}${Values scaled to solar circle.}\\
$^{\rm {b}}${Scaled Intensity of FUV interstellar radiation field.}\\
$^{\rm {c}}${Scaled primary cosmic ray ionization rate}\\
$^{\rm {d}}${Scaled primary EUV and soft X-ray ionization rate.}\\
$^{\rm {e}}${Scaled gas phase metallicity, $Z_g^\prime$, and 
dust/PAH abundances, $Z_d^\prime$.}\\

\newpage
{\tablenum{3}\label{tbl:physicalcond}}
\begin{figure}[hb]
\centering
\includegraphics{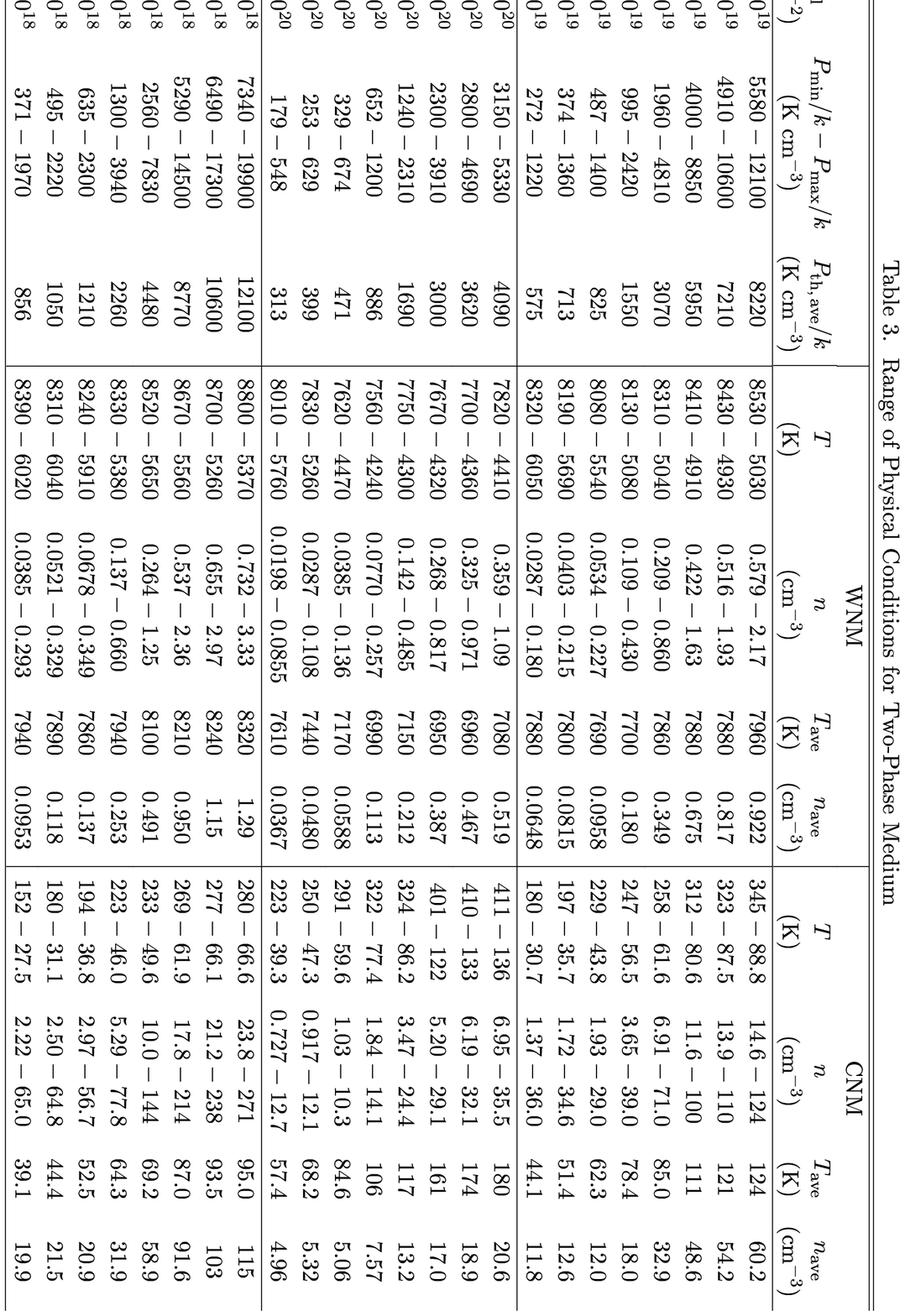}
\end{figure}

\clearpage


{\small
\tablenum{4}
\noindent
\begin{tabular}{lccccc}
\multicolumn{6}{c}{Table 4.\ \ Dependence on Model Parameters$^{\rm {a}}$
              \label{tbl:depmodelparam}} \\ \hline
{Model} & {Standard}$^{\rm b}$ & 
Low $\phi_{\rm PAH}$ $^{\rm c}$ &
High $\phi_{\rm PAH}$ $^{\rm d}$ & 
Low $n_{\rm PAH}/n$ $^{\rm e}$ &
Low $G_0$$^{\rm f}$ \\ \hline \hline
$P_{\rm min}$ (K cm$^{-3}$), $N=N_{\rm cl}$\tablenotemark{g} & 
           1960 & 1560 & 2270 & 1580 & 1460\\
$P_{\rm max}$ (K cm$^{-3}$), $N=N_{\rm cl}$\tablenotemark{g} & 
           4810 & 3150 & 5970 & 3920 & 3980\\
$\pthave$\tablenotemark{h}\,\,  (K cm$^{-3}$), 
           $N=N_{\rm cl}$\tablenotemark{g} & 
           3070 & 2220 & 3680 & 2490 & 2410\\
$T_{\rm ave}$\tablenotemark{i} (K), $N=N_{\rm cl}$\tablenotemark{g} & 
            85 & 94 & 96 & 79 & 76 \\
$n_{\rm ave}$\tablenotemark{j} (K), $N=N_{\rm cl}$\tablenotemark{g} & 
             33 & 21 & 35 & 29 & 29 \\
$n\Lambda$\tablenotemark{k}\,\, (erg s$^{-1}$ H$^{-1}$),
      $N=N_{\rm cl}$\tablenotemark{g} & $4.1(-26)$  & $3.2(-26)$ &
       $5.0(-26)$ & $3.3(-26)$ & $3.1(-26)$ \\
${\rm C~I^*/C~I_{\rm tot}}$\tablenotemark{l}\,\, ,
      $N=N_{\rm cl}$\tablenotemark{g} & 0.190 & 0.145 &
       0.205 & 0.168 & 0.167 \\
$T_{\rm ave}$\tablenotemark{i}\,\, (K), 
         $1\times 10^{20}$ (cm$^{-2}$)\tablenotemark{m}&
                        71 & 67 & 84 & 64 & 65 \\
$n\Lambda$\tablenotemark{k} (erg s$^{-1}$ H$^{-1}$),
      $1\times 10^{20}$ (cm$^{-2}$)\tablenotemark{m} & 
       $3.6(-26)$  & $2.6(-26)$ &
       $4.3(-26)$ & $2.8(-26)$ & $2.7(-26)$ \\
${\rm C~I^*/C~I_{\rm tot}}$\tablenotemark{l}\,\, ,
      $1\times 10^{20}$ (cm$^{-2}$)\tablenotemark{m} & 0.201 & 0.162 &
       0.213 & 0.178 & 0.172 \\ 
C~I/C~II\tablenotemark{n}\,\, , 
          $1\times 10^{20}$ (cm$^{-2}$)\tablenotemark{m} &
          2.3(-3) & 7.1(-4) & 6.1(-3) & 1.5(-3) & 3.7(-3) \\ \hline
\end{tabular}

\noindent
$^{\rm a}${$a(-b)$ means $a\times 10^{-b}$.}\\
$^{\rm b}${Standard model with $\phi_{\rm PAH} = 0.5$,
     $n_{\rm PAH}/n = 6\times 10^{-7}$, $G_0 = 1.7$,
     $N_{\rm cl} = 1\times 10^{19}$ cm$^{-2}$.}\\
$^{\rm c}${Low PAH collision rates with
      $\phi_{\rm PAH} = 0.25$, $n_{\rm PAH}/n = 6\times 10^{-7}$, $G_0 = 1.7$,
      $N_{\rm cl} = 1\times 10^{19}$ cm$^{-2}$.}\\
$^{\rm {d}}${High PAH collision rates with
       $\phi_{\rm PAH} = 1.0$, $n_{\rm PAH}/n = 6\times 10^{-7}$, $G_0 = 1.7$,
      $N_{\rm cl} = 1\times 10^{19}$ cm$^{-2}$.}\\
$^{\rm {e}}${Low PAH abundance with
      $\phi_{\rm PAH} = 0.5$, $n_{\rm PAH}/n = 4\times 10^{-7}$, $G_0 = 1.7$,
       $N_{\rm cl} = 1\times 10^{19}$ cm$^{-2}$.}\\
$^{\rm f}${Low FUV field with
       $\phi_{\rm PAH} = 0.5$, $n_{\rm PAH}/n = 6\times 10^{-7}$, $G_0 = 1.1$,
       $N_{\rm cl} = 1\times 10^{19}$ cm$^{-2}$.}\\
$^{\rm {g}}${Model result at cloud depth of
          $N_{\rm cl} = 1\times 10^{19}$ cm$^{-2}$ and $\pthave$.}\\
$^{\rm {h}}${$\pthave=\sqrt{P_{\rm max}\times P_{\rm min}}$\,\,.}\\
$^{\rm {i}}${Temperature of CNM at $\pthave$.}\\
$^{\rm {j}}${Density of CNM at $\pthave$.}\\
$^{\rm {k}}${Gas cooling rate per hydrogen atom 
from [C~II] 158 $\mu$m line emission in CNM.}\\
$^{\rm {l}}${${\rm C~I^*/C~I_{\rm tot}}$ population ratio.}\\
$^{\rm {m}}${Model result at cloud interior at depth of
         $1\times 10^{20}$ cm$^{-2}$ and $\pthave$.}\\
$^{\rm {n}}${C~I/C~II abundance ratio.}\\
}

\newpage
\tablenum{5}
\begin{tabular}{lcccc}
\multicolumn{5}{c}{Table 5.\ \ Parameters in Galactic Disk Potential Model\tablenotemark{a}
              \label{tbl:potential}} \\ \hline\hline
{Component} & {$\Sigma_d$} & {$R_m$} & {$R_d$} & {$z_d$} \\
{} & {($M_\odot$ pc$^{-2}$)\tablenotemark{b}}  & {(kpc)} & {(kpc)} &
{(pc)} \\ \hline

H~I ($R \le 13$ kpc) & 7.94 & 1.0 & 1000 & 178\\
H~I ($R > 13$ kpc) & 571 & 10 & 4.00 & 324\\
${\rm H_2}$ & 57.5 & 3.3 & 2.89 & 63.4 \\
H~II   & 1.39     & 0  & 30.0 & 880 \\
Thin star disk ($R \le R_0$) & 1058 & 0 & 2.55 & 180 \\
Thin star disk ($R  >  R_0$) & 1058 & 0 & 2.55 & $z_d^{*}$\tablenotemark{c} \\
Thick star disk & 70.6 & 0 & 2.55 & 1000 \\ \hline
\end{tabular}

\noindent
$^{\rm {a}}${Disk potential model based on \cite{deh98}. Each component has a
density distribution given by the form
 $\rho (R) = \Sigma_d (2z_d)^{-1} \exp ( -[R_m/R] - [R/R_d] - [|z|/z_d])$.
Bulge and halo components are the same as \cite{deh98} model 2b.} \\
$^{\rm {b}}${Includes He mass.} \\
$^{\rm {c}}${Stellar disk height given by
$z_d^* = 180 \times \Sigma (R_0)/\Sigma(R)$ pc.} \\

\newpage
\centerline{Figures}

\figcaption[HIsurf.ps]{Azimuthally averaged H~I surface density
$\Sigma_{\rm H\, I}$ ({\em dash}) and H$_2$ surface density
$\Sigma_{\rm H_2}$ ({\em dot}) in the
Galactic disk versus Galactocentric radius $R$. Total H~I plus
H$_2$ is shown as a {\em solid} line. Mass does not include He.
\label{HIsurf}}

\figcaption[HIz.ps]{Azimuthally averaged H~I half width to half maximum
height $H_z^{\rm H\, I}$
versus Galactocentric radius $R$.
\label{HIzfig} }

\figcaption[HImean.ps]{Mean H~I density in the Galactic midplane versus
Galactocentric radius $R$. The value
at $R=R_0$ is taken to be $\langle n_{\rm H\, I}\rangle = 0.57$ 
cm$^{-3}$ and scaled by
$\langle n_{\rm H\, I} \rangle\propto 
\Sigma_{\rm H\, I}/H_z^{\rm H\, I}$ at other
radii.\label{fig:HImean}}

\figcaption[Opac.ps]{Calculated FUV opacity in the Galactic midplane versus
Galactocentric radius $R$.\label{fig:Opac}}

\figcaption[FUVf.ps]{Calculated FUV field in the Galactic midplane versus
Galactocentric
radius $R$ normalized to the value at $R=R_0$. At $R=R_0$ the field
strength
is equal to $4\pi J^{\rm FUV}(R_0 ) = 2.7\times 10^{-3}$ ergs cm$^{-2}$
s$^{-1}$. \label{fig:FUVf}}

\figcaption[CR.ps]{Primary cosmic-ray ({\em solid}) and EUV plus
X-ray ({\em dot})
ionization rates versus Galactocentric
radius $R$ normalized to the value at $R=R_0$. At $R=R_0$ the primary
cosmic-ray ionization rate is taken to be $\zeta_{\rm CR}(R_0) =
1.8\times 10^{-17}$  cm$^{-3}$ s$^{-1}$.
The EUV and X-ray rate depends
on the adopted value of the WNM cloud column, $N_{\rm cl}$.
For $N_{\rm cl} = 1.0\times 10^{20}$ cm$^{-2}$,
$\zeta_{\rm XR}(R_0) = 8.9\times 10^{-19}$ s$^{-1}$;
$N_{\rm cl} = 1.0\times 10^{19}$ cm$^{-2}$,
$\zeta_{\rm XR}(R_0) = 1.6\times 10^{-17}$ s$^{-1}$;
$N_{\rm cl} = 3.0\times 10^{18}$ cm$^{-2}$,
$\zeta_{\rm XR}(R_0) = 7.6\times 10^{-17}$ s$^{-1}$.
Note that the EUV and X-ray rate always exceeds the cosmic ray 
rate because of the effects of secondary ionizations. Typical
values for $N_{\rm cl}$ are of order $10^{19}$ cm$^{-2}$.
\label{CosR} }

\figcaption[whmPnRfig.ps]{Phase diagrams showing thermal pressure $P/k$
versus hydrogen nucleus density $n$ at Galactocentric 
radii $R = 3$, 5, 8.5, 11, 15, and 18 kpc. 
Curves apply to the WNM/CNM
boundary at a depth of $1\times 10^{19}$ cm$^{-2}$ through the WNM.
Gas is thermally stable to isobaric perturbations where $d P/d n > 0$. 
\label{whmPnRfig}}

\figcaption[]{Thermal pressure $P/k$
versus hydrogen nucleus  density $n$ at $R = 8.5$ kpc 
showing the effects of varying
the collision rate parameter $\phi_{\rm PAH}$ and the PAH abundances.
For the standard PAH abundances ($n_{\rm PAH}/n = 6\times 10^{-7}$),
curves are shown for
$\phi_{\rm PAH} = 0.25$ ({\em dash}),
$\phi_{\rm PAH} = 0.5$ ({\em solid}),
and $\phi_{\rm PAH} = 1.0$ ({\em dot}). Our standard model uses
$\phi_{\rm PAH} = 0.5$. Also shown is a curve for $\phi_{\rm PAH} = 0.5$
and a low PAH abundance of $n_{\rm PAH}/n = 4\times 10^{-7}$ 
({\em long dash}).
\label{fig:phipah}}

\figcaption[whmPnRNfig.ps]{Thermal pressure $P/k$ versus hydrogen nucleus
density $n$ at Galactocentric
radius $R=8.5$ kpc. Curves are shown for various values of WNM atomic column
density $N_{\rm cl}$ and apply to the WNM/CNM boundary.
\label{whmPnRNfig}}

\figcaption[whmhcR.ps]{{\em Upper Panels}: Heating and Cooling curves 
versus hydrogen
nucleus  density $n$ at
various Galactic distances, $R$. Heating rates
({\em dash}); Photoelectric heating from small grains and PAHs (PE);
EUV and X ray (XR); Cosmic ray (CR); photoionization of C (C~I). Cooling rates
({\em solid}); C~II
158 $\mu$m fine-structure (C~II); O~I 63 $\mu$m fine-structure (O I);
Recombination onto
small grains and PAHs (Rec); Ly$\alpha$ plus metastable
transitions (Ly$\alpha$); C~I fine-structure 609 $\mu$m (C I$^*$);
C I fine-structure 370 $\mu$m (C~I$^{**}$). {\em Lower Panels}:
Gas temperature T ({\em solid}) and electron fraction $n_e/n$ ({\em dash})
versus hydrogen nucleus density $n$. ($a$) $R=5$ kpc. ($b$) $R=8.5$ kpc.
($c$) $R=11$ kpc. ($d$) $R=17$ kpc.
\label{fig:whmhcR}}

\figcaption[IRfig.ps]{Infrared intensity in the Galactic midplane versus
Galactic
longitude. COBE observations ({\em solid}) from \cite{sod94}. Calculated dust
emission
({\em dot}) using the FUV opacity from Fig.\ \ref{fig:Opac}, and the   
FUV field from Fig.\ \ref{fig:FUVf}.\label{fig:IRfig}}

\figcaption[PWNM.ps]{Limiting thermal pressures in the Galactic midplane 
versus Galactocentric
radius $R$. $P_{\rm WNM'}$ ({\em solid}) is the thermal pressure produced
by the H~I  layer assuming all of the gas is in the form of WNM and
supported in hydrostatic equilibrium by thermal pressure. $P_{\rm max}$
({\em short dash}) and $P_{\rm min}$ ({\em long dash}) are the maximum and
minimum pressure range for a two-phase medium. $\pthave$ ({\em dash}) is
the mean pressure $\pthave=(P_{\rm min}\times P_{\rm max})^{1/2}$. 
Panels show the effects of different WNM columns.
($a$) $N_{\rm cl} = 1.0\times 10^{19}$ cm$^{-2}$. ($b$)
$N_{\rm cl} = 1.0\times 10^{20}$ cm$^{-2}$. ($c$)
$N_{\rm cl} = 3.0\times 10^{18}$ cm$^{-2}$. ($d$) $N_{\rm cl} = 1.0\times
10^{19}$ cm$^{-2}$ curves with $\langle P_{\rm WNM} \rangle$ added.
$\langle P_{\rm WNM} \rangle$ is the
thermal pressure that would be present in the Galactic midplane if all of
the H~I had a temperature of 8000 K and a density given by the mean
density shown in Fig.\ \ref{fig:HImean} (see \S~\ref{subsub:turbsupwnm}). 
\label{fig:PWNM}}

\figcaption[TOYsm.ps]{Region of validity for the analytic solution
of $\pmin$ (eq.~[\ref{eq:pmin}]) as a function of the dust abundance
$\zd$, gas phase metal  abundance $\zg$, ionization rate $\zio$, and
FUV radiation field $\gs$. All parameters are scaled to
their value in the solar neighborhood.
The valid region is shown for the
case of dust and metal abundances scaling linearly with the elemental
abundance ($Z^\prime = \zd = \zg$). The shaded region shows the range
in which the analytic solution is good to within $\pm 50$\% of the
numerical results. Also shown are the more restrictive conditions
derived in Appendix~\ref{appen:toymodel} and given in equations
(\ref{eq:con1}) through (\ref{eq:con3}):
({\em solid}) PAHs are positively charged and  the destruction
of ${\rm H}^+$ is  dominated by reactions with
${\rm PAH^-}$ rather than with ${\rm PAH^0}$
(eq.~[\ref{eq:con1}]);
({\em dot}) Electrons are supplied by ${\rm H^+}$ rather than
${\rm C^+}$ (eq.~[\ref{eq:con2}]);
({\em dash}) the destruction of ${\rm H}^+$ is  dominated by reactions with
${\rm PAH^-}$ rather than by electron recombination
(eq.~[\ref{eq:con3}]).
The analytic
solution is good to  within $\pm 45$\% of the numerical results within
this region.
\label{fig:toy} }

\newpage
\plotone{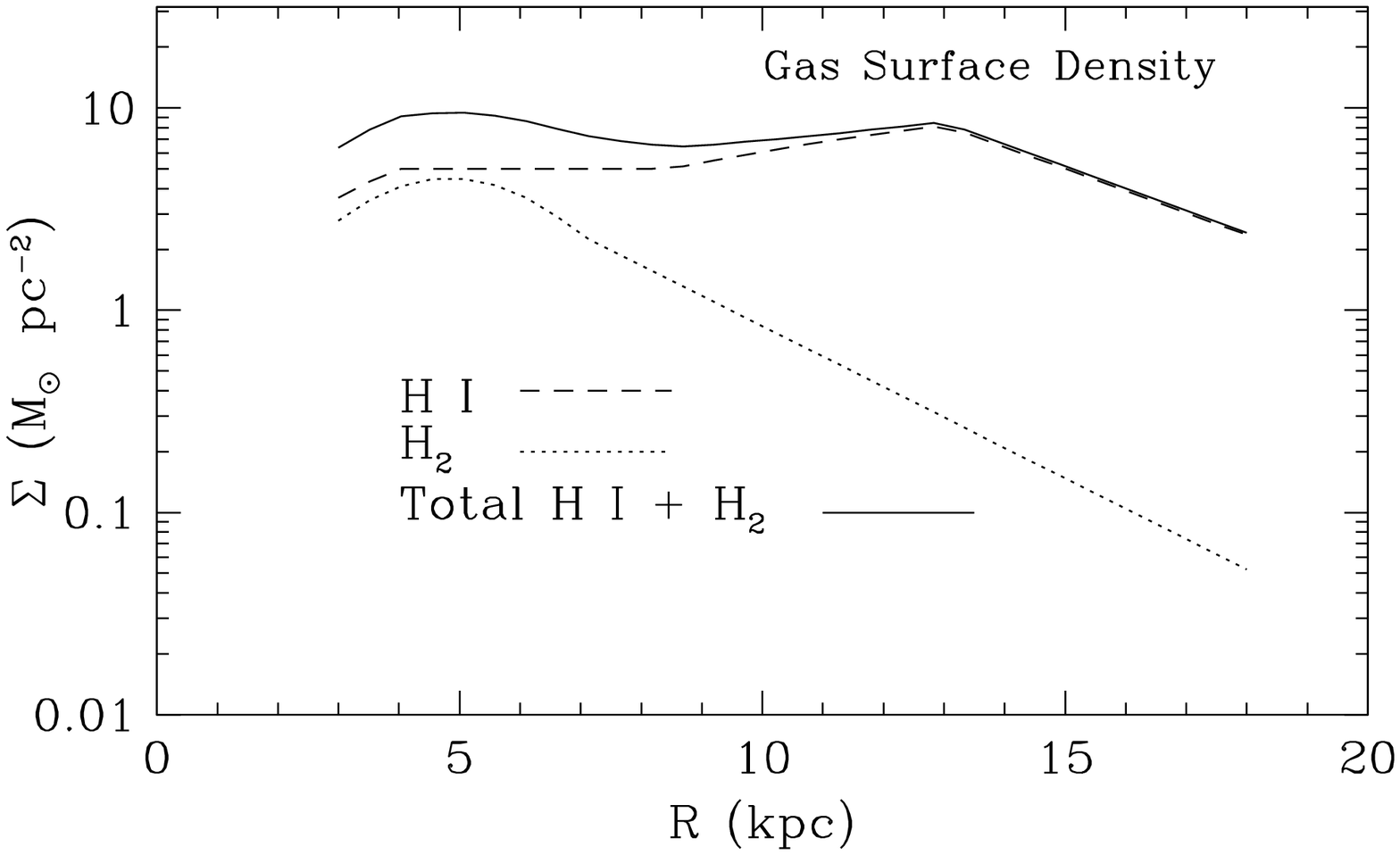}
\newpage
\plotone{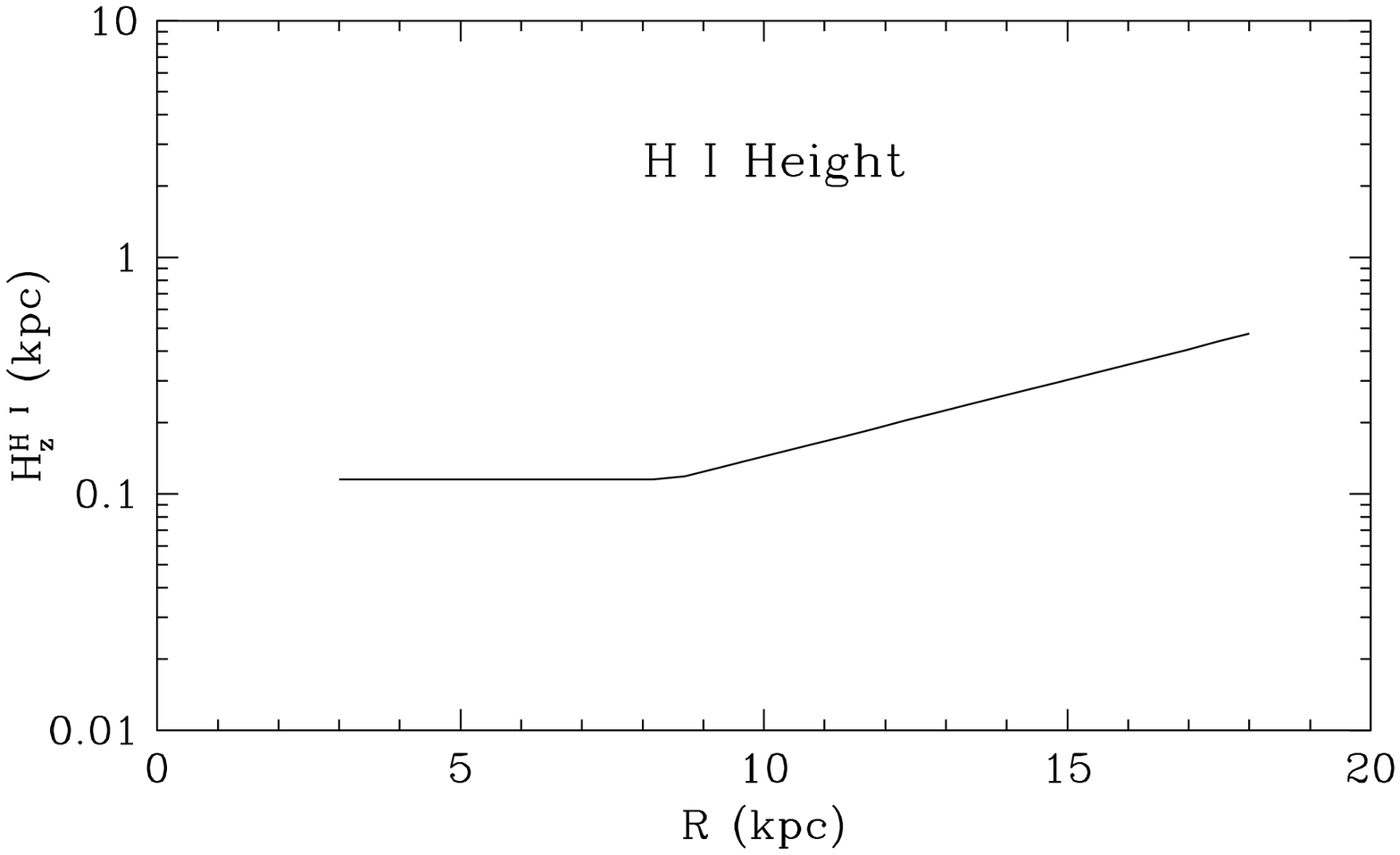}
\newpage
\plotone{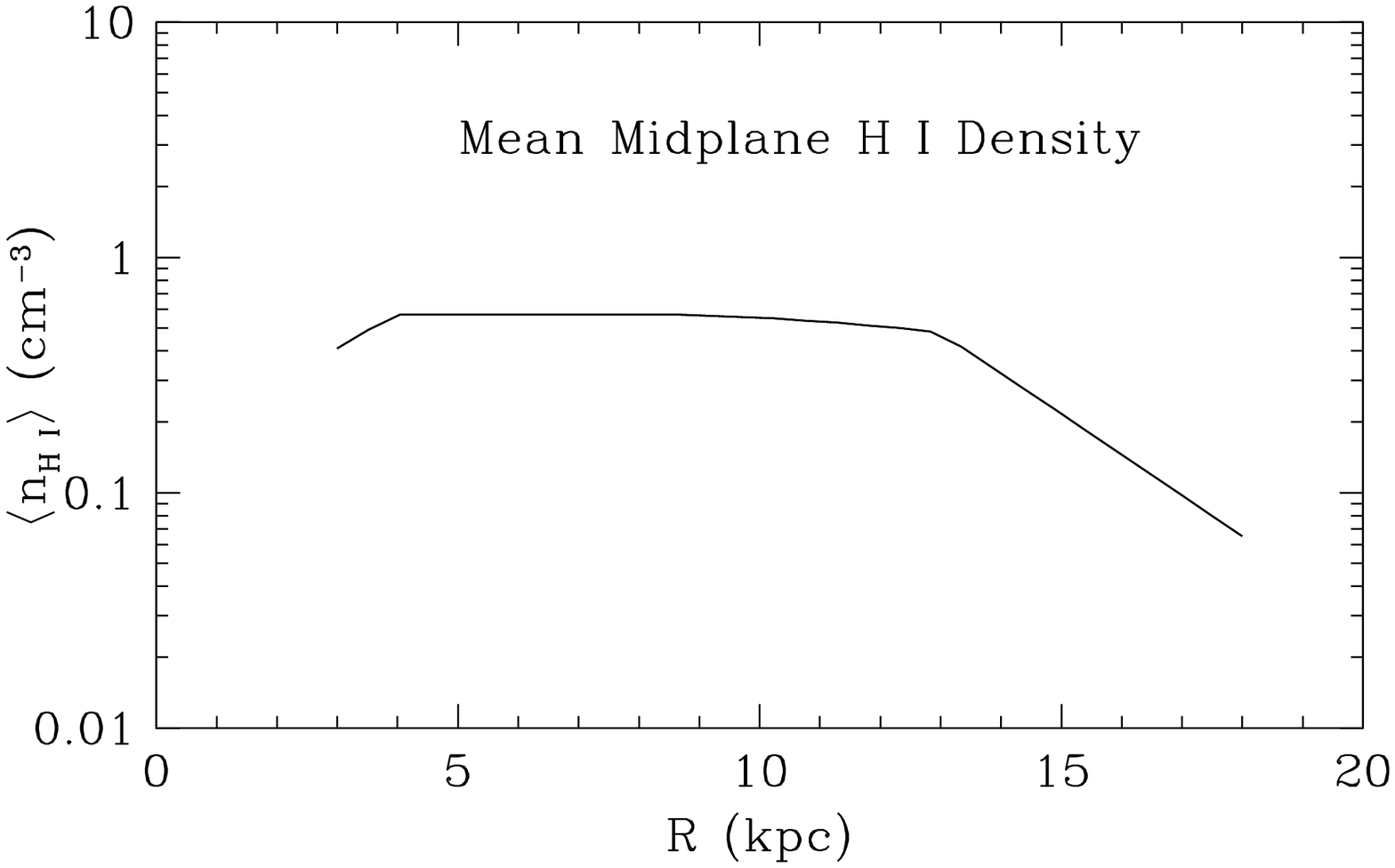}
\newpage
\plotone{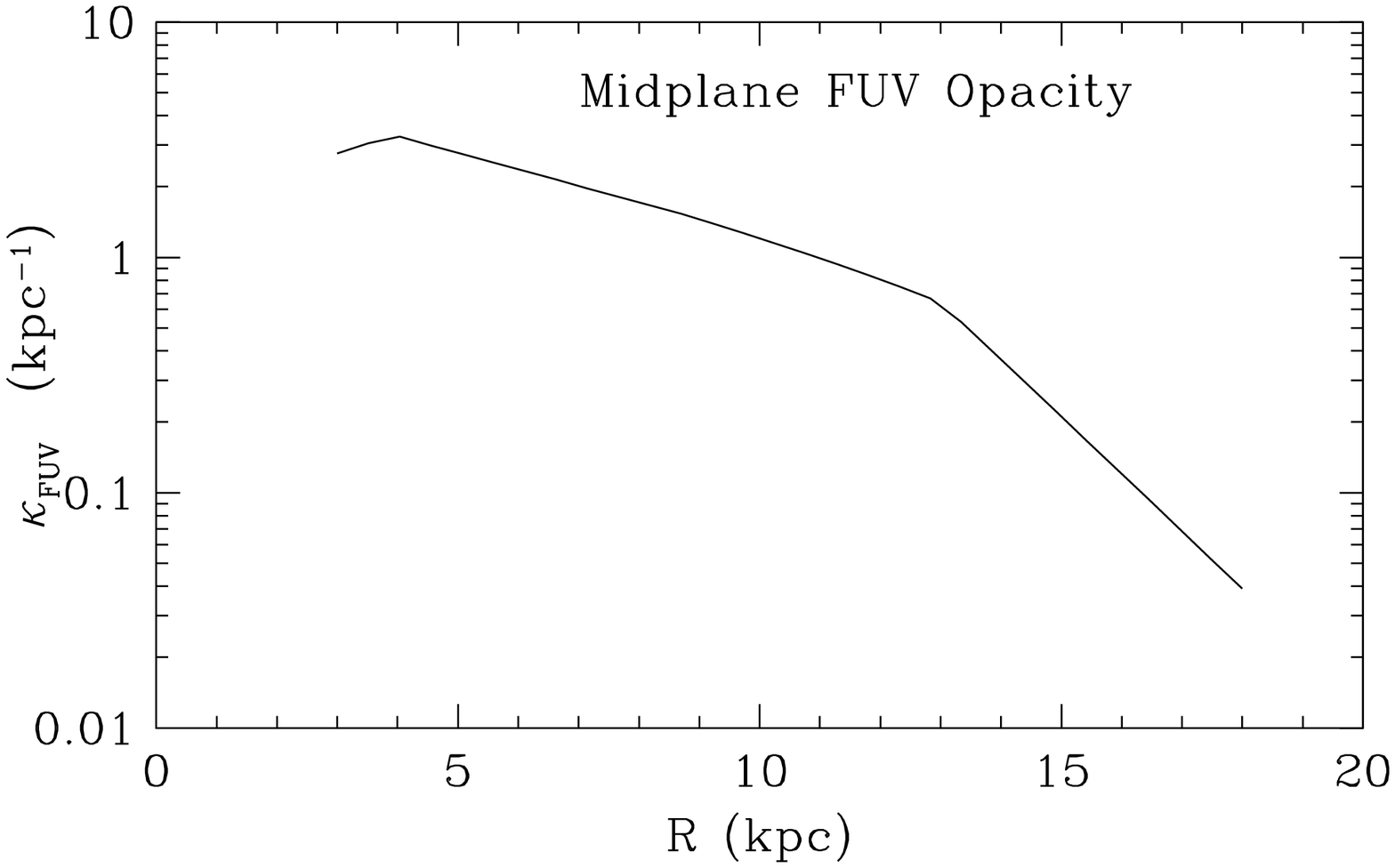}
\newpage
\plotone{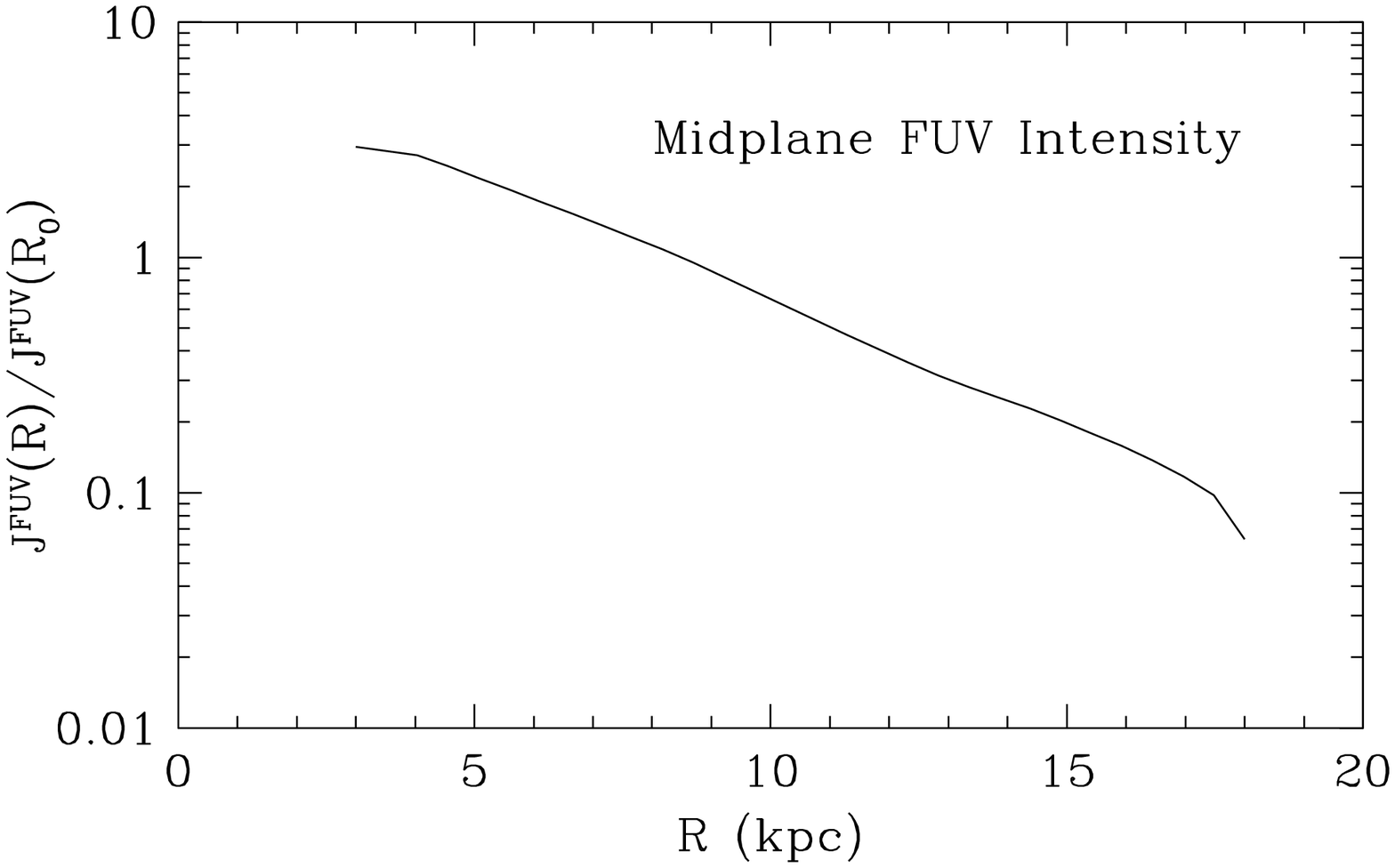}
\newpage
\plotone{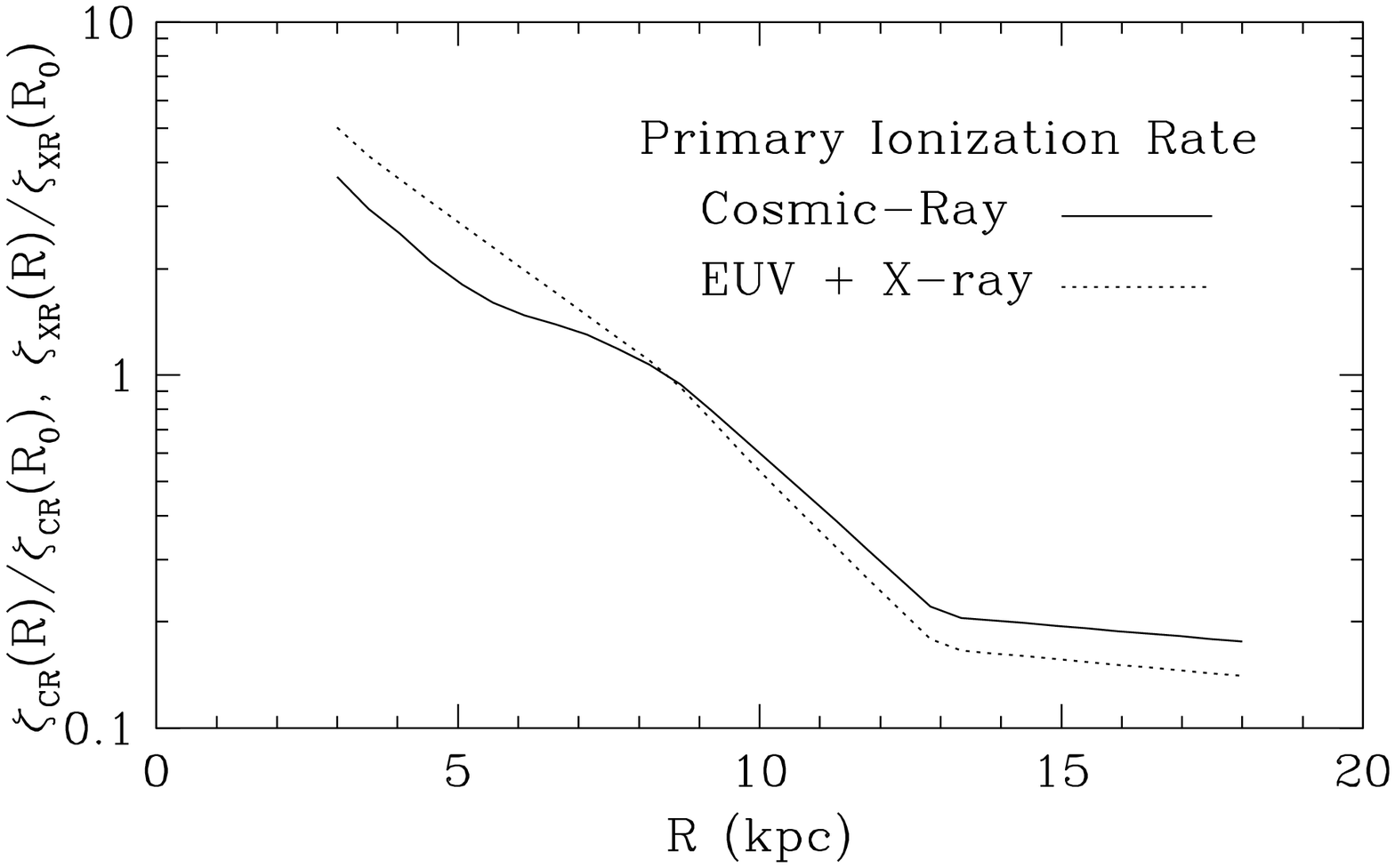}
\newpage
\plotone{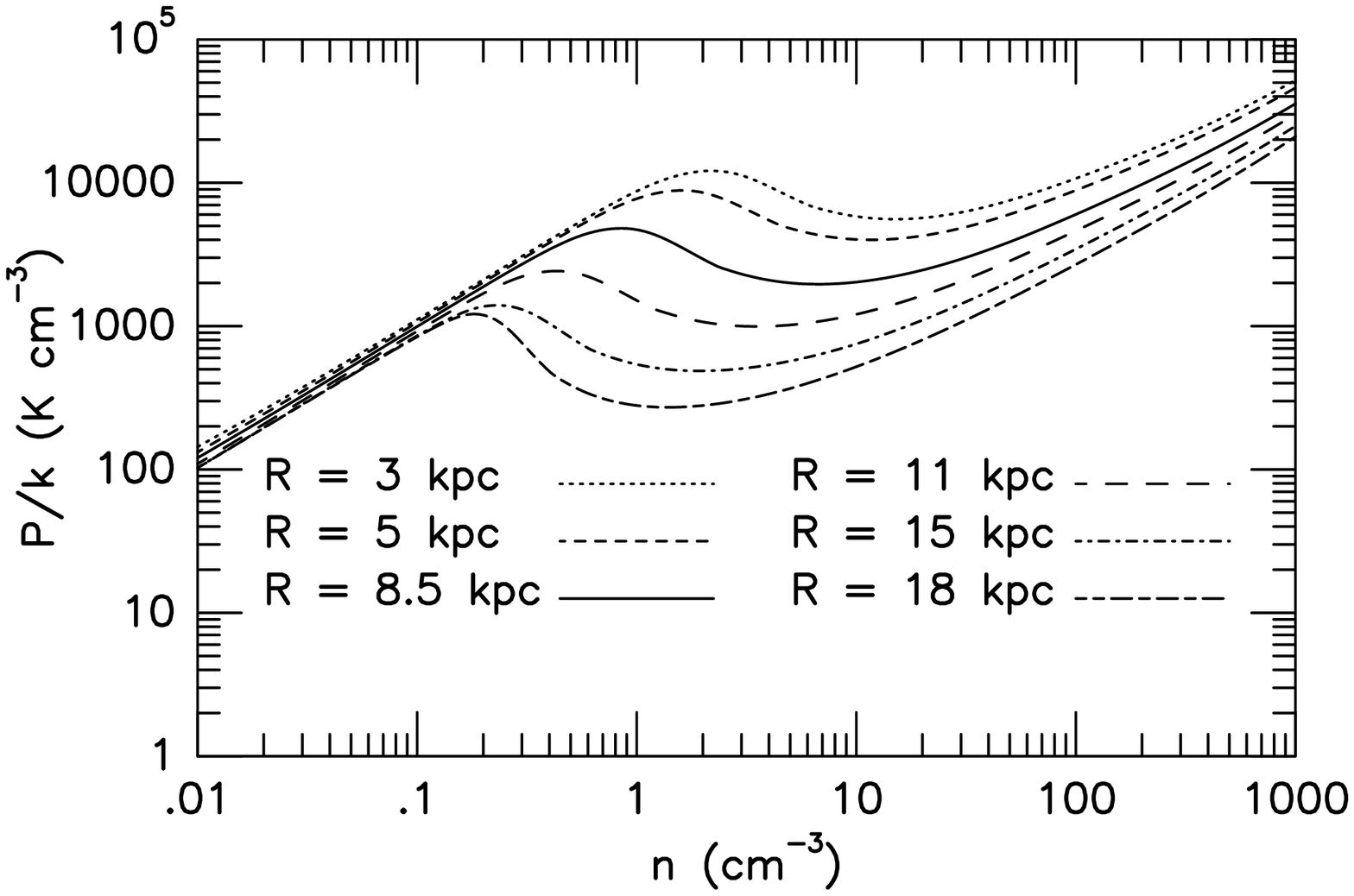}
\newpage
\plotone{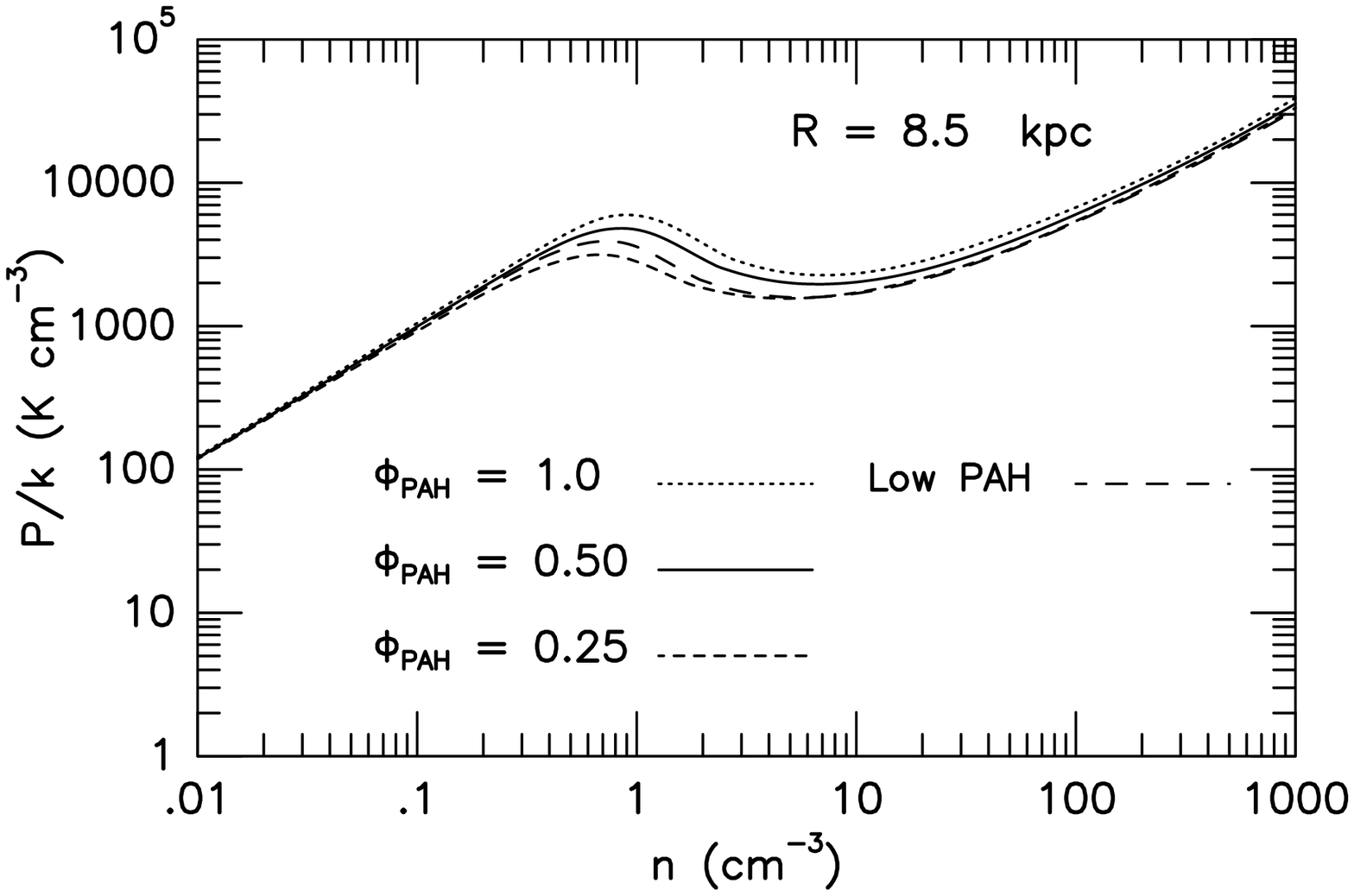}
\newpage
\plotone{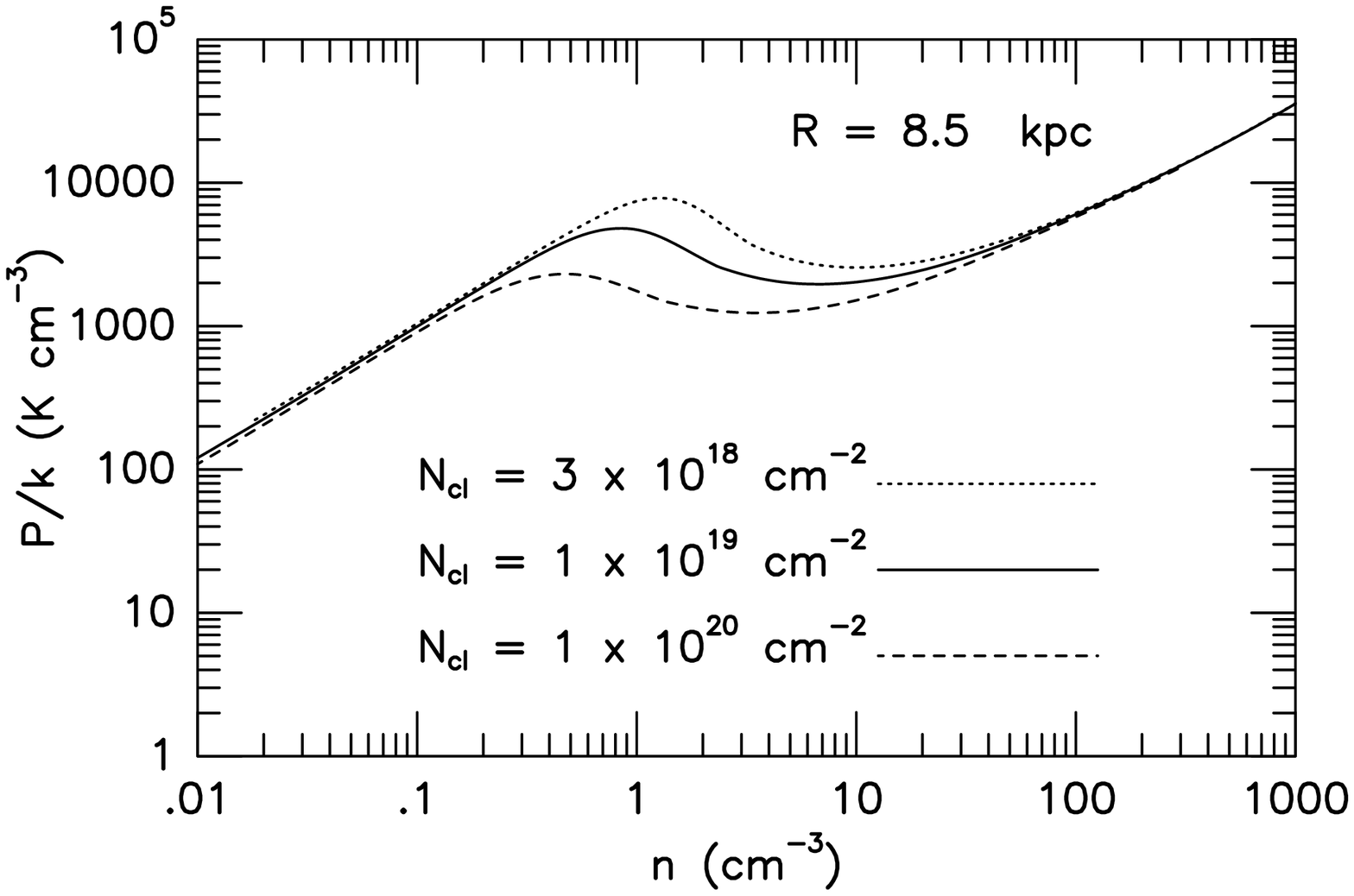}
\newpage
\plotone{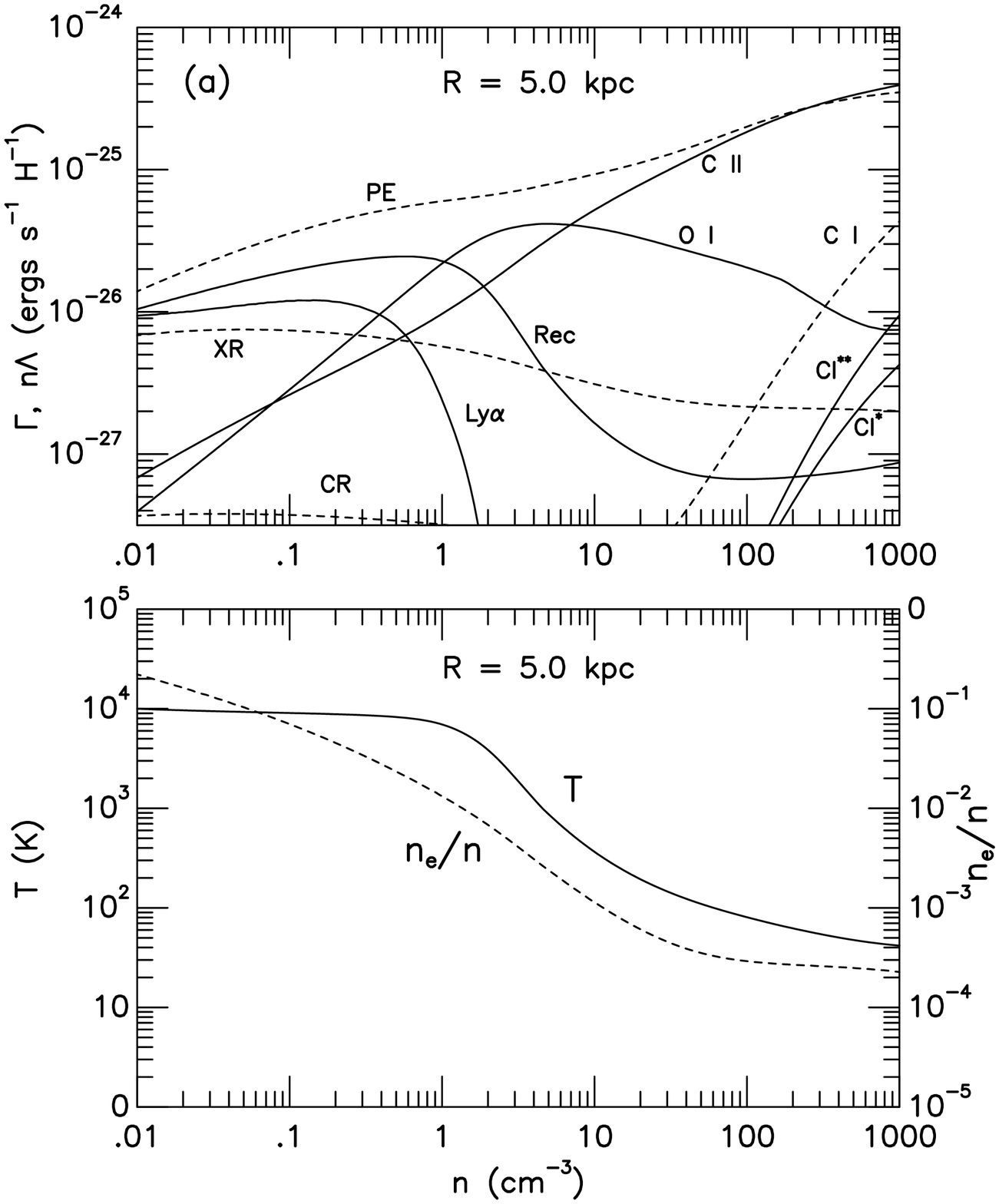}
\newpage
\plotone{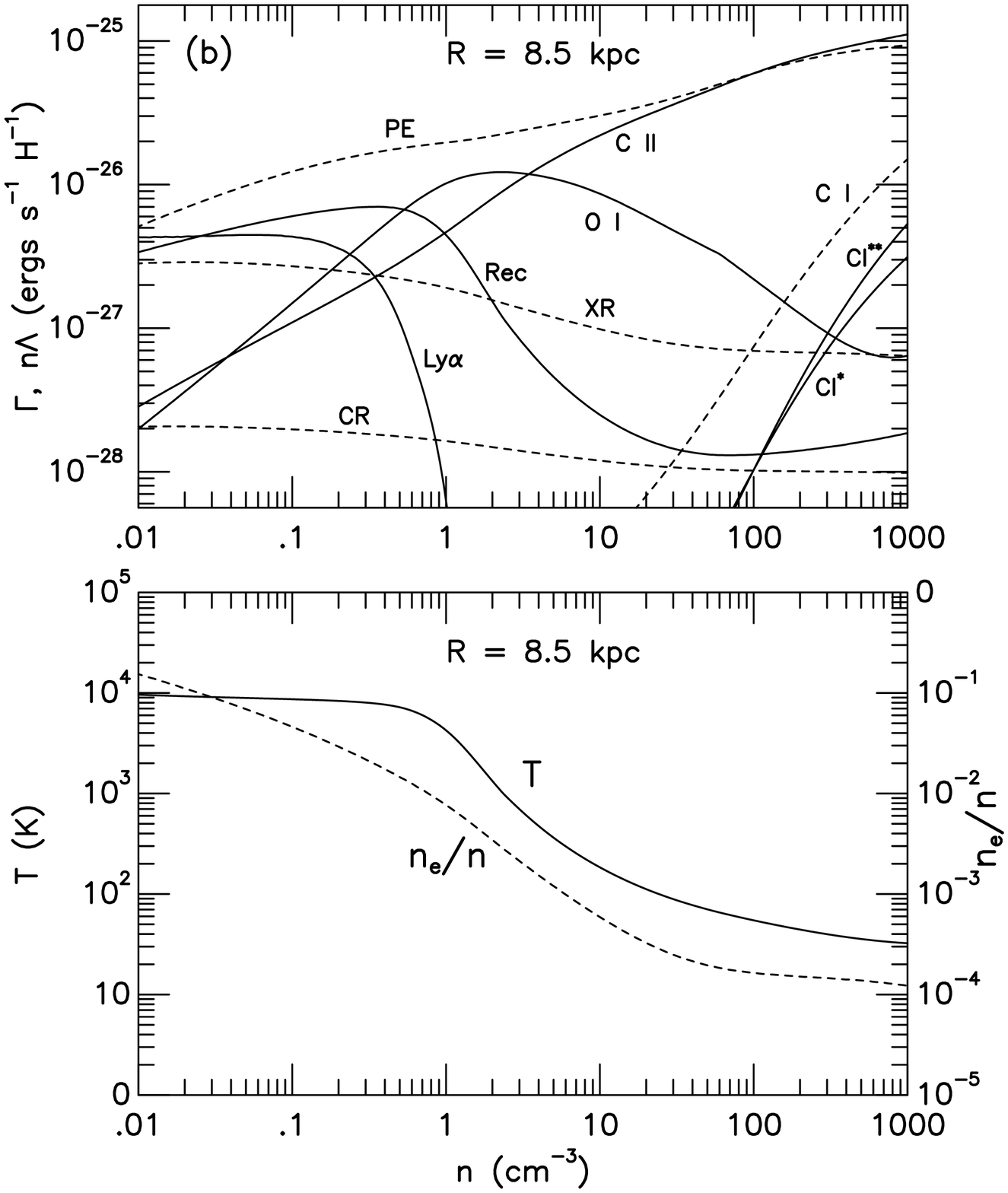}
\newpage
\plotone{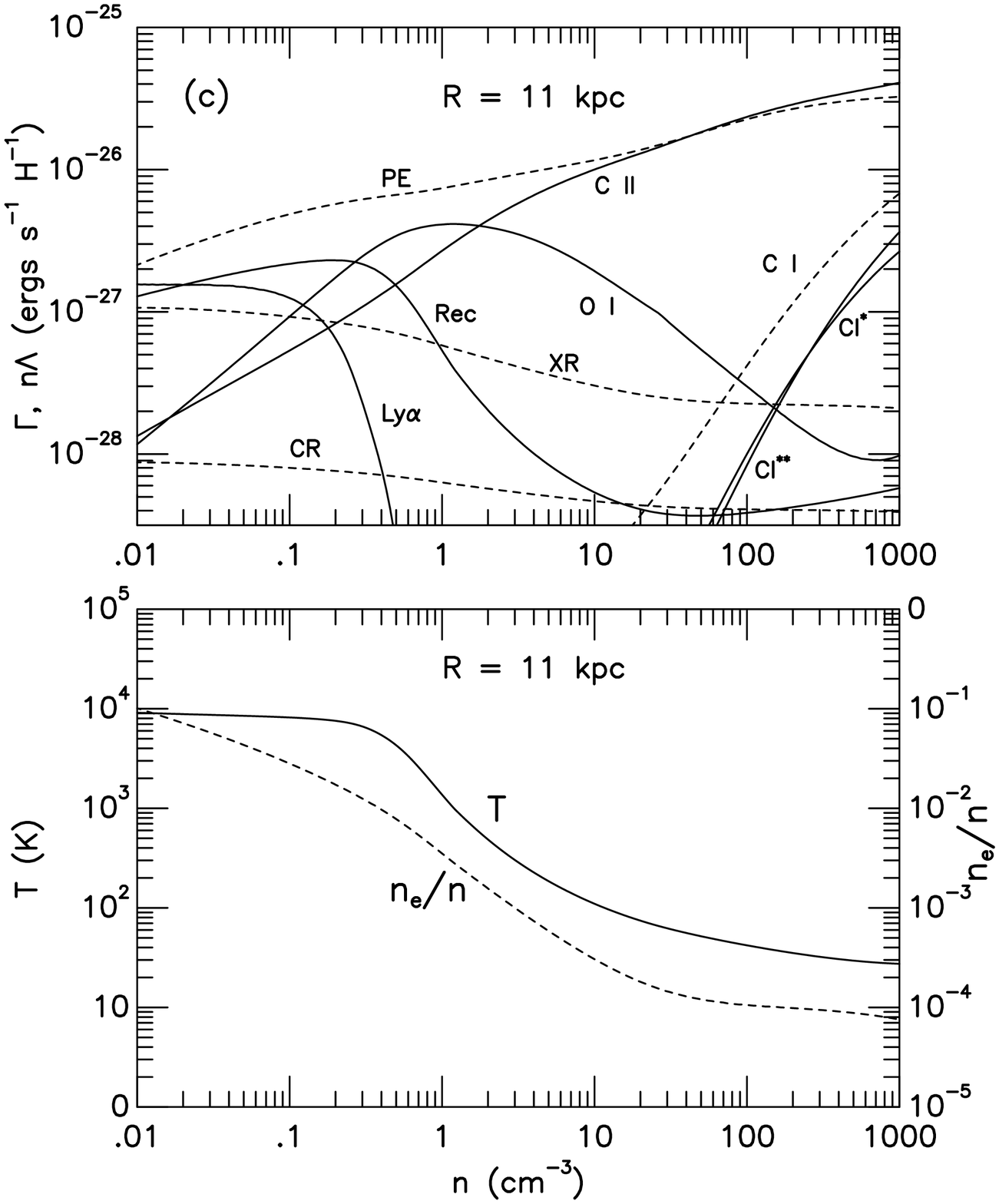}
\newpage
\plotone{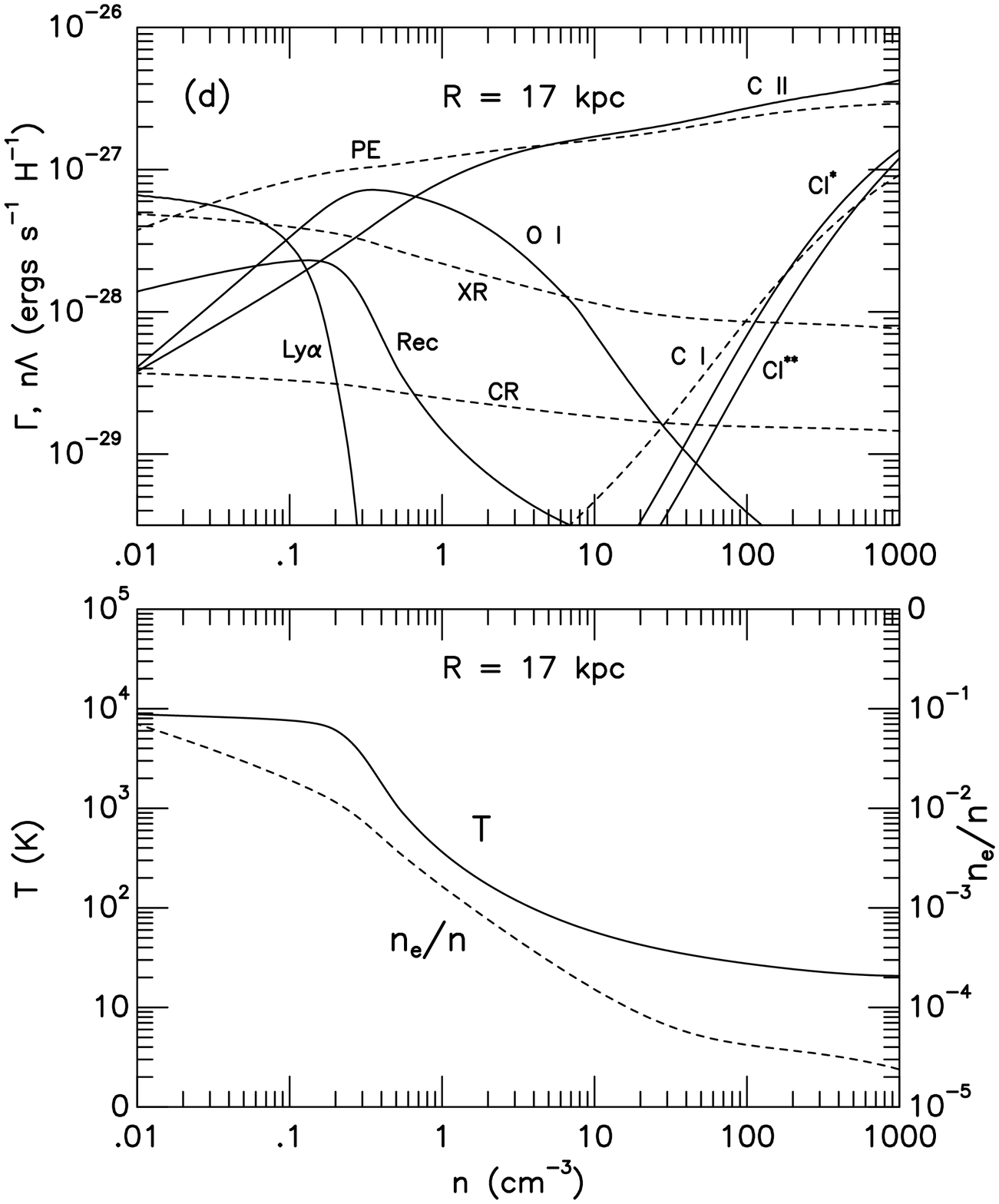}
\newpage
\plotone{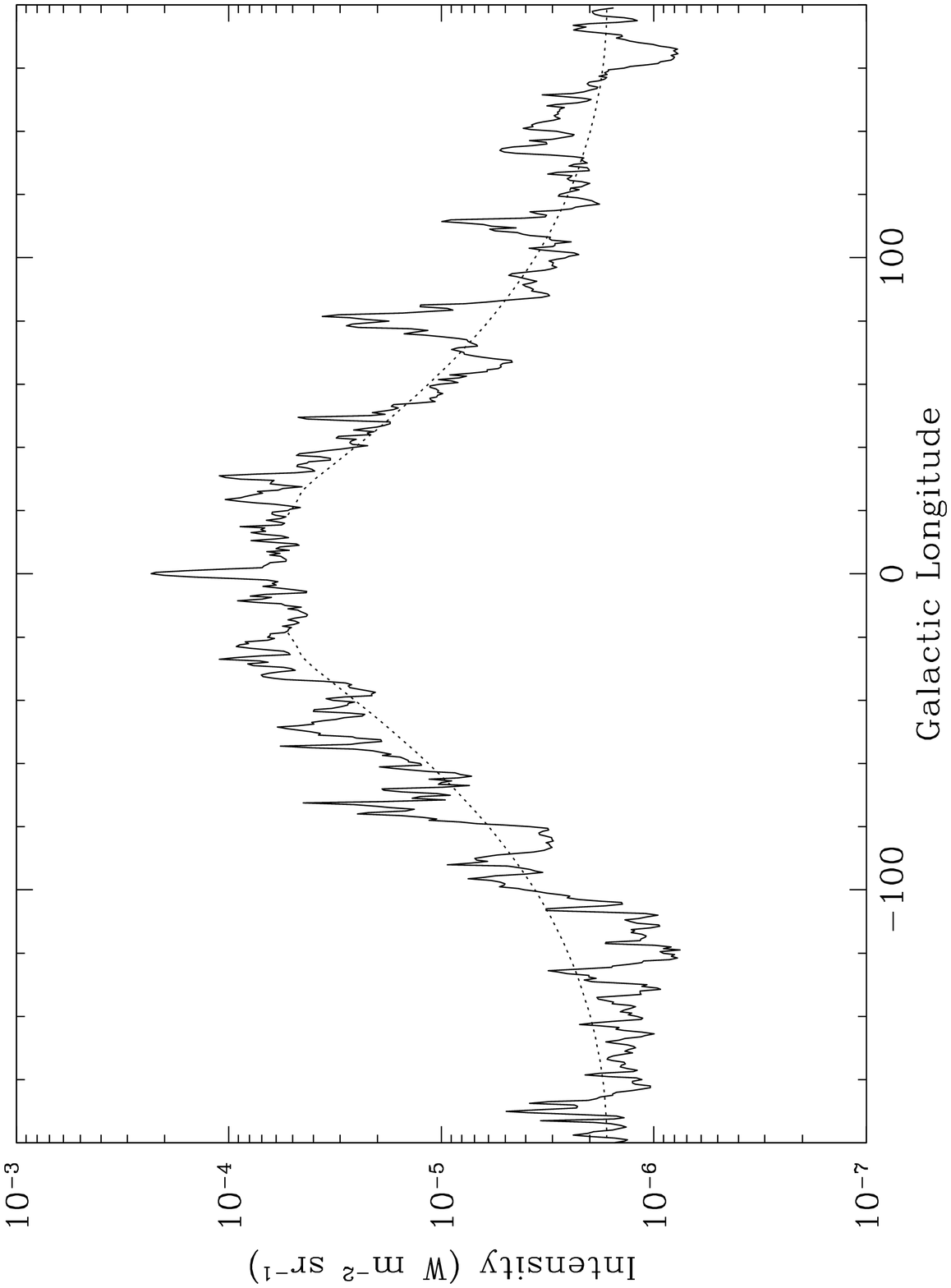}
\newpage
\plotone{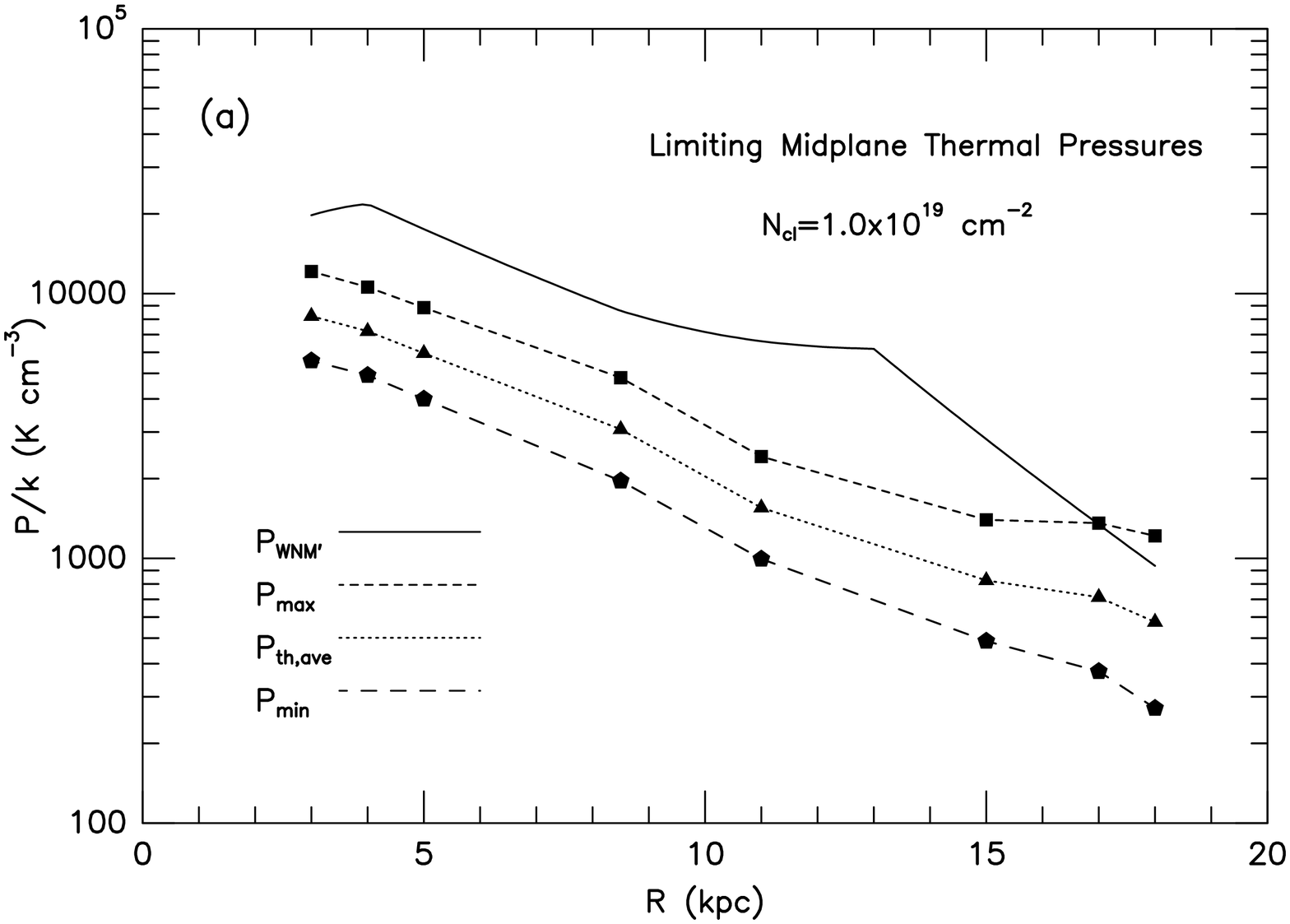}
\newpage
\plotone{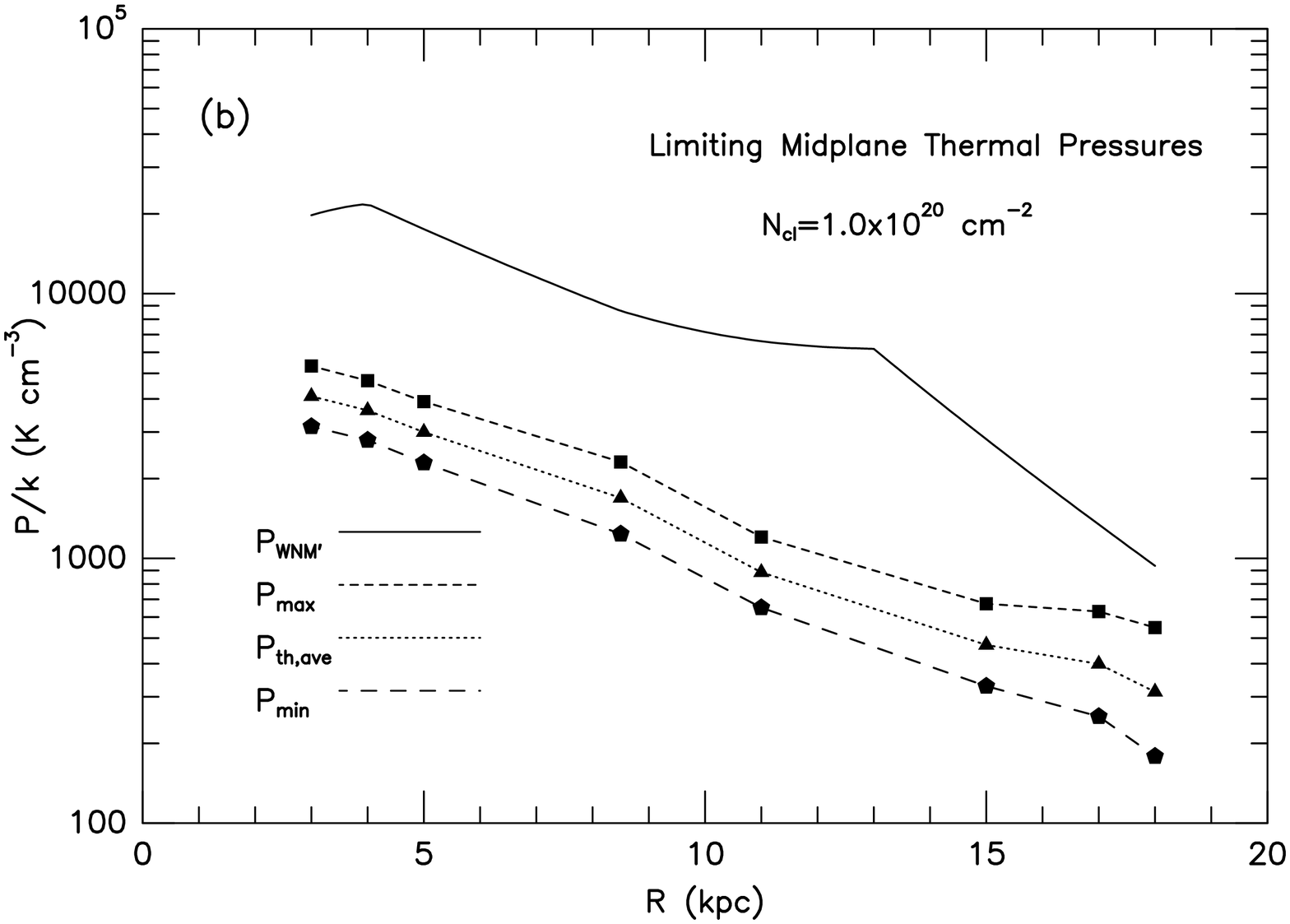}
\newpage
\plotone{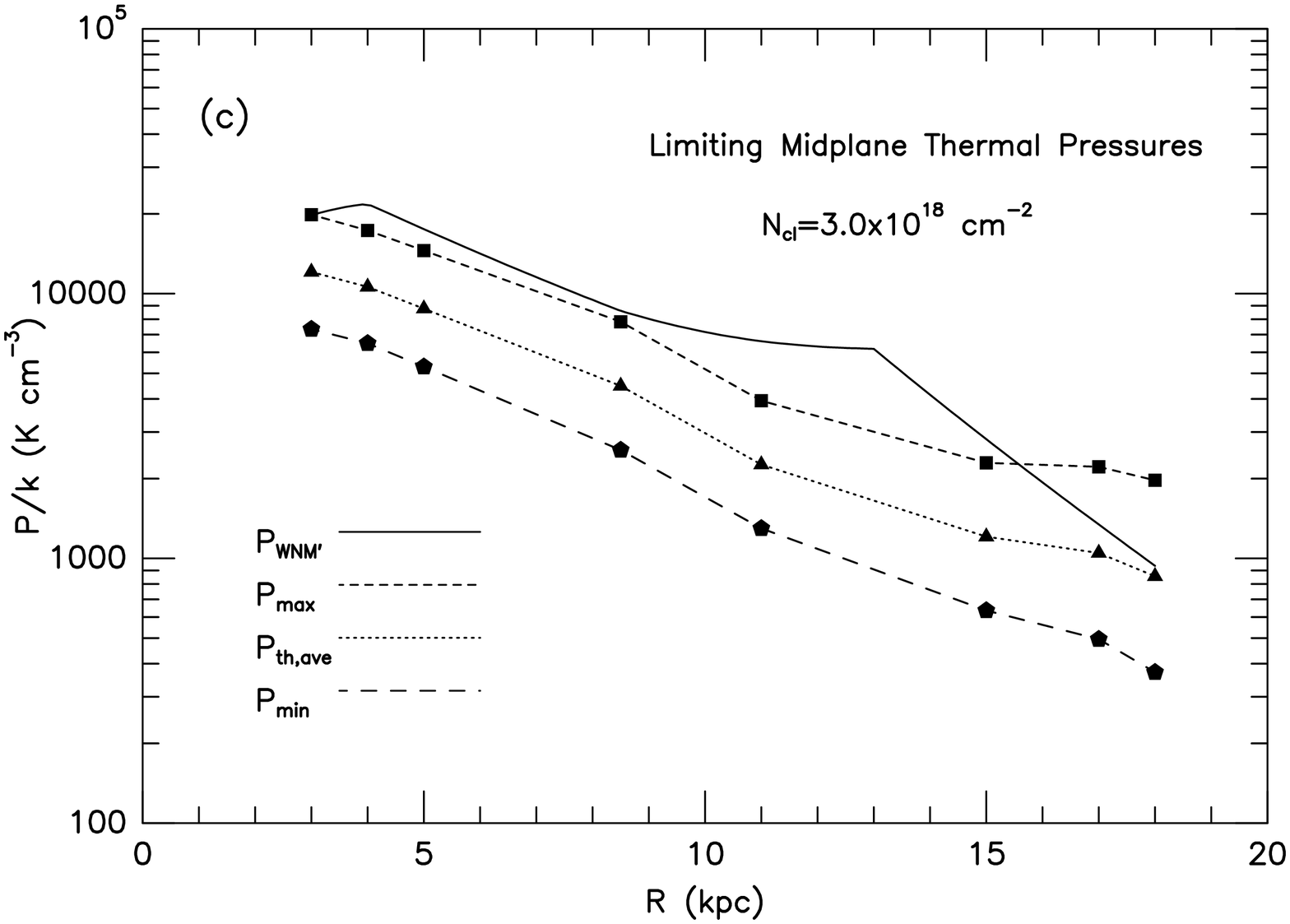}
\newpage
\plotone{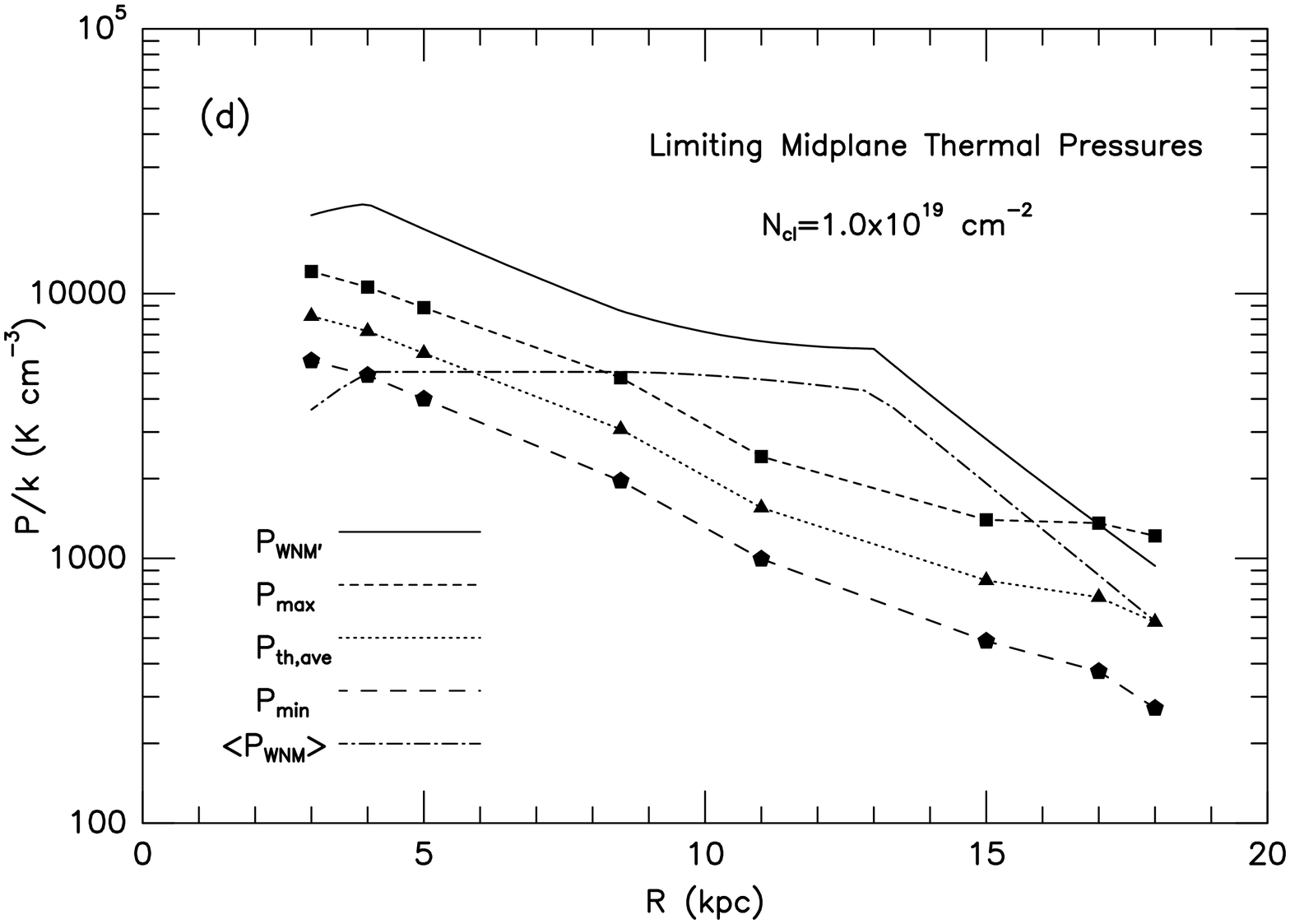}
\newpage
\plotone{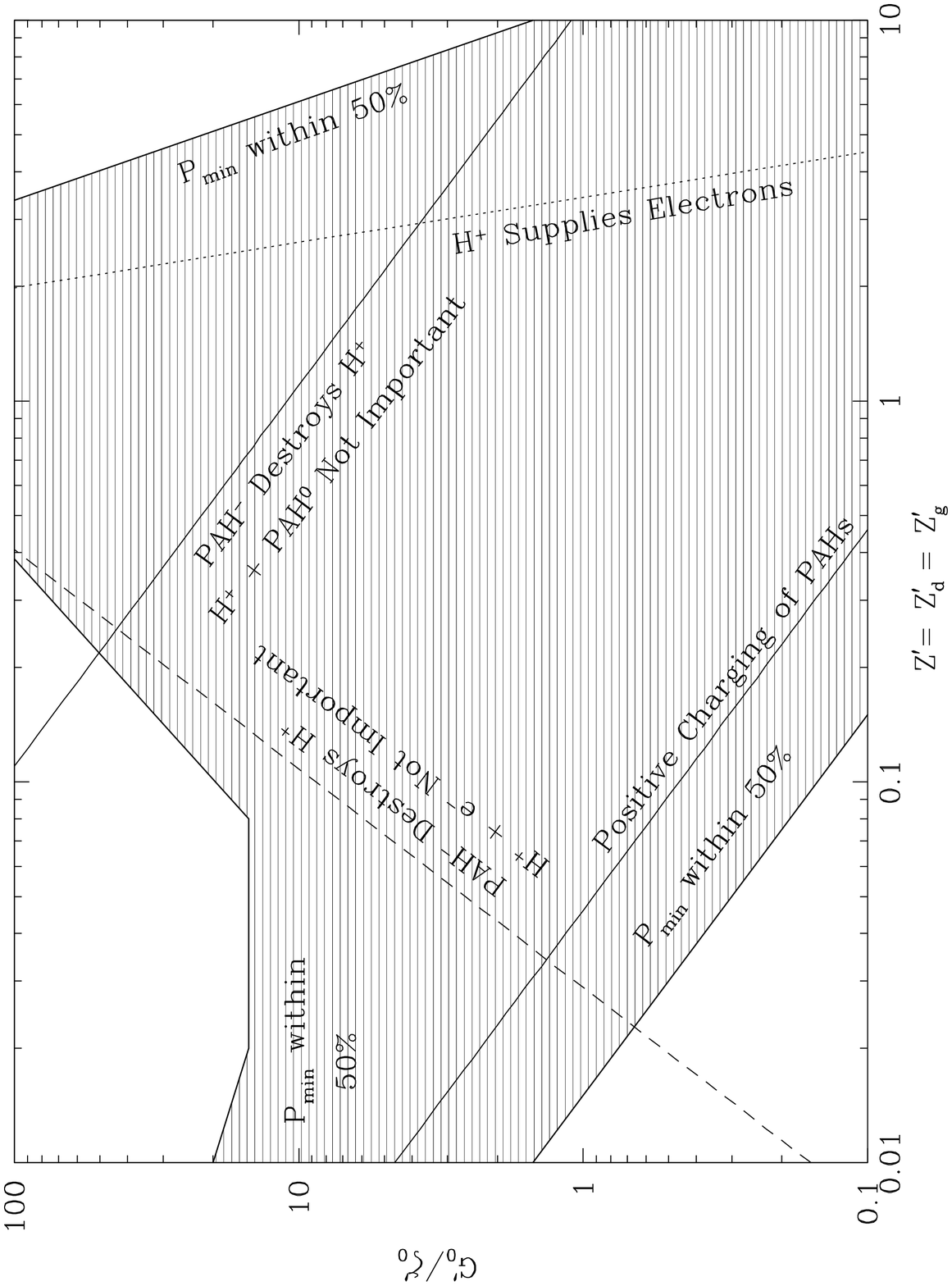}

\end{document}